\documentclass[10pt,twoside]{article}
\usepackage{amssymb}
\usepackage{graphicx}

\begin{document}

\title{Bayesian Analysis of the (Generalized) Chaplygin Gas and Cosmological
Constant Models using the 157 gold SNe Ia Data}
\author{R. Colistete Jr.\thanks{%
e-mail: \texttt{colistete@cce.ufes.br}} and J. C. Fabris\thanks{%
e-mail: \texttt{fabris@cce.ufes.br}} \\
\\
\mbox{\small Universidade Federal do Esp\'{\i}rito Santo,
Departamento
de F\'{\i}sica}\\
\mbox{\small Av. Fernando Ferrari s/n - Campus de Goiabeiras, CEP
29060-900, Vit\'oria, Esp\'{\i}rito Santo, Brazil}}
\date{\today}
\maketitle

\begin{abstract}
The generalized Chaplygin gas model (GCGM) contains 5 free parameters that
must be constrained using the different observational data. These parameters
are: the Hubble constant $H_0$, the parameter $\bar A$ related to the sound
velocity, the equation of state parameter $\alpha$, the curvature parameter $%
\Omega_{k0}$, and the Chaplygin gas density parameter
$\Omega_{c0}$. The pressureless matter parameter $\Omega_{m0}$ may
be obtained as a dependent quantity. Here, these parameters are
constrained through the type Ia supernovae data. The ``gold
sample'' of 157 supernovae data is used. Negative and large
positive values for $\alpha$ are taken into account. The analysis
is made by employing the Bayesian statistics and the prediction
for each parameter is obtained by marginalizing on the remained
ones. This procedure leads to the following predictions: $\alpha =
- 0.75^{+4.04}_{-0.24}$, $H_0
= 65.00^{+1.77}_{-1.75}$, $\Omega_{k0} = - 0.77^{+1.14}_{-5.94}$, $%
\Omega_{m0} = 0.00^{+1.95}_{-0.00}$, $\Omega_{c0} = 1.36^{+5.36}_{-0.85}$, $%
\bar A = 1.000^{+0.000}_{-0.534}$. Through the same analysis the
specific case of the ordinary Chaplygin gas model (CGM), for which
$\alpha = 1$, is studied. In this case, there are now four free
parameters and the predictions for them are: $H_0 =
65.01^{+1.81}_{-1.71}$, $\Omega_{k0} = - 2.73^{+1.53}_{-0.97}$,
$\Omega_{m0} = 0.00^{+1.22}_{-0.00}$, $\Omega_{c0} =
1.34^{+0.94}_{-0.70}$, $\bar A = 1.000^{+0.000}_{-0.270}$. To
complete the analysis the $\Lambda$CDM, with its three free
parameters, is considered. For all these models, particular cases
are considered where one or two parameters are fixed. The age of
the Universe, the deceleration parameter and the moment the
Universe begins to accelerate are also evaluated. The quartessence
scenario, that unifies the description for dark matter and dark
energy, is favoured. A closed (and in some cases a flat) and
accelerating Universe is also preferred. The CGM case $\alpha = 1$
is far from been ruled out, and it is even preferred in some
particular cases. In most of the cases the $\Lambda$CDM is
disfavoured with respect to GCGM and CGM.
\end{abstract}

\section{Introduction}

One of the most intriguing observational recent results in
cosmology concerns the possible accelerating phase of the Universe
today. This result comes from the observational programs of the
type Ia supernovae, carried out since the second half of the last
decade \cite{riess,mutter}. Type Ia supernovae seem to be
excellent standard candles: their detonation mechanism occurs
under very specific conditions and their absolute magnitude can
be, in principle, easily inferred. For a more detailed discussion
on this problem, see reference \cite{filippenko}. The crossing of
these results with those coming from the anisotropy of the cosmic
microwave background radiation \cite {spergel} leads to a scenario
where the matter content of the Universe is composed essentially
of $70\%$ of an unclustered component of negative pressure, the
dark energy, and $30\%$ of a clustered component of zero pressure,
the cold dark matter. The most natural candidate for dark energy
seems to be the cosmological constant, since it can be connected
with the vacuum energy in quantum field theory \cite{weinberg}.
But, the small value resulting from observations for the energy
density of the cosmological constant term, $\rho _{\Lambda
}=10^{-47}\,GeV^{4}$, leads to a discrepancy of about $120$ orders
of magnitude with the theoretically predicted value
\cite{carroll}. Hence, other possibilities have been exploited in
the literature, like quintessence, a kind of an inflaton
self-interacting field adapted to the present phase of the
Universe \cite{stein1,stein2}. More recently, the Chaplygin gas
model (CGM) has been evoked \cite{pasquier}. This model is based
on a string inspired configuration that leads to a specific
equation of state where pressure is negative and varies with the
inverse of the density \cite{jack}. This model has been
generalized, giving birth to the generalized Chaplygin gas model
(GCGM), where now the pressure varies with a power of the inverse
of the density \cite{berto}. These proposals have many advantages,
among which we can quote the following: in spite of presenting a
negative pressure, the sound velocity is positive, what assures
stability \cite{fabris}; these models can unify the description of
dark energy and dark matter, since the fluid can clusters at small
scale, remaining a smooth component at large scales \cite{berto};
the CG has an interesting connection with string theory
\cite{jack}. Some criticisms have been addressed to the GCGM (CGM)
mainly connected with its features related to the power spectrum
for the agglomerated matter \cite{sand}. However, in our opinion,
this specific criticism is not conclusive, since the introduction
of baryons may alleviate the objections presented against the
cosmological scenarios based on the GCGM (GCM) \cite{avelino}.

In order to test these different models for dark energy (and dark
matter) it is essential to compare their predictions against the
different observational results available now. These observational
results refer mainly to the type Ia supernovae, the anisotropy of
the CMBR, the X-ray data of clusters and super-clusters of
galaxies, the gravitational lenses and the matter power spectrum.
Each observational test constrains differently the free parameters
of the models; the crossing of the informations coming from those
observational data can strongly restrict or support a specific
cosmological model. In carrying out this task, a statistical
analysis must be applied for each particular observational
test. One important point is how to perform this statistical
analysis: the final conclusions may, in some cases, depend on the
statistical framework (Bayesian, frequentist, etc.), as well as on
the parameters that are allowed to be free, and how these
parameters are constrained (through a joint probability for two
parameters, minimizing the error function or through a
marginalization of all parameters excepted one, etc.). In some
cases, the different procedures adopted may lead to quite
different conclusions on the best value for a given set of
parameters. The choice of the observational data sample may of
course be important as well.

In preceding works \cite{colistete1,colistete2} we have tested the ordinary
CGM and the GCGM against the type Ia supernovae observational data. The
comparison between theoretical prediction and observational data, in what
concerns type Ia supernovae, is made essentially by computing the
luminosity function $D_{L}$ and re-expressing it as the distance moduli
\begin{equation}
\mu _{0}=5\log \left( \frac{D_{L}}{M\!pc}\right) +25\quad ,  \label{dm}
\end{equation}
in terms of which the observational data are given. In those
works, we employed a Bayesian statistical analysis considering all
possible free parameters, which were in number of $4$ for the CGM
and $5$ for the GCGM. In reference \cite{colistete1}, the same
analysis has been also performed for the model with cosmological
constant and cold dark matter ($\Lambda $CDM), in order to allow a
proper comparison between the different models. The final
estimation for each parameter was obtained through the
marginalization on all other parameters. One limitation of these
previous works was the use of a sub-sample of 26 supernovae data.
This sample seems to be very restrictive when we remember that
there is today the ``gold'' sample, with 157 supernovae, and the
``silver'' sample with $192$ supernovae \cite{riessa}. Samples of
up to $230$ supernovae are now available \cite{tonry}. But the 26
supernovae of those works have a very good quality, and they lead
to a quite small value for the $\chi ^{2}$ fitting parameter:
using the ``gold'' sample the value for $\chi ^{2}$ increases a
little. This reflects, for example, the fact that those larger
samples contain supernovae whose observational status is not very
well established: for example, they contain supernovae with the
almost the same redshift $z$ but with different values for the
luminosity distances, without a superposition of the error bars.
However, the recent works on the subject employ enlarged samples
of supernovae data for obvious reasons: they contain supernovae
with $z\geq 1$, leading to a better estimation of the deceleration
parameter today $q_{0}$; moreover, it is expected that the
dispersion in the parameters estimations can be narrowed when more
supernovae are used. Hence, in this work, we return back to the
problem of estimating parameters but now using the ``gold''
sample. We will do it for the GCGM, the CGM and for the $\Lambda
$CDM model. The statistical method will be the same as in
references \cite{colistete1,colistete2}.

The type Ia supernovae data have been used in many works to
constraint the parameters of the GCGM
\cite{makler}--\cite{bento2004}. In these works, however,
restrictions on some free parameters were introduced by, for
example, fixing the curvature of the spatial section equal to zero
or using a specific value for the Hubble constant as suggested by
the spectrum of the anisotropy of the cosmic microwave background
radiation (CMBR). Moreover, a more simplified statistical
treatment has been frequently employed, like the $\chi ^{2}$
statistics. In almost all of them the parameter $\alpha $ has been
restricted to the interval $0\leq \alpha \leq 1$. This last
restriction has been alleviated in a recent works
\cite{makler,maklerbis,berto2,ygong,bento2004}. For example ref.
\cite{berto2}, where the $\chi ^{2}$ statistics and quartessence
scenario (that unifies dark matter and dark energy into a single
fluid, the Chaplygin gas, so the dark matter content is null) have
been employed, values of $\alpha $ greater than $1$ have been
allowed. The best value found for the parameter $\alpha $ is
$3.75$ for the flat case and $2.87$ for the non-flat case.

The main reason to treat again this problem is to give a complete analysis
of the GCGM, as well as of the CGM and $\Lambda $CDM, using all free
parameters. The gold sample with 157 supernovae will be used. The parameter $%
\alpha $ will be allowed to take any positive or negative value. The
Bayesian statistics will be employed, and the predicted values for each
parameter will be obtained through the marginalization process. This last
step leads to results quite different from those obtained through, for
example, the $\chi ^{2}$ statistics. In fact, even if we agree with the
authors of references \cite{berto2,bento2004} that the minimization method
leads to $\alpha$ positive (in our case we find $\alpha =5.66$ when all
five free parameters are free), the marginalization on the other parameters
indicates that the best value for this parameter is negative, although the
dispersion is quite large and high positive values can not be excluded.

In treating three theoretical models (GCGM, CGM and $\Lambda$CDM) we intend to give a
more unified description of all these different cases. In what concerns the
GCGM, the free parameters to be estimated are the value of the Hubble
constant today $H_0$, the parameters $\alpha$, $\bar A$ (connected with the
sound velocity of the fluid), the curvature parameter $\Omega_{k0}$ and the
GCG parameter $\Omega_{c0}$ (alternatively, the ordinary matter parameter $%
\Omega_{m0}$). In the CGM, the parameter $\alpha$ is fixed to unity. For the $%
\Lambda$CDM model the parameters are $H_0$, $\Omega_{c0}$ (i.e., $%
\Omega_{\Lambda0}$) and $\Omega_{m0}$ (or the curvature parameter $\Omega_{k0}$).
In both cases, we also estimate the age of the Universe $t_0$, the deceleration
parameter $q_0$ and the value of the scale factor at the moment the Universe
begins to accelerate $a_i$ (which can also be expressed in terms of the
redshift $z_i$).

We will exemplify the subtleties connected with the statistical analysis by
displaying the analysis based only on the better $\chi^2$ value, on the
joint probability of two different parameters and by marginalizing on all
parameters excepting one. In order also to better compare with the
literature, we will also consider some particular cases where the spatial
curvature is fixed as flat, or the ordinary matter is fixed equal to $%
\Omega_{m0} = 0.04$ (inspired on the nucleosynthesis results) or $%
\Omega_{m0} = 0$ or fixing two of the parameters.

Our results indicate the following values for the parameters for each model.
For the GCGM we find: $\alpha = - 0.75^{+4.04}_{-0.24}$, $H_0 =
65.00^{+1.77}_{-1.75}$, $\Omega_{k0} = - 0.77^{+1.14}_{-5.94}$, $\Omega_{m0}
= 0.00^{+1.95}_{-0.00}$, $\Omega_{c0} = 1.36^{+5.36}_{-0.85}$, $\bar A =
1.000^{+0.000}_{-0.534}$. For the CGM ($\alpha = 1$) the estimations give: $H_0
= 65.01^{+1.81}_{-1.71}$, $\Omega_{k0} = - 2.73^{+1.53}_{-0.97}$, $%
\Omega_{m0} = 0.00^{+1.22}_{-0.00}$, $\Omega_{c0} = 1.34^{+0.94}_{-0.70}$, $%
\bar A = 1.00^{+0.00}_{-0.270}$. Finally, for the $\Lambda$CDM the
results give: $H_0 = 65.00^{+1.78}_{-1.74}$, $\Omega_{k0} = -
2.12^{+1.96}_{-1.61}$, $\Omega_{m0} = 1.01^{+1.08}_{-0.85}$,
$\Omega_{c0} = 1.36^{+0.92}_{-0.78}$. Hence, our analysis indicate
that the traditional Chaplygin gas model can not be ruled out, at
least in what concerns type Ia supernovae data, and that the
$\Lambda$CDM case is not the preferred one. However, the
dispersions are large enough so that no definitive conclusion can
be made. Only through the crossing with other tests a more
restrictive scenario can come up. However, in crossing the
different tests, a uniform statistical procedure must be used.
This work intends to be the initial step in this program.

This paper is organized as follows. In next section, the model and the
different relevant quantities are set up. In section $3$ we present the
parameter estimation. In section $4$ we present our conclusions.

\section{Definition of the models and of the relevant quantities}

The GCGM is obtained through the introduction of a perfect fluid with an
equation of state given by
\begin{equation}
p = - \frac{A}{\rho^\alpha} \quad ,
\end{equation}
where $A$ and $\alpha$ are constants. When $\alpha = 1$ we
re-obtain the equation of state for the CGM. Henceforth, we will
use mainly the term \textit{generalized Chaplygin gas model},
keeping in mind that when $\alpha = 1$ we have the traditional
\textit{Chaplygin gas model}. In principle, the parameter $\alpha$
is restricted in such a way that $0 \leq \alpha \leq 1$. However,
we will allow $\alpha$ to take negative and large positive values.
Negative values and values greater than $1$ can be potentially
dangerous since they can lead to an imaginary sound velocity and a
sound velocity greater than the velocity of light, respectively.
But, these problems appear more dramatically at perturbative
level. In this case, however, a fundamental description for the
GCG must be employed (for example, using self-interacting scalar
fields) what avoids those drawbacks. The supernovae data tests
mainly the background model, and in this sense to enlarge the
possible values of $\alpha$ does not bring any difficulty.

In the GCGM we introduce also pressureless matter in order to take into
account the presence of baryons in the Universe and also in order to verify
if the unified scenario (where no dark matter is present) is favoured by the
data. Hence, the dynamics of the Universe is driven by the Friedmann's
equation
\begin{eqnarray}
\biggr(\frac{\dot a}{a}\biggl)^2 + \frac{k}{a^2} &=& \frac{8\pi G}{3}\biggr(%
\rho_m + \rho_c\biggl) \quad , \label{be1} \\
\dot\rho_m + 3 \frac{\dot a}{a}\rho_m &=& 0 \quad , \label{be2} \\
\dot\rho_c + 3\frac{\dot a}{a}\biggr(\rho_c - \frac{A}{\rho_c^\alpha}\biggl) %
&=& 0 \quad \label{be3} ,
\end{eqnarray}
where $\rho_m$ and $\rho_c$ stand for the pressureless matter and Chaplygin
gas component, respectively. As usual, $k = 0, 1, - 1$ indicates a flat,
closed and open spatial section.

The equations expressing the conservation law for each fluid (\ref{be2},\ref
{be3}) lead to
\begin{equation}
\rho_m = \frac{\rho_{m0}}{a^3} \quad , \quad \rho_c = \biggr\{A + \frac{B}{%
a^{3(1 + \alpha)}}\biggl\}^{1/(1 + \alpha)} \quad ,
\end{equation}
The value of the scale factor today is taken equal to unity, $a_0 = 1$.
Hence, $\rho_{m0}$ and $\rho_{c0} = \biggr\{A + B\biggl\}^{1/(1 + \alpha)}$
are the pressureless matter and GCG densities today. Eliminating from the
last relation the parameter $B$, the GCG density at any time can be
re-expressed as
\begin{equation}
\rho_c = \rho_{c0}\biggr\{\bar A + \frac{1 - \bar A}{a^{3(1 + \alpha)}} %
\biggl\}^{1/(1 + \alpha)} \quad ,
\end{equation}
where $\bar A = A/\rho_{c0}$. This parameter $\bar A$ is connected with the
sound velocity for the Chaplygin gas today by the relation $v_s^2 =
\alpha\bar A$.

The luminosity distance is given by \cite{weinbergb,coles}
\begin{equation}
d_L = \frac{a_0^2}{a}r_1 \quad ,
\end{equation}
$r_1$ being the co-moving coordinate of the source. Using the expression for
the propagation of light
\begin{equation}
ds^2 = 0 = dt^2 - \frac{a^2dr^2}{1 - kr^2} \quad ,
\end{equation}
and the Friedmann's equation (\ref{be1}), we can re-express the luminosity
distance as
\begin{equation}
d_L = (1 + z)S[f(z)] \quad ,
\end{equation}
where
\begin{equation}
S(x) = x \quad (k = 0) \quad ,\quad S(x) = \sin x \quad (k = 1)
\quad , \quad S(x) = \sinh x \quad (k = - 1)\quad ,
\end{equation}
and
\begin{equation}
f(z) = \frac{1}{H_0}\int_0^z \frac{d\,z^{\prime
}}{\{\Omega_{m0}(z^{\prime}+ 1)^3 + \Omega_{c0}[\bar A +
(z^{\prime}+ 1)^{3(1+\alpha)}(1 - \bar A)]^{1/(1+\alpha)} -
\Omega_{k0}(z^{\prime}+ 1)^2\}^{1/2}} \quad ,
\end{equation}
with the definitions
\begin{equation}
\Omega_{m0} = \frac{8\pi G}{3}\frac{\rho_{m0}}{H_0^2} \quad , \quad
\Omega_{c0} = \frac{8\pi G}{3}\frac{\rho_{c0}}{H_0^2} \quad , \quad
\Omega_{k0} = - \frac{k}{H_0^2} \quad ,
\end{equation}
such that the condition $\Omega_{m0} +\Omega_{c0} + \Omega_{k0} = 1$ holds.
The final equations have been also expressed in terms of the redshift $z = -
1 + \frac{1}{a}$.

The age of the Universe and the value of the decelerated parameter $q_0 = -
\frac{a\ddot a}{\dot a^2}$ are given by
\begin{eqnarray}
\frac{t_{0}}{T}&=&\int_{0}^{z}\frac{d\,z^{\prime }}{(1+z^{\prime })\{\Omega
_{m0}(z^{\prime }+1)^{3}+\Omega _{c0}[\bar{A}+(z^{\prime }+1)^{3(1+\alpha
)}(1-\bar{A})]^{1/(1+\alpha )}-\Omega _{k0}(z^{\prime }+1)^{2}\}^{1/2}}\quad
, \\
q_{0} &=& \frac{\Omega _{m0}+\Omega _{c0}(1-3\bar{A})}{2} ,
\end{eqnarray}
where $T=(100/H_{0})\times 10^{10}$, so that $t_{0}$ has units of
years.

The value $a_{i}$\ of the scale factor signs the start of the
recent accelerating phase of the Universe, it is given by the roots of the
equation
\begin{equation}
\ddot{a}=a(\dot{H}+H^{2})\quad ,
\end{equation}
and is related to the redshift value $z_{i}$ such that $\frac{a_{i}}{a_{0}}=%
\frac{1}{1+z_{i}}$ or $z_{i}=-1+\frac{a_{0}}{a_{i}}$.

Following the same lines sketched above, we can obtain the corresponding
expressions for the $\Lambda$CDM model. However, it is easier just to insert
in the above relations the condition $\bar A = 1$ in order to recover the $%
\Lambda$CDM case.

In order to compare the theoretical results with the observational data, we
must compute the distance moduli, as given by relation (\ref{dm}). A crucial aspect
of the present work is the employment of the Bayesian statistics, which will
be outlined in the next section. The first step in this sense is to compute
the quality of the fitting through the least squared fitting quantity $\chi
^{2}$ defined by
\begin{equation}
\chi ^{2}=\sum_{i}\frac{\left( \mu _{0,i}^{o}-\mu _{0,i}^{t}\right) ^{2}}{%
\sigma _{\mu _{0},i}^{2}}\quad .  \label{Chi2}
\end{equation}
In this expression, $\mu _{0,i}^{o}$ is the measured value, $\mu _{0,i}^{t}$
is the value calculated through the model described above, $\sigma _{\mu%
_{0},i}^{2}$ is the measurement error and includes the dispersion in the
distance modulus due to the dispersion in galaxy redshift due to peculiar
velocities, following ref. \cite{riessa}.

\section{Parameter estimations}

To constraint the five independent parameters for the GCGM, the four
independent parameters for the CGM and the three parameters for
$\Lambda $CDM, we use the Bayesian statistical analysis. The method
and its motivation are described in detail in ref. \cite{colistete1}.
Since there is no prior constraint, the probability of the set of
distance moduli $\mu_{0}$ conditional on the values of a set of
parameters $\{a_{i}\}$ is given by a product of Gaussian functions:
\begin{equation}
p(\mu _{0}|\{a_{i}\})=\prod_{i}\frac{1}{\sqrt{2\pi \,\sigma _{\mu
_{0},i}^{2}}}\exp \biggr[-\frac{\left( \mu _{0,i}^{o}-\mu _{0,i}^{t}\right) ^{2}}%
{2 \,\sigma _{\mu _{0},i}^{2}}\biggl]
\label{pd1}
\end{equation}
This probability distribution must be normalized. Evidently, when,
for a set of values of the parameters, the $\chi ^{2}$ is minimum
the probability is maximum. This is a valuable information but is
not enough to constraint the parameters. In table $1$ the values
of the parameters for the maximum probability (minimum $\chi
^{2}$) is given, using the gold sample of supernovae, for the GCGM
with five free parameters and for other cases where the baryonic,
the curvature or both are fixed. The same estimations are
presented in tables $2$ and $3$ for the CGM and $\Lambda $CDM.
Note first that the minimum values for $\chi ^2$ using the ``gold
sample'' are higher than the corresponding ones using the
restricted sample of 26 supernovae \cite{colistete1,colistete2}.
Note also that from this point of view, the best value for the
parameter $\alpha $ is much bigger than $1$. These results must be
compared with a more complete analysis to be presented below. For
more details about the $\chi ^2$ minimization process, see refs.
\cite{colistete1,colistete2}.

\begin{table}[t]
\begin{center}
\begin{tabular}{|c|c|c|c|c|c|c|}
\hline\hline
&  &  &  &  &  &  \\[-7pt]
& GCGM & GCGM : & GCGM : & GCGM : & GCGM : $k=0$, & GCGM : $k=0$, \\
&  & $k=0$ & $\Omega _{m0}=0$ & $\Omega _{m0}=0.04$ & $\Omega _{m0}=0$ & $%
\Omega _{m0}=0.04$ \\[2pt] \hline
&  &  &  &  &  &  \\[-7pt]
$\chi _{\nu }^{2}$ & $1.1089$ & $1.1094$ & $1.1089$ & $1.1093$ & $1.1094$ & $%
1.1097$ \\[2pt] \hline
&  &  &  &  &  &  \\[-7pt]
$\alpha $ & $5.66$ & $2.85$ & $5.73$ & $5.24$ & $2.85$ & $3.07$ \\%
[2pt] \hline
&  &  &  &  &  &  \\[-7pt]
$H_{0}$ & $65.03$ & $65.13$ & $65.02$ & $65.01$ & $65.13$ & $65.09$ \\%
[2pt] \hline
&  &  &  &  &  &  \\[-7pt]
$\Omega _{k0}$ & $0.179$ & $0$ & $0.181$ & $0.166$ & $0$ & $0$ \\[2pt] \hline
&  &  &  &  &  &  \\[-7pt]
$\Omega _{m0}$ & $0.00$ & $0.00$ & $0$ & $0.04$ & $0$ & $0.04$ \\[2pt] \hline
&  &  &  &  &  &  \\[-7pt]
$\Omega _{c0}$ & $0.821$ & $1.00$ & $0.819$ & $0.830$ & $1$ & $0.96$ \\%
[2pt] \hline
&  &  &  &  &  &  \\[-7pt]
$\bar{A}$ & $0.991$ & $0.929$ & $0.992$ & $0.991$ & $0.929$ & $0.949$ \\%
[2pt] \hline
&  &  &  &  &  &  \\[-7pt]
$t_{0}$ & $13.81$ & $13.55$ & $13.81$ & $13.73$ & $13.55$ & $13.55$ \\%
[2pt] \hline
&  &  &  &  &  &  \\[-7pt]
$q_{0}$ & $-0.810$ & $-0.894$ & $-0.894$ & $-0.799$ & $-0.894$ & $-0.866$ \\%
[2pt] \hline
&  &  &  &  &  &  \\[-7pt]
$a_{i}$ & $0.762$ & $0.753$ & $0.762$ & $0.756$ & $0.760$ & $0.751$ \\%
[2pt] \hline\hline
\end{tabular}
\end{center}
\caption{The best-fitting parameters, i.e., when $\protect\chi _{\protect\nu
}^{2}$ is minimum, for each type of spatial section and matter content of
the generalized Chaplygin gas model. $H_{0}$ is given in $km/M\!pc.s$, $\bar{%
A}$ in units of $c$, $t_{0}$ in $Gy$ and $a_{i}$ in units of $a_{0}$. }
\label{tableBestFitGCG}
\end{table}

\begin{table}[h!]
\begin{center}
\begin{tabular}{|c|c|c|c|c|c|c|}
\hline\hline
&  &  &  &  &  &  \\[-7pt]
& CGM & CGM : & CGM : & CGM : & CGM : $k=0$, & CGM : $k=0$, \\
&  & $k=0$ & $\Omega _{m0}=0$ & $\Omega _{m0}=0.04$ & $\Omega _{m0}=0$ & $%
\Omega _{m0}=0.04$ \\[2pt] \hline
&  &  &  &  &  &  \\[-7pt]
$\chi _{\nu }^{2}$ & $1.1107$ & $1.1141$ & $1.1107$ & $1.1108$ & $1.1141$ & $%
1.1147$ \\[2pt] \hline
&  &  &  &  &  &  \\[-7pt]
$H_{0}$ & $65.08$ & $64.73$ & $65.06$ & $65.06$ & $64.73$ & $64.70$ \\%
[2pt] \hline
&  &  &  &  &  &  \\[-7pt]
$\Omega _{k0}$ & $-0.38$ & $0$ & $-0.36$ & $-0.38$ & $0$ & $0$ \\[2pt] \hline
&  &  &  &  &  &  \\[-7pt]
$\Omega _{m0}$ & $0.00$ & $0.00$ & $0$ & $0.04$ & $0$ & $0.04$ \\[2pt] \hline
&  &  &  &  &  &  \\[-7pt]
$\Omega _{c0}$ & $1.38$ & $1.00$ & $1.36$ & $1.34$ & $1$ & $0.96$ \\%
[2pt] \hline
&  &  &  &  &  &  \\[-7pt]
$\bar{A}$ & $0.761$ & $0.811$ & $0.764$ & $0.778$ & $0.811$ & $0.834$ \\%
[2pt] \hline
&  &  &  &  &  &  \\[-7pt]
$t_{0}$ & $13.29$ & $14.01$ & $13.34$ & $13.32$ & $14.01$ & $14.03$ \\%
[2pt] \hline
&  &  &  &  &  &  \\[-7pt]
$q_{0}$ & $-0.888$ & $-0.717$ & $-0.876$ & $-0.872$ & $-0.717$ & $-0.701$ \\%
[2pt] \hline
&  &  &  &  &  &  \\[-7pt]
$a_{i}$ & $0.734$ & $0.699$ & $0.732$ & $0.732$ & $0.699$ & $0.695$ \\%
[2pt] \hline\hline
\end{tabular}
\end{center}
\caption{The best-fitting parameters, i.e., when $\protect\chi _{\protect\nu
}^{2}$ is minimum, for each type of spatial section and matter content of
the traditional Chaplygin gas model. $H_{0}$ is given in $km/M\!pc.s$, $\bar{%
A}$ in units of $c$, $t_{0}$ in $Gy$ and $a_{i}$ in units of $a_{0}$. }
\label{tableBestFitCG}
\end{table}

\begin{table}[t]
\begin{center}
\begin{tabular}{|c|c|c|c|c|}
\hline\hline
&  &  &  &  \\[-7pt]
& $\Lambda CDM$ & $\Lambda CDM$ : & $\Lambda CDM$ : & $\Lambda CDM$ : \\
&  & $k=0$ & $\Omega _{m0}=0$ & $\Omega _{m0}=0.04$ \\[2pt] \hline
&  &  &  &  \\[-7pt]
$\chi _{\nu }^{2}$ & $1.1123$ & $1.1279$ & $1.1481$ & $1.1449$ \\[2pt] \hline
&  &  &  &  \\[-7pt]
$H_{0}$ & $65.00$ & $64.32$ & $64.05$ & $64.10$ \\[2pt] \hline
&  &  &  &  \\[-7pt]
$\Omega _{k0}$ & $-1.34$ & $0$ & $0.60$ & $0.56$ \\[2pt] \hline
&  &  &  &  \\[-7pt]
$\Omega _{m0}$ & $1.00$ & $0.31$ & $0$ & $0.04$ \\[2pt] \hline
&  &  &  &  \\[-7pt]
$\Omega _{c0}$ & $1.34$ & $0.69$ & $0.40$ & $0.44$ \\[2pt] \hline
&  &  &  &  \\[-7pt]
$t_{0}$ & $13.02$ & $14.87$ & $18.41$ & $17.17$ \\[2pt] \hline
&  &  &  &  \\[-7pt]
$q_{0}$ & $-0.839$ & $-0.537$ & $-0.400$ & $-0.420$ \\[2pt] \hline
&  &  &  &  \\[-7pt]
$a_{i}$ & $0.720$ & $0.607$ & $0$ & $0.357$ \\[2pt] \hline\hline
\end{tabular}
\end{center}
\caption{The best-fitting parameters, i.e., when $\protect\chi _{\protect\nu
}^{2}$ is minimum, for each type of spatial section and matter content of
the $\Lambda$CDM model. $H_{0}$ is given in $km/M\!pc.s$, $t_{0}$ in $Gy$
and $a_{i}$ in units of $a_{0}$. }
\label{tableBestFitLCDM}
\end{table}

\begin{figure}[t]
\begin{center}
\includegraphics[scale=0.8]{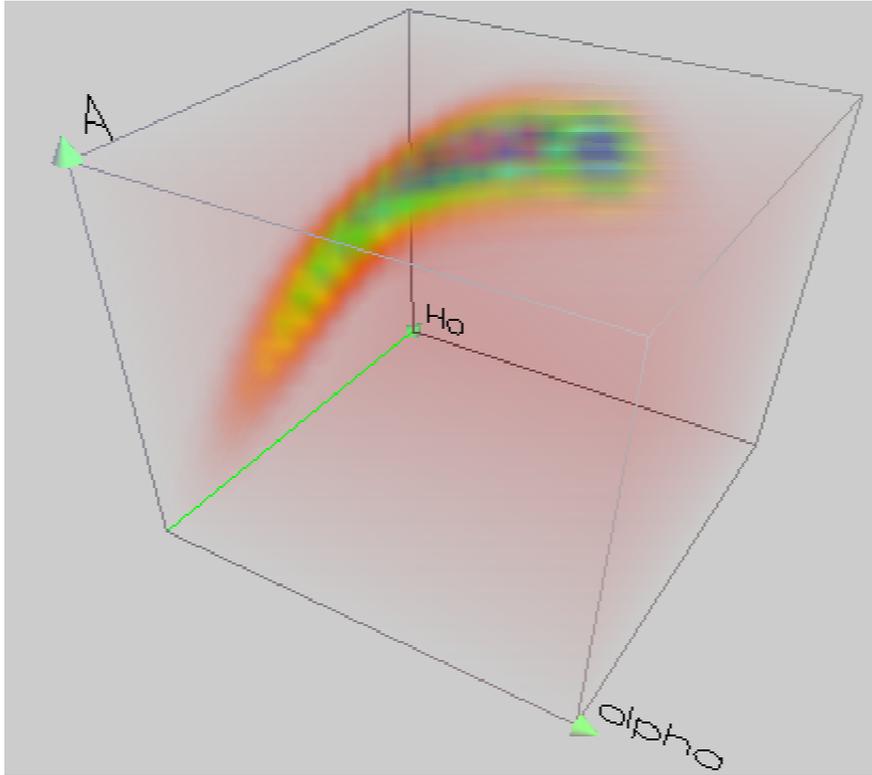}
\end{center}
\caption{{\protect\footnotesize The graphics of the joint PDF as function of
$(\protect\alpha ,H_{0},\bar{A})$ for the generalized Chaplygin gas model
when fixing\ $k=0$ and $\Omega _{m0}=0.04$. It is a $3D$ density plot where
the function (here the PDF) is rendered as semi-transparent colourful gas,
or we can say that the plot is made by voxels (volume elements) analogous to
pixels (picture elements). Where the PDF is minimum, the transparency is
total and the colour is red, where the PDF is maximum then the voxel is
opaque and the colour is violet; mid-range values are semi-transparent and
coloured between red and violet (red, orange, yellow, green, blue, violet).
The parameter ranges are $-1<\protect\alpha \leqslant 9$, $62\leqslant
H_{0}\leqslant 68$ and $0.5\leqslant \bar{A}\leqslant 1$, with resolutions $%
0.25\times 0.5\times 0.01$, respectively. The joint PDF of $(\protect\alpha
,H_{0},\bar{A})$ for other cases ($5$ free parameters, $4$ free parameters,
etc.)\ show a similar $3D$ shape. }}
\label{figPdf3Dm04}
\end{figure}

\begin{table}[t]
\begin{center}
\begin{tabular}{|c|c|c|c|c|c|c|}
\hline\hline
&  &  &  &  &  &  \\[-7pt]
& GCGM & GCGM : & GCGM : & GCGM : & GCGM : $k=0$, & GCGM : $k=0$, \\
&  & $k=0$ & $\Omega_{m0}=0$ & $\Omega_{m0}=0.04$ & $\Omega_{m0}=0$ & $%
\Omega_{m0}=0.04$ \\[2pt] \hline
&  &  &  &  &  &  \\[-7pt]
$\alpha$ & $-0.75_{-0.24}^{+4.04}$ & $1.18_{-2.18}^{+4.12}$ & $%
-0.05_{-0.79}^{+5.36}$ & $-0.10_{-0.76}^{+5.10}$ & $1.57_{-1.97}^{+4.86}$ & $%
1.52_{-1.97}^{+4.53}$ \\[2pt] \hline
&  &  &  &  &  &  \\[-7pt]
$H_{0}$ & $65.00_{-1.75}^{+1.77}$ & $64.51_{-1.56}^{+1.64}$ & $%
64.92_{-1.82}^{+1.80}$ & $64.91_{-1.81}^{+1.80}$ & $64.92_{-1.65}^{+1.55}$ &
$64.82_{-1.61}^{+1.54}$ \\[2pt] \hline
&  &  &  &  &  &  \\[-7pt]
$\Omega_{k0}$ & $-0.77_{-5.94}^{+1.14}$ & $0$ & $0.00_{-2.94}^{+0.66}$ & $%
0.06_{-3.03}^{+0.66}$ & $0$ & $0$ \\[2pt] \hline
&  &  &  &  &  &  \\[-7pt]
$\Omega_{m0}$ & $0.00_{-0.00}^{+1.95}$ & $0.00_{-0.00}^{+0.30}$ & $0$ & $%
0.04 $ & $0$ & $0.04$ \\[2pt] \hline
&  &  &  &  &  &  \\[-7pt]
$\Omega_{c0}$ & $1.36_{-0.85}^{+5.36}$ & $1.00_{-0.30}^{+0.0}$ & $%
1.00_{-0.66}^{+2.94}$ & $1.02_{-0.66}^{+3.03}$ & $1$ & $0.96$ \\[2pt] \hline
&  &  &  &  &  &  \\[-7pt]
$\bar{A}$ & $1.000_{-0.534}^{+0.000}$ & $0.989_{-0.245}^{+0.011}$ & $%
0.988_{-0.488}^{+0.012}$ & $0.988_{-0.522}^{+0.012}$ & $%
0.987_{-0.293}^{+0.013}$ & $0.987_{-0.277}^{+0.012}$ \\[2pt] \hline
&  &  &  &  &  &  \\[-7pt]
$t_{0}$ & $12.63_{-1.19}^{+2.04}$ & $13.70_{-0.88}^{+1.82}$ & $%
13.28_{-1.46}^{+2.31}$ & $13.16_{-1.40}^{+2.48}$ & $13.40_{-0.73}^{+1.18}$ &
$13.54_{-0.64}^{+1.48}$ \\[2pt] \hline
&  &  &  &  &  &  \\[-7pt]
$q_{0}$ & $-0.818_{-0.459}^{+0.381}$ & $-0.711_{-0.247}^{+0.298}$ & $%
-0.787_{-0.446}^{+0.431}$ & $-0.828_{-0.480}^{+0.461}$ & $%
-0.878_{-0.062}^{+0.388}$ & $-0.881_{-0.059}^{+0.370}$ \\[2pt] \hline
&  &  &  &  &  &  \\[-7pt]
$a_{i}$ & $0.746_{-0.115}^{+0.057}$ & $0.721_{-0.176}^{+0.121}$ & $%
0.762_{-0.171}^{+0.087}$ & $0.765_{-0.195}^{+0.102}$ & $%
0.770_{-0.132}^{+0.065}$ & $0.757_{-0.152}^{+0.066}$ \\[2pt] \hline
&  &  &  &  &  &  \\[-7pt]
$p(\alpha>0)$ & $41.98\,\%$ & $87.65\,\%$ & $81.59\,\%$ & $79.34\,\%$ & $%
96.40\,\%$ & $95.52\,\%$ \\[2pt] \hline
&  &  &  &  &  &  \\[-7pt]
$p(\alpha=1)$ & $19.58\,\%$ & $91.09\,\%$ & $59.93\,\%$ & $55.78\,\%$ & $%
69.98\,\%$ & $72.70\,\%$ \\[2pt] \hline
&  &  &  &  &  &  \\[-7pt]
$p(\Omega_{k0}<0)$ & $96.29\,\%$ & $-$ & $70.35\,\%$ & $75.19\,\%$ & $-$ & $-
$ \\[2pt] \hline
&  &  &  &  &  &  \\[-7pt]
$p(\Omega_{k0}=0)$ & $37.04\,\%$ & $-$ & $3.35\,\sigma$ & $90.80\,\%$ & $-$
& $-$ \\[2pt] \hline
&  &  &  &  &  &  \\[-7pt]
$p(\bar{A} \neq 1)$ & $0.00\,\%$ & $59.15\,\%$ & $99.96\,\%$ & $98.57\,\%$ &
$100\,\%$ & $100\,\%$ \\[2pt] \hline
&  &  &  &  &  &  \\[-7pt]
$p(q_{0}<0)$ & $6.61\,\sigma$ & $6.99\,\sigma$ & $100\,\%$ & $100\,\%$ & $%
100\,\%$ & $100\,\%$ \\[2pt] \hline
&  &  &  &  &  &  \\[-7pt]
$p(a_{i}<1)$ & $6.88\,\sigma$ & $7.49\,\sigma$ & $100\,\%$ & $100\,\%$ & $%
100\,\%$ & $100\,\%$ \\[2pt] \hline\hline
\end{tabular}
\end{center}
\caption{The estimated parameters for the generalized Chaplygin gas model
(GCGM) and some specific cases of spatial section and matter content. We use
the Bayesian analysis to obtain the peak of the one-dimensional marginal
probability and the $2\,\protect\sigma $ credible region for each parameter.
$H_{0}$ is given in $km/M\!pc.s$, $\bar{A}$ in units of $c$, $t_{0}$ in $Gy$
and $a_{i}$ in units of $a_{0}$.}
\label{tableParEstGCG}
\end{table}

\begin{table}[t]
\begin{center}
\begin{tabular}{|c|c|c|c|c|c|c|}
\hline\hline
&  &  &  &  &  &  \\[-7pt]
& CGM & CGM : & CGM : & CGM : & CGM : $k=0$, & CGM : $k=0$, \\
&  & $k=0$ & $\Omega _{m0}=0$ & $\Omega _{m0}=0.04$ & $\Omega _{m0}=0$ & $%
\Omega _{m0}=0.04$ \\[2pt] \hline
&  &  &  &  &  &  \\[-7pt]
$H_{0}$ & $65.01_{-1.71}^{+1.81}$ & $64.48_{-1.53}^{+1.53}$ & $%
64.93_{-1.77}^{+1.81}$ & $64.93_{-1.76}^{+1.80}$ & $64.67_{-1.52}^{+1.53}$ &
$64.64_{-1.51}^{+1.52}$ \\[2pt] \hline
&  &  &  &  &  &  \\[-7pt]
$\Omega _{k0}$ & $-2.73_{-0.97}^{+1.53}$ & $0$ & $-0.22_{-1.10}^{+0.77}$ & $%
-0.24_{-1.09}^{+0.73}$ & $0$ & $0$ \\[2pt] \hline
&  &  &  &  &  &  \\[-7pt]
$\Omega _{m0}$ & $0.00_{-0.00}^{+1.22}$ & $0.00_{-0.00}^{+0.29}$ & $0$ & $%
0.04$ & $0$ & $0.04$ \\[2pt] \hline
&  &  &  &  &  &  \\[-7pt]
$\Omega _{c0}$ & $1.34_{-0.70}^{+0.94}$ & $1.00_{-0.29}^{+0.00}$ & $%
1.22_{-0.77}^{+1.10}$ & $1.20_{-0.73}^{+1.09}$ & $1$ & $0.96$ \\[2pt] \hline
&  &  &  &  &  &  \\[-7pt]
$\bar{A}$ & $1.000_{-0.270}^{+0.000}$ & $0.857_{-0.072}^{+0.141}$ & $%
0.746_{-0.080}^{+0.165}$ & $0.759_{-0.083}^{+0.166}$ & $%
0.812_{-0.071}^{+0.056}$ & $0.834_{-0.072}^{+0.056}$ \\[2pt] \hline
&  &  &  &  &  &  \\[-7pt]
$t_{0}$ & $12.73_{-1.14}^{+1.81}$ & $14.07_{-0.62}^{+0.77}$ & $%
13.20_{-1.31}^{+2.16}$ & $13.20_{-1.31}^{+1.99}$ & $14.02_{-0.59}^{+0.64}$ &
$14.01_{-0.57}^{+0.68}$ \\[2pt] \hline
&  &  &  &  &  &  \\[-7pt]
$q_{0}$ & $-0.883_{-0.429}^{+0.382}$ & $-0.654_{-0.113}^{+0.158}$ & $%
-0.839_{-0.454}^{+0.417}$ & $-0.860_{-0.428}^{+0.420}$ & $%
-0.715_{-0.085}^{+0.109}$ & $-0.700_{-0.082}^{+0.104}$ \\[2pt] \hline
&  &  &  &  &  &  \\[-7pt]
$a_{i}$ & $0.739_{-0.088}^{+0.068}$ & $0.698_{-0.075}^{+0.056}$ & $%
0.754_{-0.136}^{+0.081}$ & $0.724_{-0.111}^{+0.090}$ & $%
0.705_{-0.050}^{+0.045}$ & $0.709_{-0.53}^{+0.33}$ \\[2pt] \hline
&  &  &  &  &  &  \\[-7pt]
$p(\Omega _{k0}<0)$ & $3.59\,\sigma $ & $-$ & $74.27\,\%$ & $77.38\,\%$ & $-$
& $-$ \\[2pt] \hline
&  &  &  &  &  &  \\[-7pt]
$p(\Omega _{k0} = 0)$ & $0.04\,\%$ & $-$ & $64.54\,\%$ & $59.18\,\%$ & $-$ &
$-$ \\[2pt] \hline
&  &  &  &  &  &  \\[-7pt]
$p(\bar{A}\neq 1)$ & $0.00\,\%$ & $96.02\,\%$ & $3.33\,\sigma $ & $%
3.23\,\sigma $ & $100\,\%$ & $100\,\%$ \\[2pt] \hline
&  &  &  &  &  &  \\[-7pt]
$p(q_{0}<0)$ & $7.59\,\sigma $ & $100\,\%$ & $100\,\%$ & $100\,\%$ & $100\,\%
$ & $100\,\%$ \\[2pt] \hline
&  &  &  &  &  &  \\[-7pt]
$p(a_{i}<1)$ & $7.66\,\sigma $ & $100\,\%$ & $100\,\%$ & $100\,\%$ & $100\,\%
$ & $100\,\%$ \\[2pt] \hline\hline
\end{tabular}
\end{center}
\caption{The estimated parameters for the traditional Chaplygin gas model
(CGM) and some specific cases of spatial section and matter content. We use
the Bayesian analysis to obtain the peak of the one-dimensional marginal
probability and the $2\,\protect\sigma $ credible region for each parameter.
$H_{0}$ is given in $km/M\!pc.s$, $\bar{A}$ in units of $c$, $t_{0}$ in $Gy$
and $a_{i}$ in units of $a_{0}$. }
\label{tableParEstCG}
\end{table}

\begin{table}[t]
\begin{center}
\begin{tabular}{|c|c|c|c|c|}
\hline\hline
&  &  &  &  \\[-7pt]
& $\Lambda$CDM & $\Lambda$CDM : & $\Lambda$CDM : & $\Lambda$CDM : \\
&  & $k=0$ & $\Omega _{m0}=0$ & $\Omega _{m0}=0.04$ \\[2pt] \hline
&  &  &  &  \\[-7pt]
$H_{0}$ & $65.00_{-1.74}^{+1.78}$ & $64.29_{-1.51}^{+1.53}$ & $%
64.02_{-1.53}^{+1.55}$ & $64.07_{-1.54}^{+1.56}$ \\[2pt] \hline
&  &  &  &  \\[-7pt]
$\Omega _{k0}$ & $-2.12_{-1.61}^{+1.96}$ & $0$ & $0.599_{-0.137}^{+0.155}$ &
$0.519_{-0.138}^{+0.155}$ \\[2pt] \hline
&  &  &  &  \\[-7pt]
$\Omega _{m0}$ & $1.01_{-0.85}^{+1.08}$ & $0.309_{-0.072}^{+0.082}$ & $0$ & $%
0.04$ \\[2pt] \hline
&  &  &  &  \\[-7pt]
$\Omega _{c0}$ & $1.36_{-0.78}^{+0.92}$ & $0.691_{-0.082}^{+0.072}$ & $%
0.401_{-0.155}^{+0.137}$ & $0.441_{-0.155}^{+0.138}$ \\[2pt] \hline
&  &  &  &  \\[-7pt]
$t_{0}$ & $12.70_{-1.22}^{+2.15}$ & $14.90_{-0.88}^{+0.82}$ & $%
18.35_{-1.04}^{+1.19}$ & $17.15_{-0.82}^{+0.85}$ \\[2pt] \hline
&  &  &  &  \\[-7pt]
$q_{0}$ & $-0.864_{-0.405}^{+0.400}$ & $-0.540_{-0.101}^{+0.136}$ & $%
-0.410_{-0.131}^{+0.164}$ & $-0.444_{-0.116}^{+0.186}$ \\[2pt] \hline
&  &  &  &  \\[-7pt]
$a_{i}$ & $0.732_{-0.114}^{+0.083}$ & $0.610_{-0.074}^{+0.077}$ & $0$ & $%
0.357_{-0.042}^{+0.048}$ \\[2pt] \hline
&  &  &  &  \\[-7pt]
$p(\Omega _{k0}<0)$ & $98.08\,\%$ & $-$ & $0\,\%$ & $0\,\%$ \\[2pt] \hline
&  &  &  &  \\[-7pt]
$p(\Omega _{k0} = 0)$ & $2.91\,\%$ & $-$ & $0\,\%$ & $0\,\%$ \\[2pt] \hline
&  &  &  &  \\[-7pt]
$p(q_{0}<0)$ & $7.57\,\sigma $ & $100\,\%$ & $5.12\,\sigma $ & $5.08\,\sigma
$ \\[2pt] \hline
&  &  &  &  \\[-7pt]
$p(a_{i}<1)$ & $7.58\,\sigma $ & $100\,\%$ & $5.12\,\sigma $ & $5.17\,\sigma
$ \\[2pt] \hline\hline
\end{tabular}
\end{center}
\caption{The estimated parameters for the $\Lambda$CDM model and some
specific cases of spatial section and matter content. We use the Bayesian
analysis to obtain the peak of the one-dimensional marginal probability and
the $2\,\protect\sigma $ credible region for each parameter. $H_{0}$ is
given in $km/M\!pc.s$, $t_{0}$ in $Gy$ and $a_{i}$ in units of $a_{0}$. }
\label{tableParEstLCDM}
\end{table}

From the probability distribution (\ref{pd1}), a joint probability
distribution for any subset of parameters can be obtained by
integrating (marginalizing) on the remaining parameters, see refs.
\cite{colistete1, colistete2}. This is a valuable information.
But, in order to estimate properly a single parameter, the
probability distribution must be marginalized on all other
parameters. This in general gives a quite different result if we
try to estimate the parameter in a  two or three-dimensional
parameter space. The reason is that, in such multidimensional
parameter space, if a parameter has a large probability density
but in a narrow region, the total contribution of this region may
be quite small compared to other large regions which have small
probability: in the marginalization process, this kind of high PDF
region contributes little to the estimation of a given parameter.
These features are exemplified in figure \ref{figPdf3Dm04} where
the probability distribution function (PDF) for the set of
parameters ($\alpha,H_0,\bar A$) in the GCGM case is displayed,
with a clearly non-Gaussian behaviour. Hence, in what follows the
estimation of a given parameter will be made by marginalizing on
all other ones.

\subsection{Estimation of $\protect\alpha$}

In the case of five free parameters, the procedure described above gives $%
\alpha =-0.75_{-0.24}^{+4.04}$. Note that this prediction differs
substantially from that extracted from the minimization of $\chi ^{2}$,
which gives a large positive best value $\alpha $, instead of a negative $%
\alpha $ when the marginalization is made. Concerning the best value when the
marginalization is made, even if the best value is negative, the dispersion
is quite high, so even large positive values are not excluded, at least at $%
2\sigma $ level. Comparing our results with other ones already published
requires some care due to the fact that usually in the works already quoted
some parameters are fixed. Moreover, the allowed values for $\alpha $ are
generally restricted to the interval $[0,1]$, except in references \cite
{makler,maklerbis,berto2,ygong,bento2004}. In reference \cite{makler}, the authors
considered $-1\leq \alpha <1$, a flat spatial section and they have fixed
the value of the Hubble constant. They found a best value around
$\alpha \sim 0.4$.

\begin{figure}[t!]
\begin{minipage}[t]{0.48\linewidth}
\includegraphics[width=\linewidth]{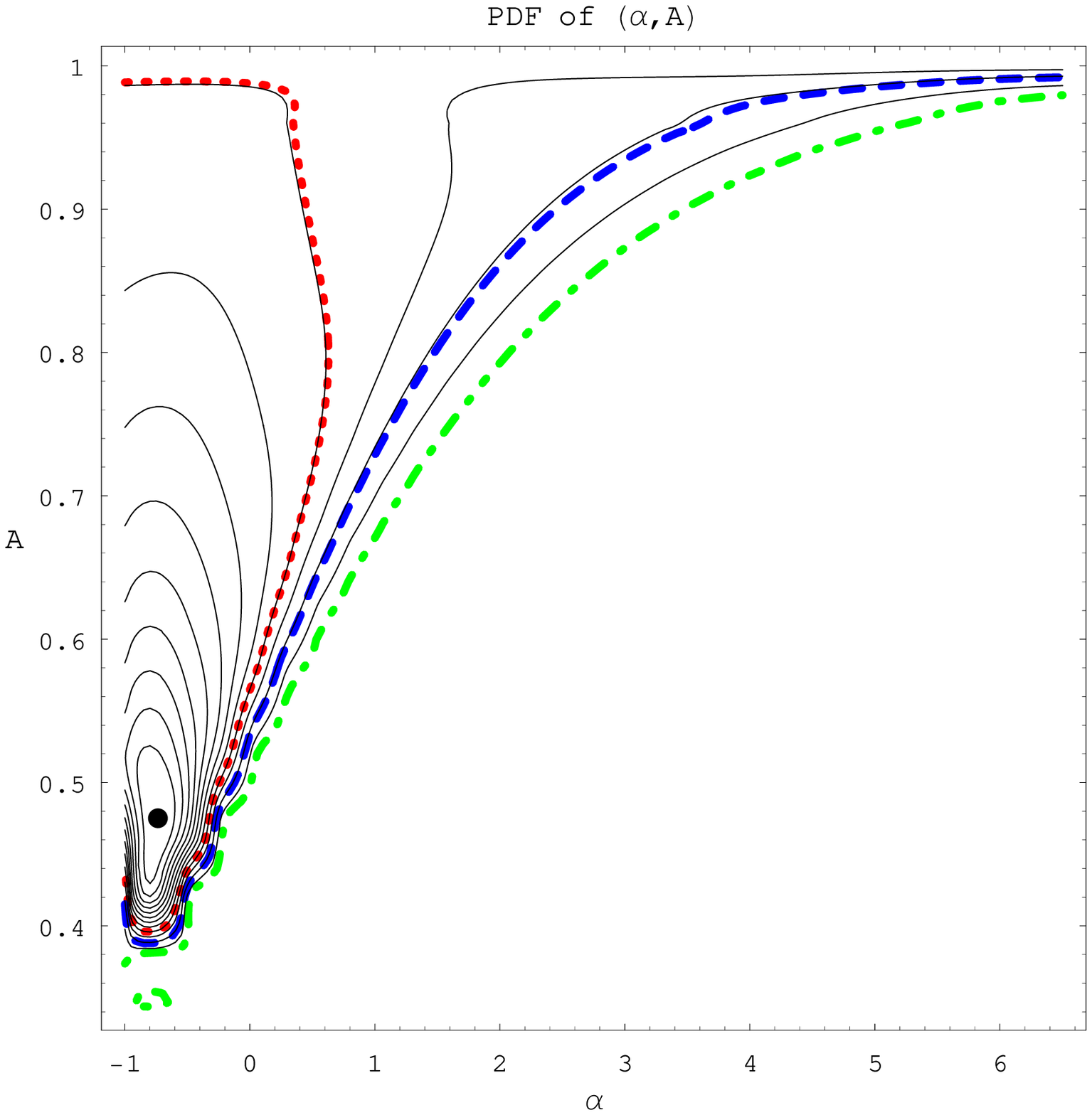}
\end{minipage} \hfill
\begin{minipage}[t]{0.48\linewidth}
\includegraphics[width=\linewidth]{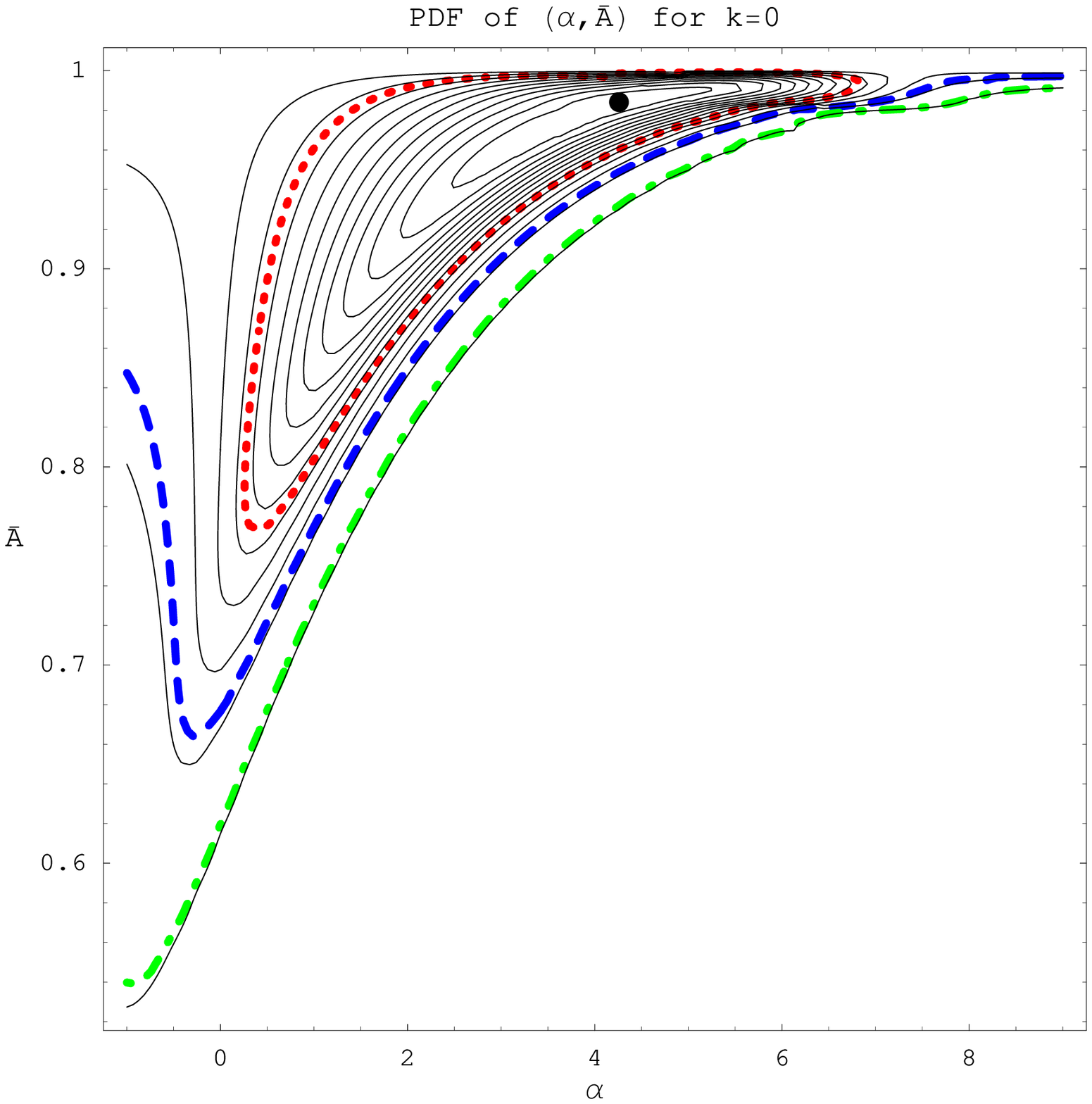}
\end{minipage} \hfill
\begin{minipage}[t]{0.48\linewidth}
\includegraphics[width=\linewidth]{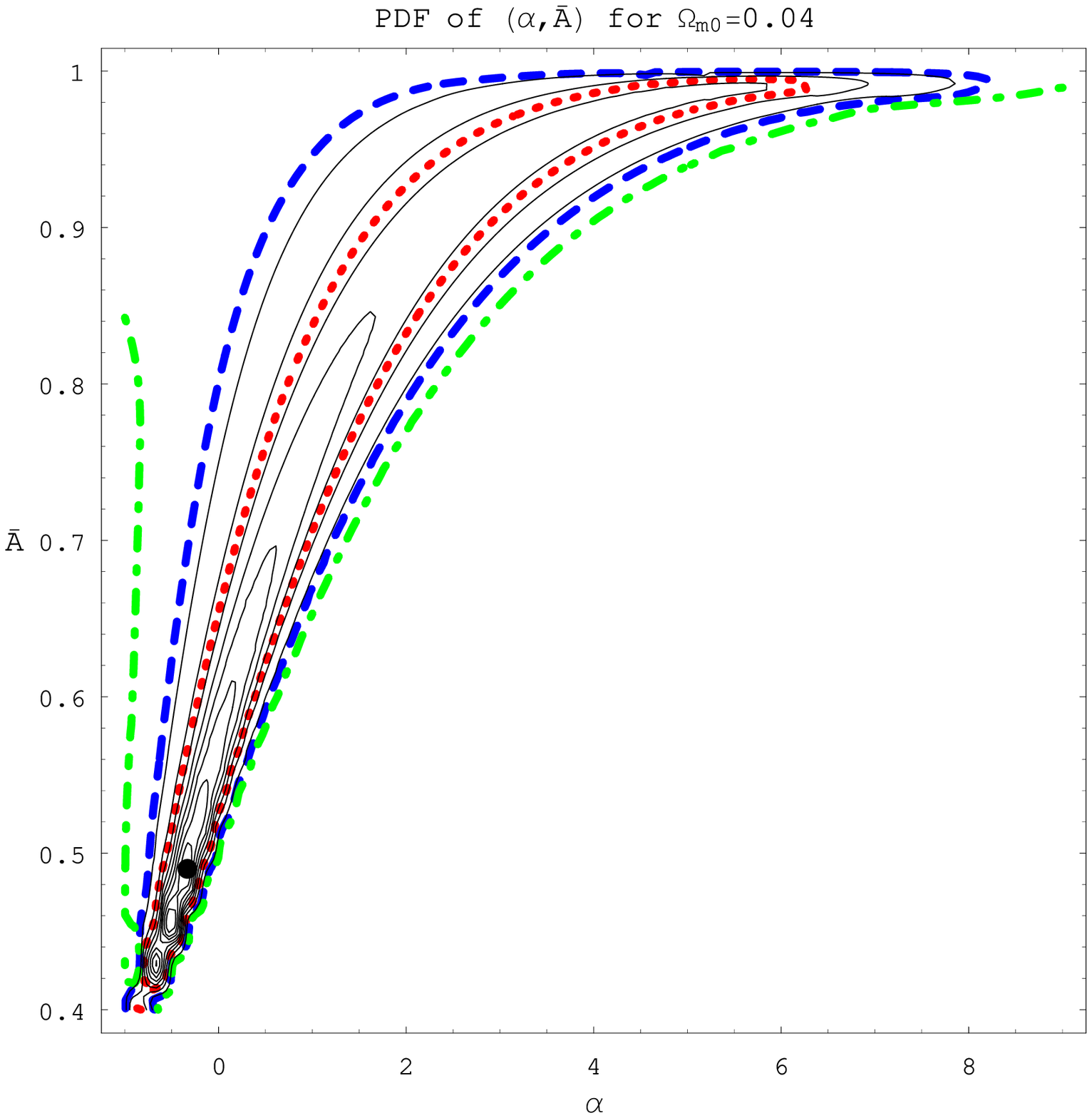}
\end{minipage} \hfill
\begin{minipage}[t]{0.48\linewidth}
\includegraphics[width=\linewidth]{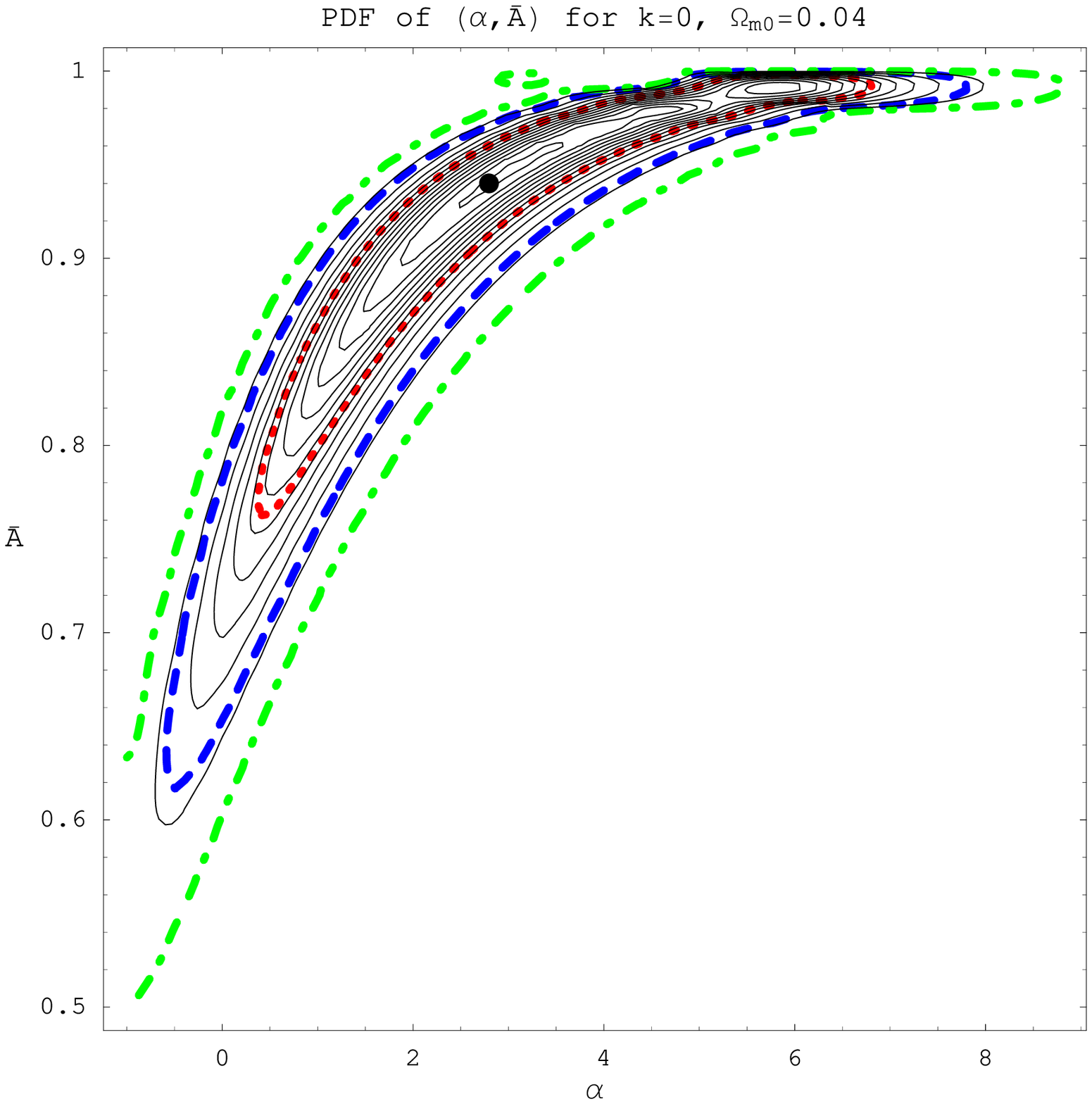}
\end{minipage} \hfill
\caption{{\protect\footnotesize The graphics of the joint PDF as function of
$(\protect\alpha,\bar{A})$ for the generalized Chaplygin gas model. The
joint PDF peak is shown by the large dot, the credible regions of $1\,%
\protect\sigma $ ($68,27\%$) by the red dotted line, the $2\,\protect\sigma $
($95,45\%$) in blue dashed line and the $3\,\protect\sigma $ ($99,73\%$) in
green dashed-dotted line. The cases for $\Omega _{m0}=0$ are not shown here
because they are similar to the ones with $\Omega _{m0}=0.04$. }}
\label{figsAlphaA}
\end{figure}

\begin{figure}[t!]
\begin{minipage}[t]{0.48\linewidth}
\includegraphics[width=\linewidth]{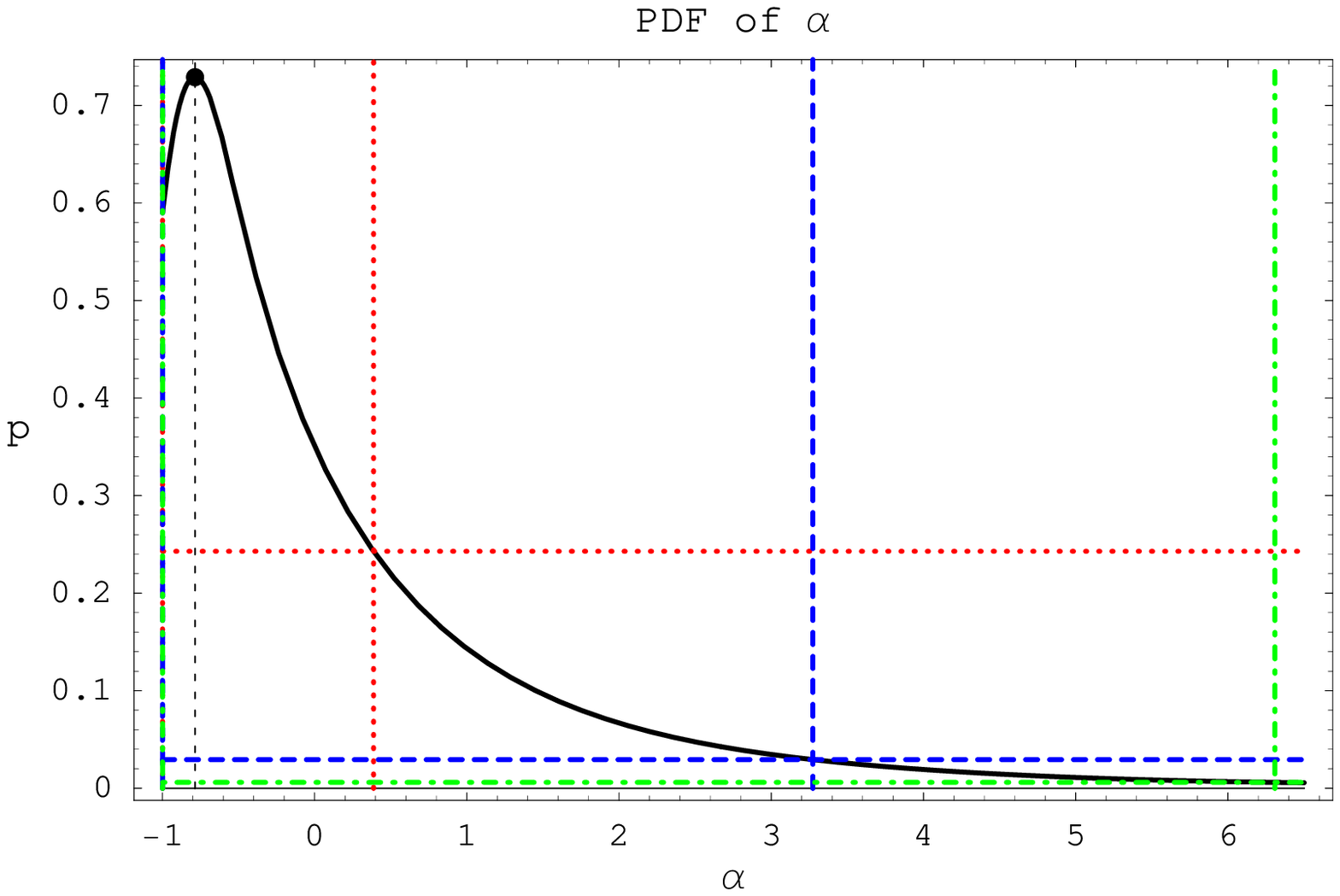}
\end{minipage} \hfill
\begin{minipage}[t]{0.48\linewidth}
\includegraphics[width=\linewidth]{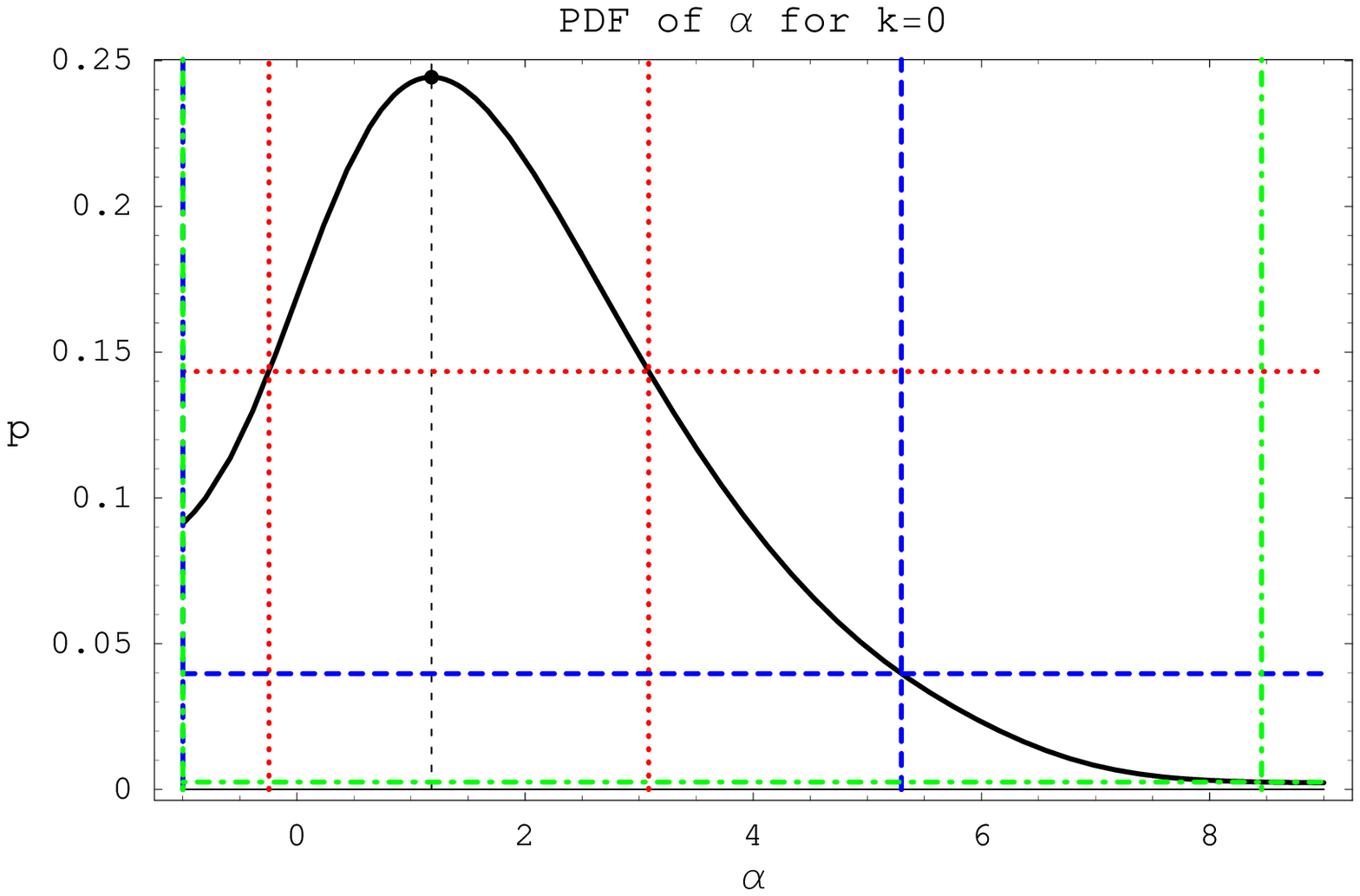}
\end{minipage} \hfill
\begin{minipage}[t]{0.48\linewidth}
\includegraphics[width=\linewidth]{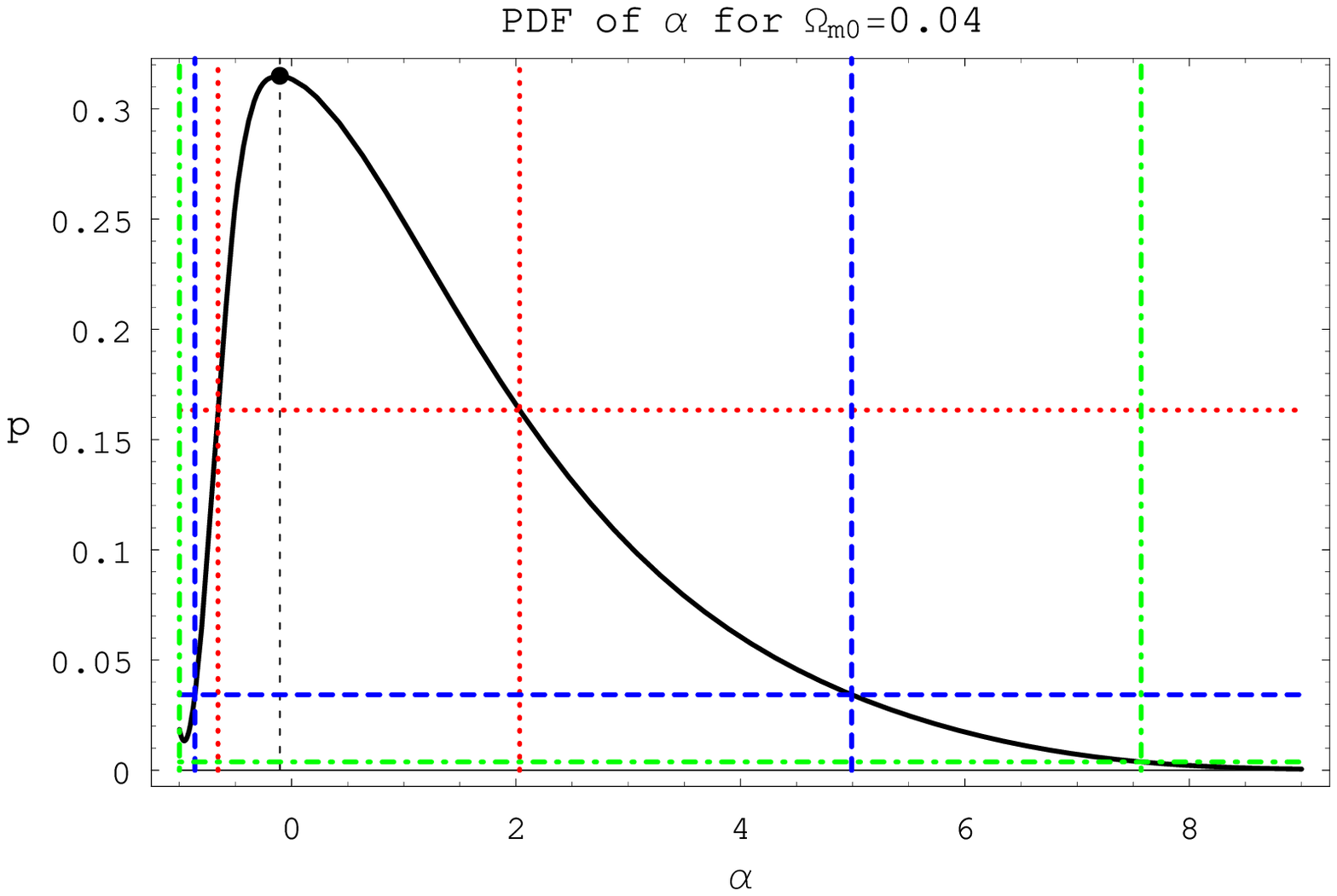}
\end{minipage} \hfill
\begin{minipage}[t]{0.48\linewidth}
\includegraphics[width=\linewidth]{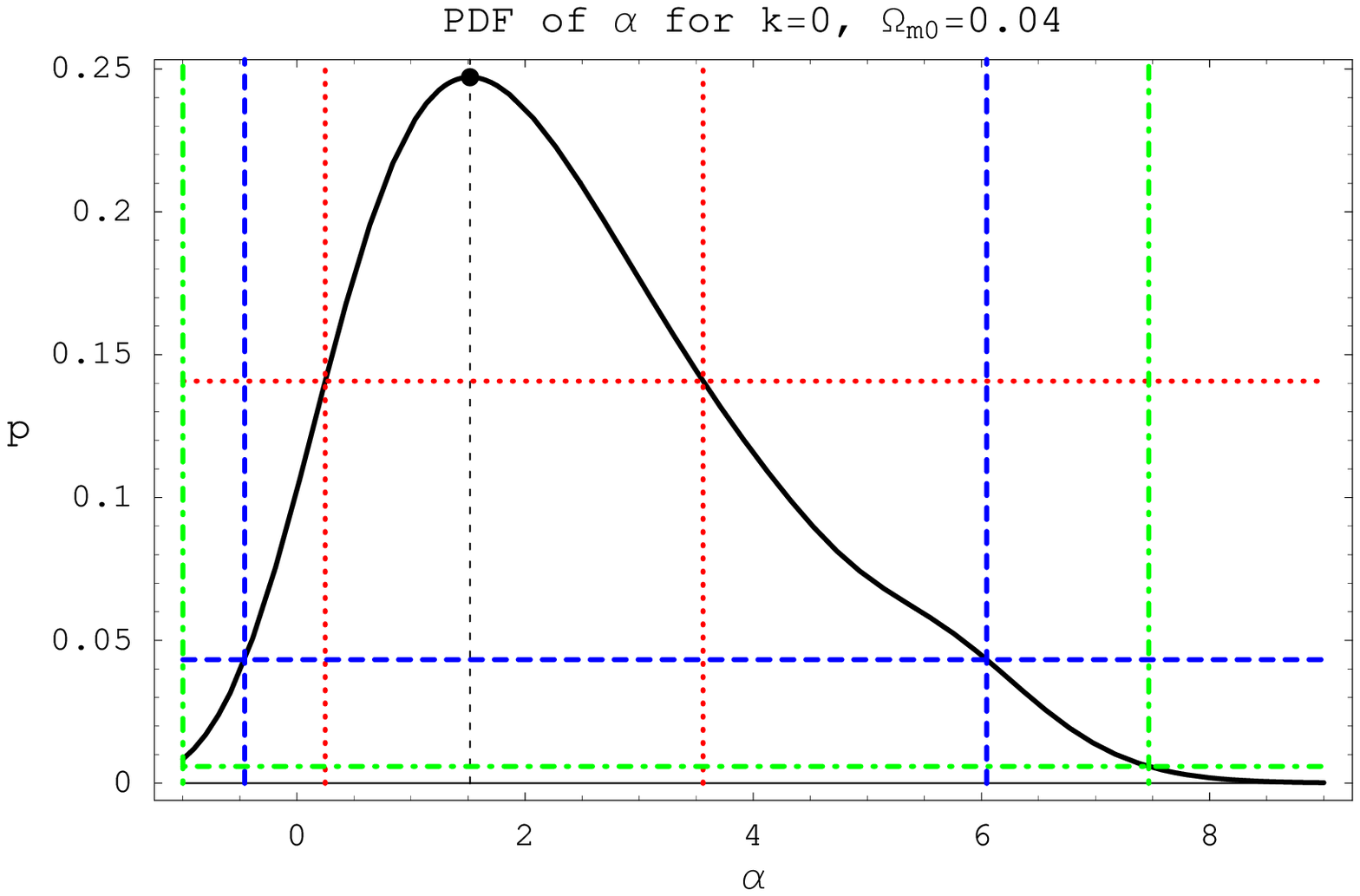}
\end{minipage} \hfill
\caption{{\protect\footnotesize The PDF of $\protect\alpha$ for the
generalized Chaplygin gas model. The solid lines are the PDF, the $1\protect%
\sigma $ ($68.27\%$) regions are delimited by red dotted lines and the $2
\protect\sigma $ ($95.45\%$) credible regions are given by blue dashed
lines. The cases for $\Omega _{m0}=0$ are not shown here because they are
similar to the ones with $\Omega _{m0}=0.04$. }}
\label{figsAlpha}
\end{figure}

If we compare with the results of ref. \cite{colistete2}, the use of a
large sample of supernovae has slightly diminished the best value, but at
same time, it has reduced considerably the dispersion. Imposing that the
space is flat, or fixing the pressureless matter equal to zero or $0.04$,
lead to a positive best value for $\alpha$; the dispersion is also affected.
For example, with the imposition that $k = 0$ and $\Omega_m = 0.04$, the
estimation leads to $\alpha = 1.52^{+4.53}_{- 1.97}$. In figure \ref{figsAlpha}
the PDF for $\alpha$ is displayed for the case of five parameters, and also for
three particular cases, where one or two parameter is fixed. Note that as
the baryonic matter or the curvature (or both) is fixed, the maximum of the
PDF is displaced to positive values. In general grounds, we can state that
even relatively large values for $\alpha$ are not excluded. Of course, for
the CGM the value of $\alpha$ is fixed to unity. However, from the analysis
for the GCGM it can be inferred that the probability to have $\alpha = 1$ is
$19.58\%$ when the five parameters are free. Restricting to null curvature
or fixing the pressureless density increases considerably this value: for
example, when the curvature is null the probability to have $\alpha = 1$ is
about $90\%$. At the same time, the probability to have $\alpha > 1$ is $%
41.98\%$ when all parameters are free, but this value can double when one or
two parameters are fixed.

\subsection{Estimation of $\bar A$}

In general the results indicate that the value of $\bar{A}$ is close to
unity. When five free parameters are considered, the marginalization of the
remaining four other parameters leads to $\bar{A}=1.000_{-0.534}^{+0.000}$.
This may suggest the conclusion that $\Lambda $CDM ($\bar{A}=1$) model is
favoured. However, the accuracy of the computation, due to the step
(between $0.01$ and $0.02$) used in the evaluation of the parameter, does not
allow this conclusion. Instead, it means the peak happens for $0.98<\bar{A}%
\leqslant 1$.

\begin{figure}[t!]
\begin{minipage}[t]{0.48\linewidth}
\includegraphics[width=\linewidth]{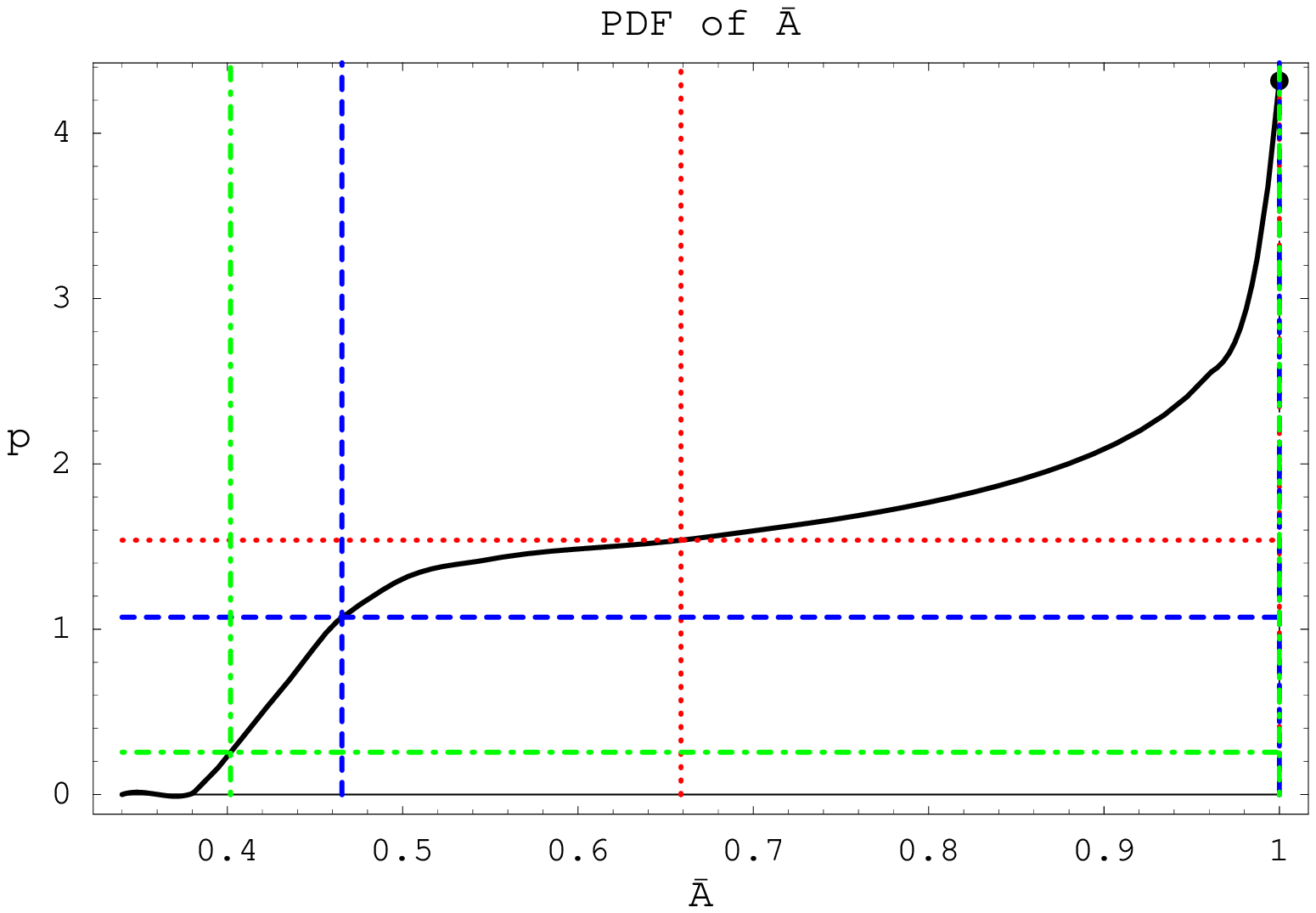}
\end{minipage} \hfill
\begin{minipage}[t]{0.48\linewidth}
\includegraphics[width=\linewidth]{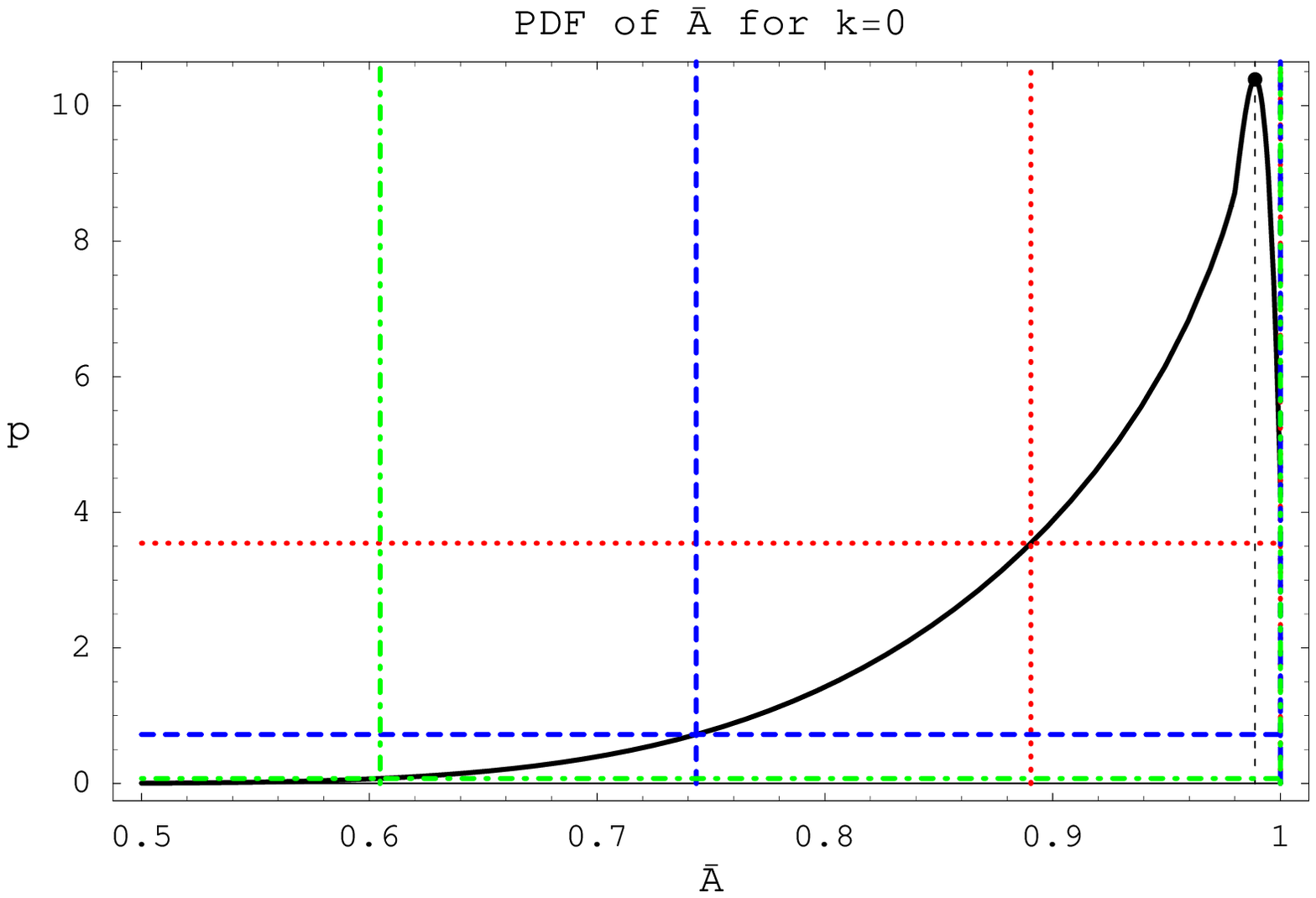}
\end{minipage} \hfill
\begin{minipage}[t]{0.48\linewidth}
\includegraphics[width=\linewidth]{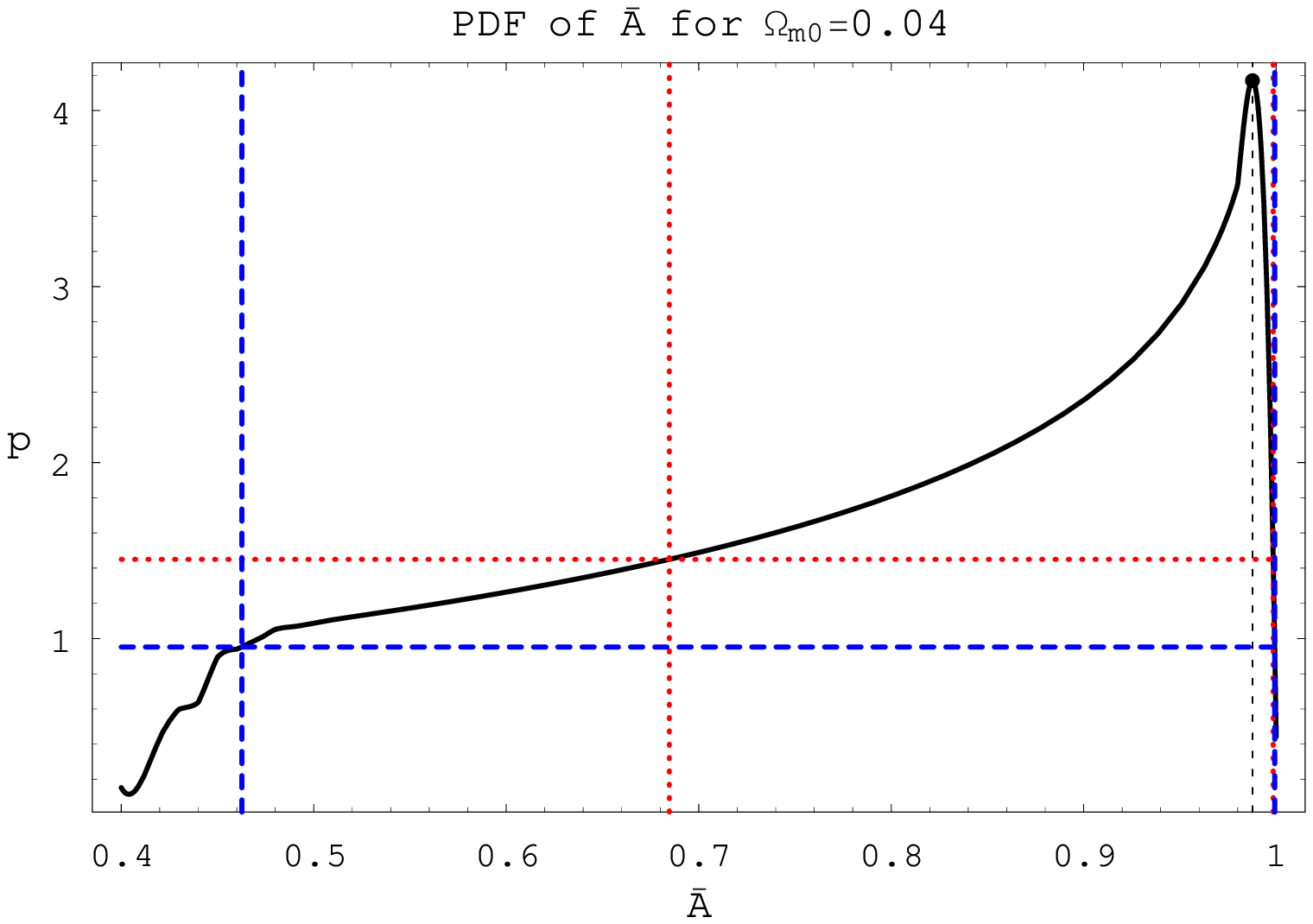}
\end{minipage} \hfill
\begin{minipage}[t]{0.48\linewidth}
\includegraphics[width=\linewidth]{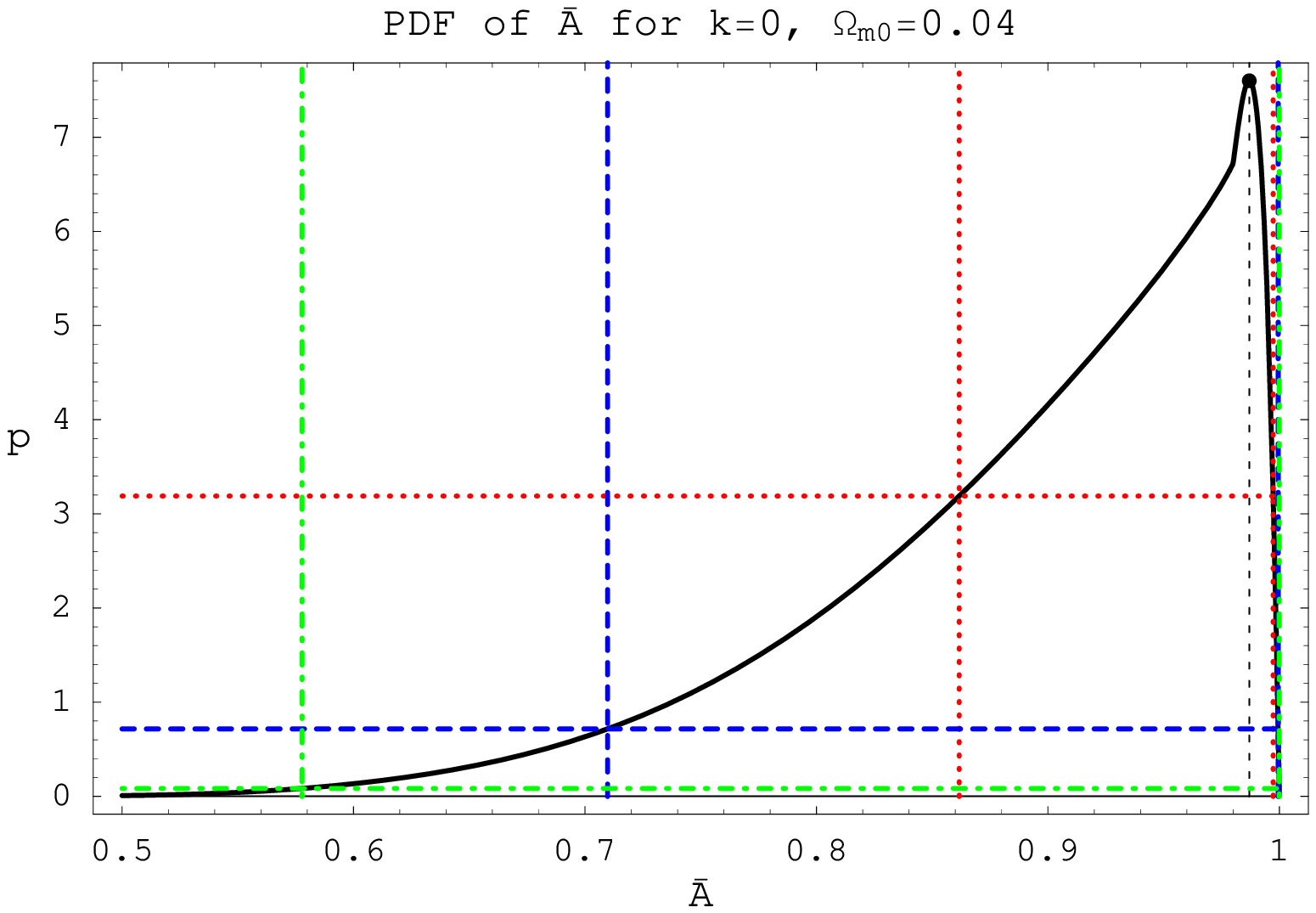}
\end{minipage} \hfill
\caption{{\protect\footnotesize The one-dimensional PDF of $\bar{A}$ for the
generalized Chaplygin gas model. The solid lines are the PDF, the $1\protect%
\sigma $ ($68.27\%$) regions are delimited by red dotted lines, the $2%
\protect\sigma $ ($95.45\%$) credible regions are given by blue dashed lines
and the $3\protect\sigma $ ($99.73\%$) regions are delimited by green
dashed-dotted lines. The cases for $\Omega _{m0}=0$ are not shown here
because they are similar to the ones with $\Omega _{m0}=0.04$. }}
\label{figsA}
\end{figure}

\begin{figure}[t!]
\begin{minipage}[t]{0.48\linewidth}
\includegraphics[width=\linewidth]{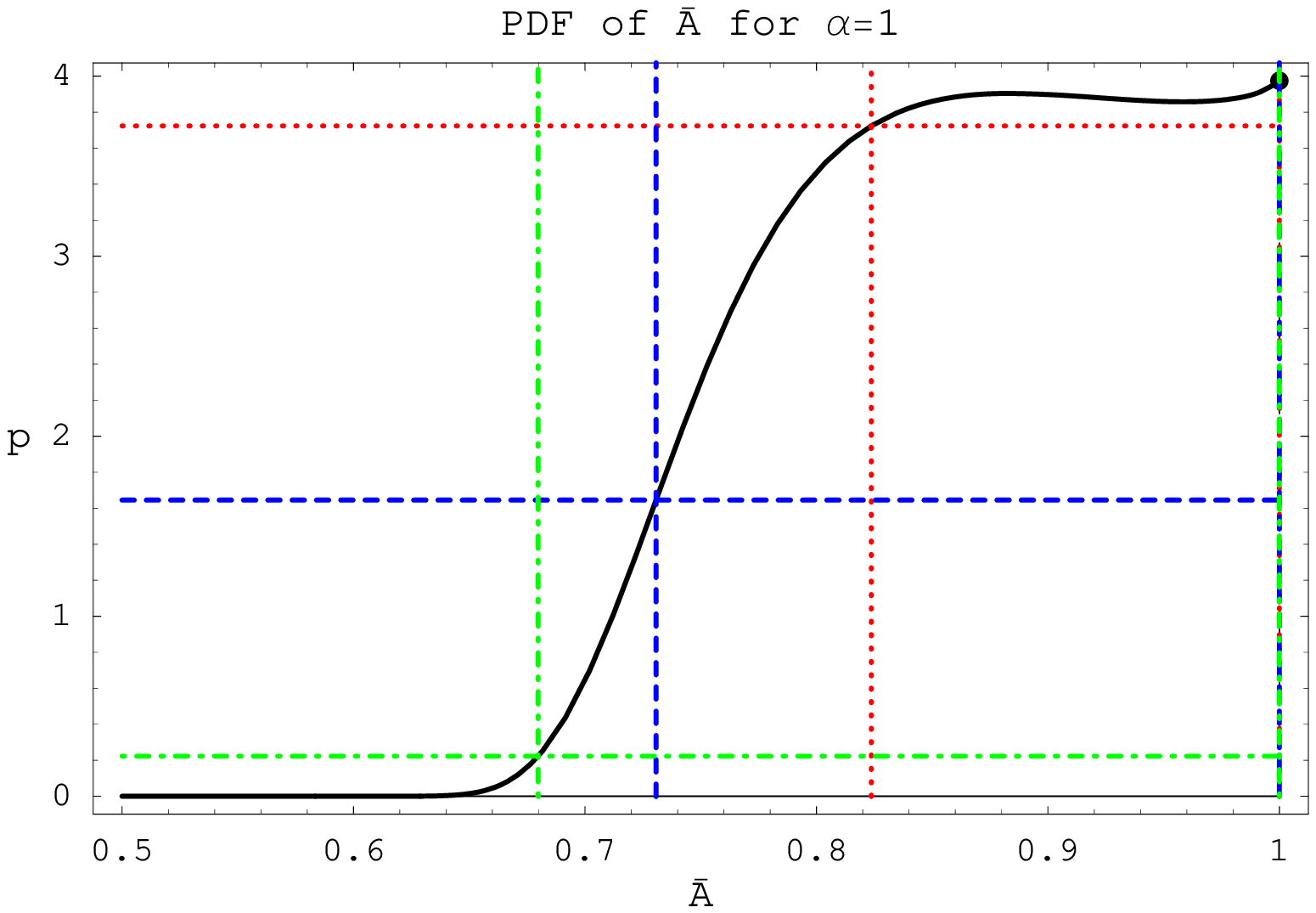}
\end{minipage} \hfill
\begin{minipage}[t]{0.48\linewidth}
\includegraphics[width=\linewidth]{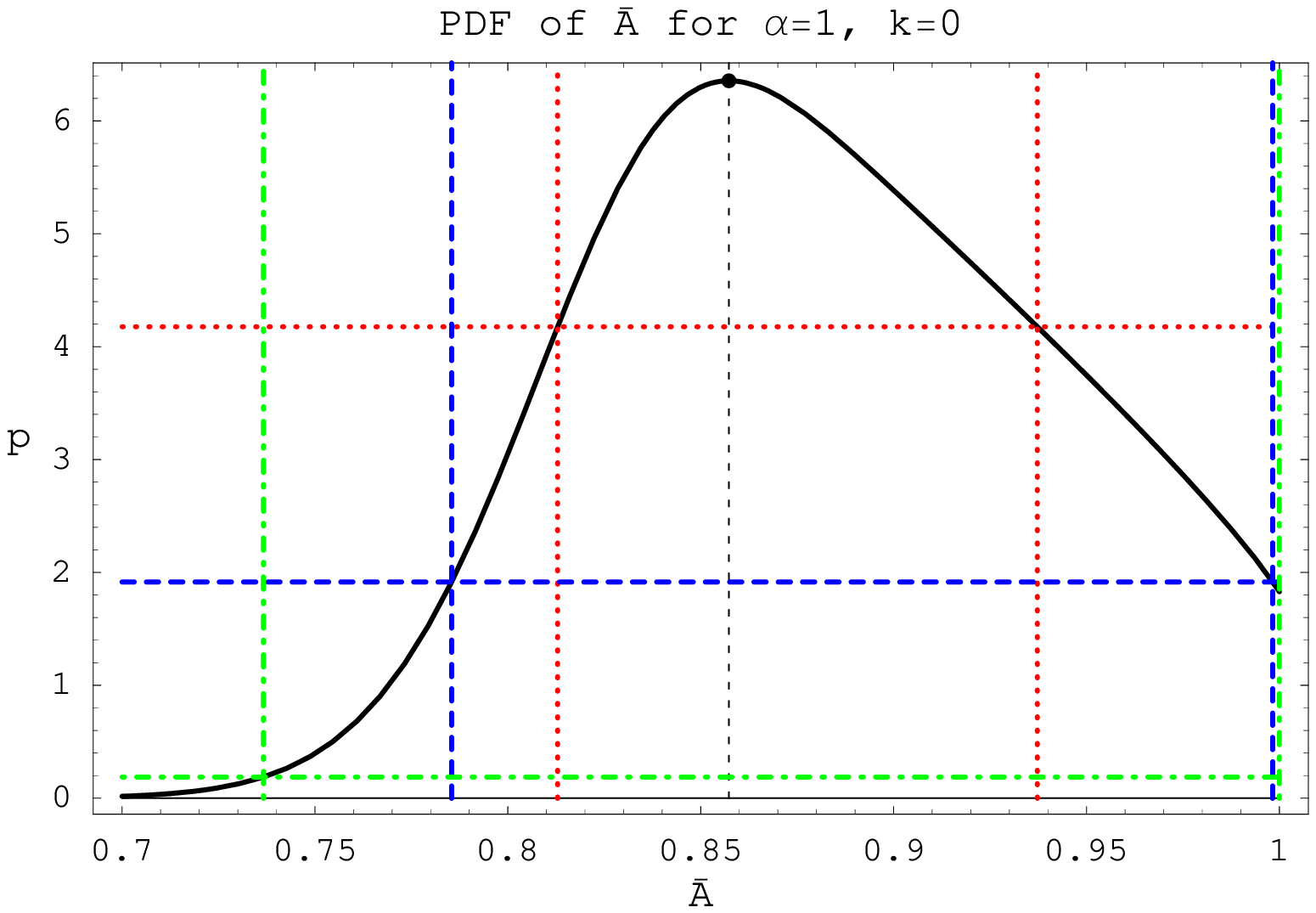}
\end{minipage} \hfill
\begin{minipage}[t]{0.48\linewidth}
\includegraphics[width=\linewidth]{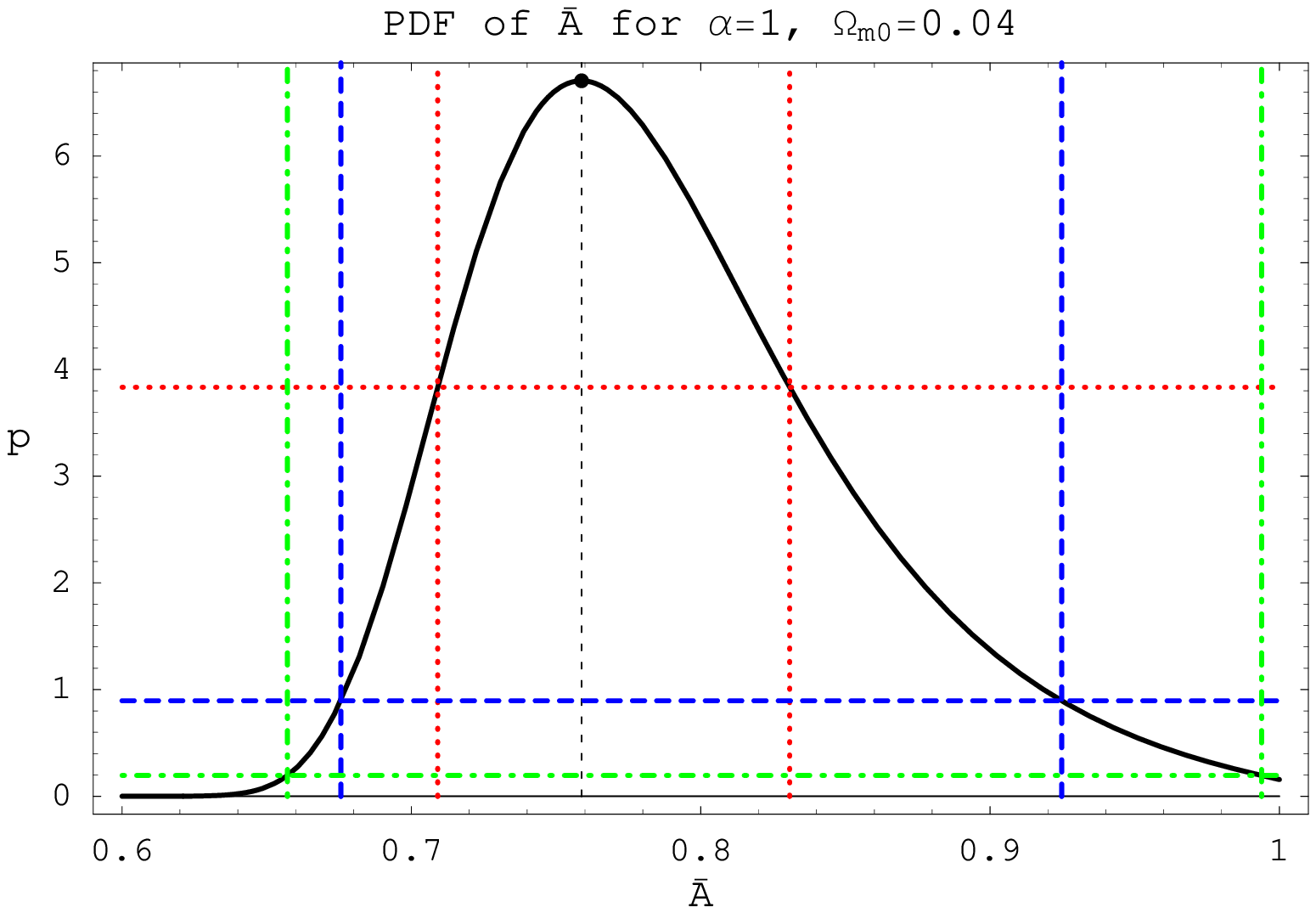}
\end{minipage} \hfill
\begin{minipage}[t]{0.48\linewidth}
\includegraphics[width=\linewidth]{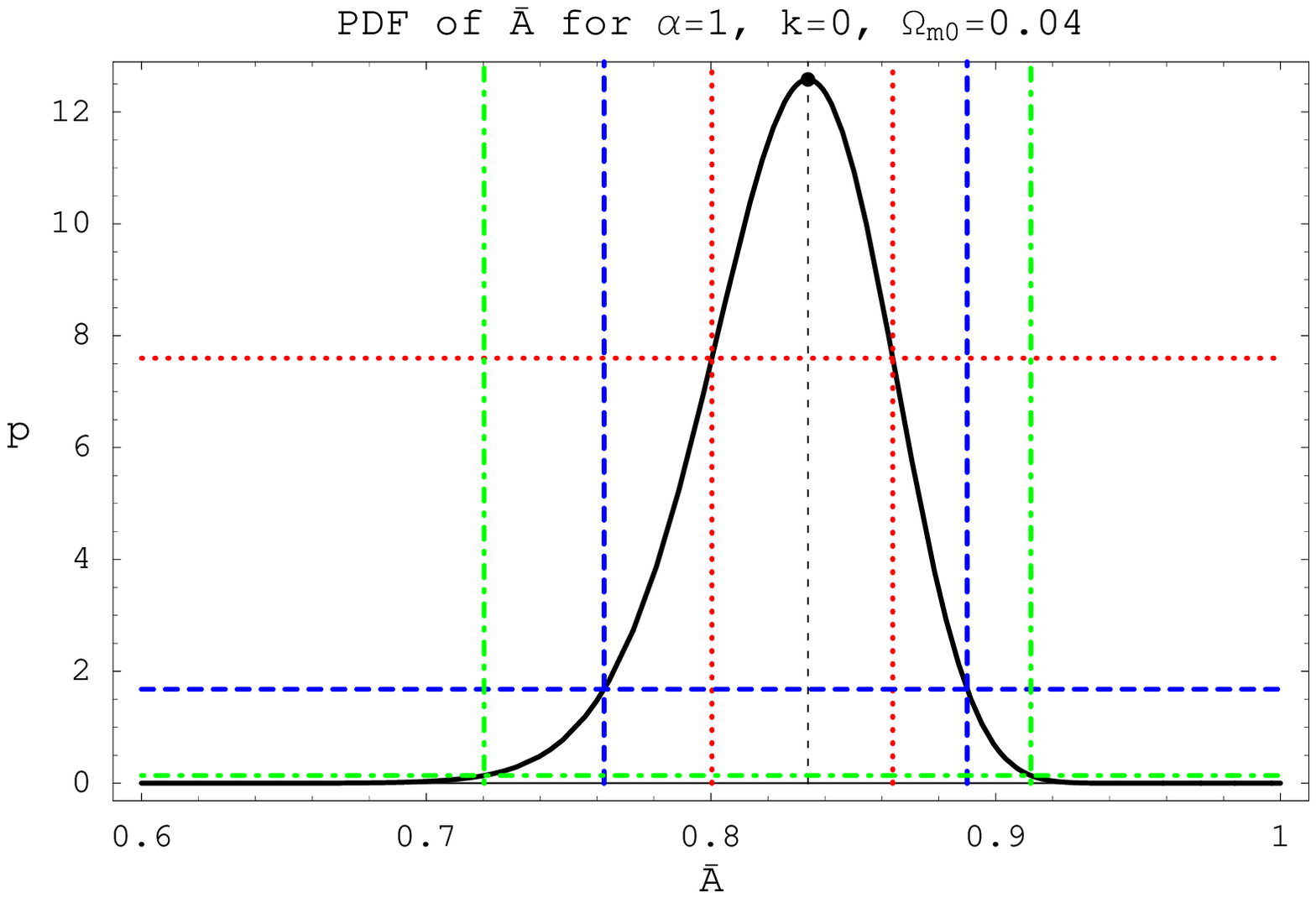}
\end{minipage} \hfill
\caption{{\protect\footnotesize The one-dimensional PDF of $\bar{A}$ for the
traditional Chaplygin gas model (CGM). The solid lines are the PDF, the $1%
\protect\sigma $ ($68.27\%$) regions are delimited by red dotted lines, the $%
2\protect\sigma $ ($95.45\%$) credible regions are given by blue dashed
lines and the $3\protect\sigma $ ($99.73\%$) regions are delimited by green
dashed-dotted lines. The cases for $\Omega _{m0}=0$ are not shown here
because they are similar to the ones with $\Omega _{m0}=0.04$. }}
\label{figsAlpha1A}
\end{figure}

In fact, fixing the curvature or the pressureless matter, the preferred value differs
slightly from unity: with $\Omega _{m}=0.04$ and $k=0$, for example, we have
$\bar{A}=0.987_{-0.272}^{+0.012}$. The situation is essentially the same for
the CGM: when all other parameters are marginalized, the best value is
unity, but with a small dispersion, $\bar{A}=1.000_{-0.270}^{+0.000}$; however,
this best value becomes smaller when one or two parameters are fixed. This
occurs in both for the GCGM and CGM. In figures \ref{figsA} and \ref{figsAlpha1A}
the PDF for $\bar{A}$ is displayed, both for the GCGM and the CGM, in the case
where the marginalization is made in all other parameters, and when the matter
density and/or the curvature is fixed. Fixing one or two parameters displaces the
maximum of probability in the direction of smaller value of $\bar{A}$. This
effect is more sensible for the CGM: in general, the best value of the CGM
is smaller than in the GCGM, and the dispersion is also smaller. Note that
the probability to have $\bar{A}\neq 1$ is zero when all parameters are
free, both for the GCGM and CGM. But, this probability can become as high as
$100\%$ when one or two parameters are fixed.

In figure \ref{figsAlphaA} the joint probabilities for $\alpha$ and $\bar A$
are displayed, with a non-Gaussian shape. This figure is an illustration of
the importance of the marginalization process because it changes the peak
values and credible regions depending if two or one-dimensional parameter
space is used. The region of highest probability, at $2\sigma$ level,
is concentrated around $\alpha = 0$ with a large dispersion for $\bar A$, as
the curvature and/or the pressureless matter is fixed, this region displaces
itself to positive values of $\alpha$ and to higher values of $\bar A$, the
opposite that happens with the PDF value of $\bar A$, due in the last case
to the marginalization procedure. The dispersion in $\bar A$ remains always
large. It is obvious, on the other hand, from the two-dimensional credible
regions that the case $\alpha = 1$ is not ruled out. Using a larger sample
of $253$ data, the authors of reference \cite{berto2} have constrained the
parameters $\alpha$ and $\bar A$ such that $[1.6,3.6]$ and $[0.856,0.946]$
at $68\%$ confidence level. But they have used the $\chi^2$ statistics and
quartessence, and in this sense their result may be considered as compatible
with ours. In reference \cite{avelino}, restricting to the flat case, and
fixing $\alpha = 1$, for example, the authors have found that $0.66 < \bar A
< 1$ at $95\%$ confidence level. This result is compatible with ours, see
table $5$.

\subsection{Estimation of $\Omega_{m0}$}

\begin{figure}[t!]
\begin{center}
\includegraphics[scale=0.8]{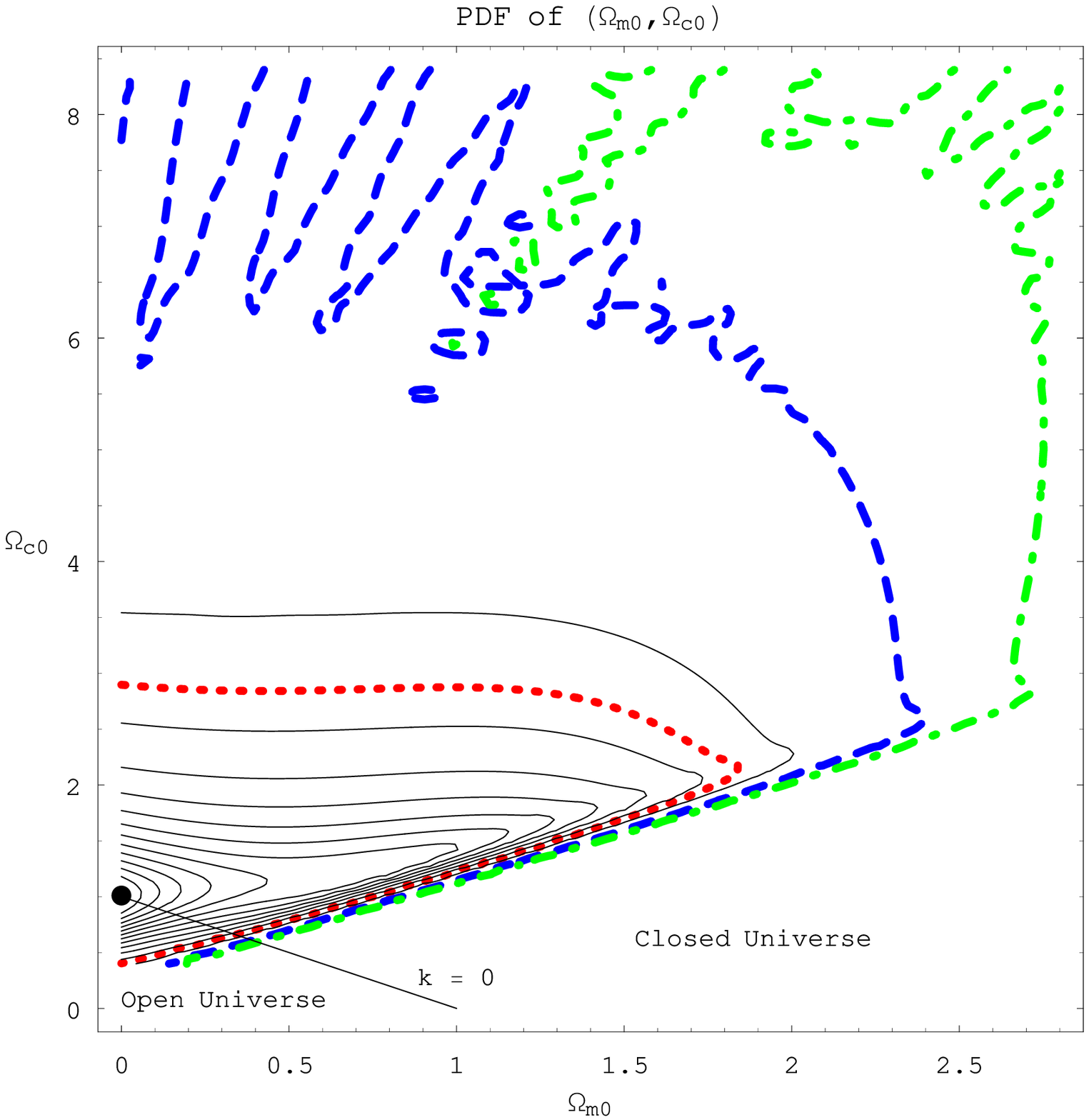}
\end{center}
\caption{{\protect\footnotesize The graphics of the joint PDF as function of
$(\Omega _{m0},\Omega _{c0})$ for the generalized Chaplygin gas model, where
$p(\Omega _{m0},\Omega _{c0})$ is a integral of $p(\protect\alpha
,H_{0},\Omega _{m0},\Omega _{c0},\bar{A})$ over the $(\protect\alpha ,H_{0},%
\bar{A})$ parameter space. It shows a different shape for the confidence
regions with respect to the $\Lambda CDM$ but similar to the CGM of ref.
\protect\cite{colistete2}. The joint PDF peak has the value $0.584$ for $%
(\Omega _{m0},\Omega _{c0})=(0.00,1.00)$ (shown by the large dot), the
credible regions of $1\,\protect\sigma $ ($68,27\%$, shown in red dotted
line), $2\,\protect\sigma $ ($95,45\%$, in blue dashed line) and $3\,\protect%
\sigma $ ($99,73\%$, in green dashed-dotted line) have PDF levels of $0.104$%
, $0.037$\ and $0.010$, respectively. As $\Omega _{k0}+\Omega _{m0}+\Omega
_{c0}=1$, the probability for a spatially flat Universe is on the line $%
\Omega _{m0}+\Omega _{c0}=1$, above it we have the region for a closed
Universe ($k>0$, $\Omega _{k0}<0$), and below, the region for an open
Universe ($k<0$, $\Omega _{k0}>0$). }}
\label{figOmegam0c0}
\end{figure}

\begin{figure}[t!]
\begin{center}
\includegraphics[scale=0.8]{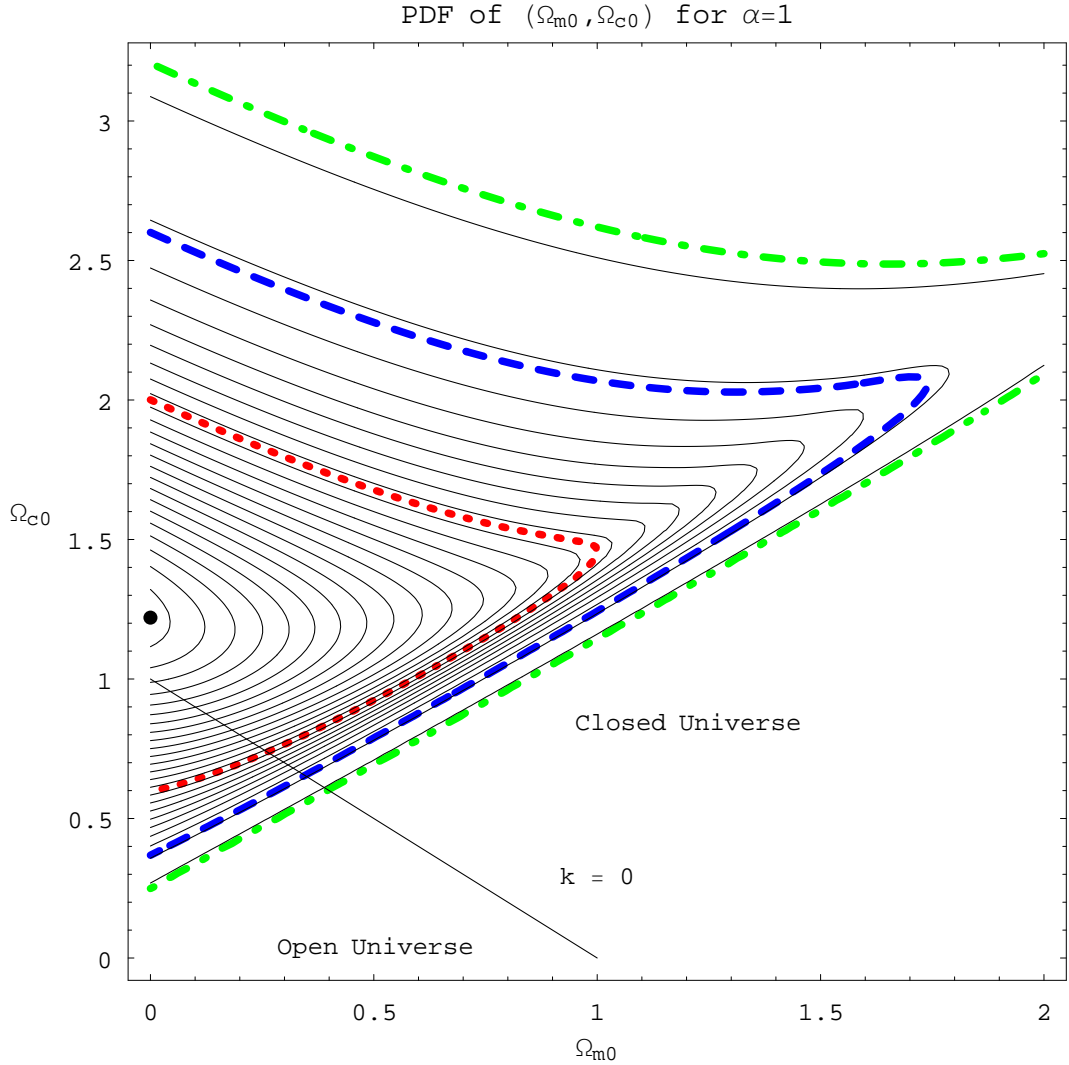}
\end{center}
\caption{{\protect\footnotesize The graphics of the joint PDF as function of
$(\Omega _{m0},\Omega _{c0})$ for the traditional Chaplygin gas model (CGM),
where $p(\Omega _{m0},\Omega _{c0})$ is a integral of $p(H_{0},\Omega
_{m0},\Omega _{c0},\bar{A})$ over the $(H_{0}, \bar{A})$ parameter space. It
shows a different shape for the confidence regions with respect to the $%
\Lambda CDM$ but similar to the CGM of ref. \protect\cite{colistete2}. The
joint PDF peak has the value $1.473$ for $(\Omega _{m0},\Omega
_{c0})=(0.00,1.22)$ (shown by the large dot), the credible regions of $1\,%
\protect\sigma $ ($68,27\%$, shown in red dotted line), $2\,\protect\sigma $
($95,45\%$, in blue dashed line) and $3\,\protect\sigma $ ($99,73\%$, in
green dashed-dotted line) have PDF levels of $0.542$, $0.103$\ and $0.009$,
respectively. As $\Omega _{k0}+\Omega _{m0}+\Omega _{c0}=1$, the probability
for a spatially flat Universe is on the line $\Omega _{m0}+\Omega _{c0}=1$,
above it we have the region for a closed Universe ($k>0$, $\Omega _{k0}<0$),
and below, the region for an open Universe ($k<0$, $\Omega _{k0}>0$). }}
\label{figAlpha1Omegam0c0}
\end{figure}

\begin{figure}[t!]
\begin{center}
\includegraphics[scale=0.8]{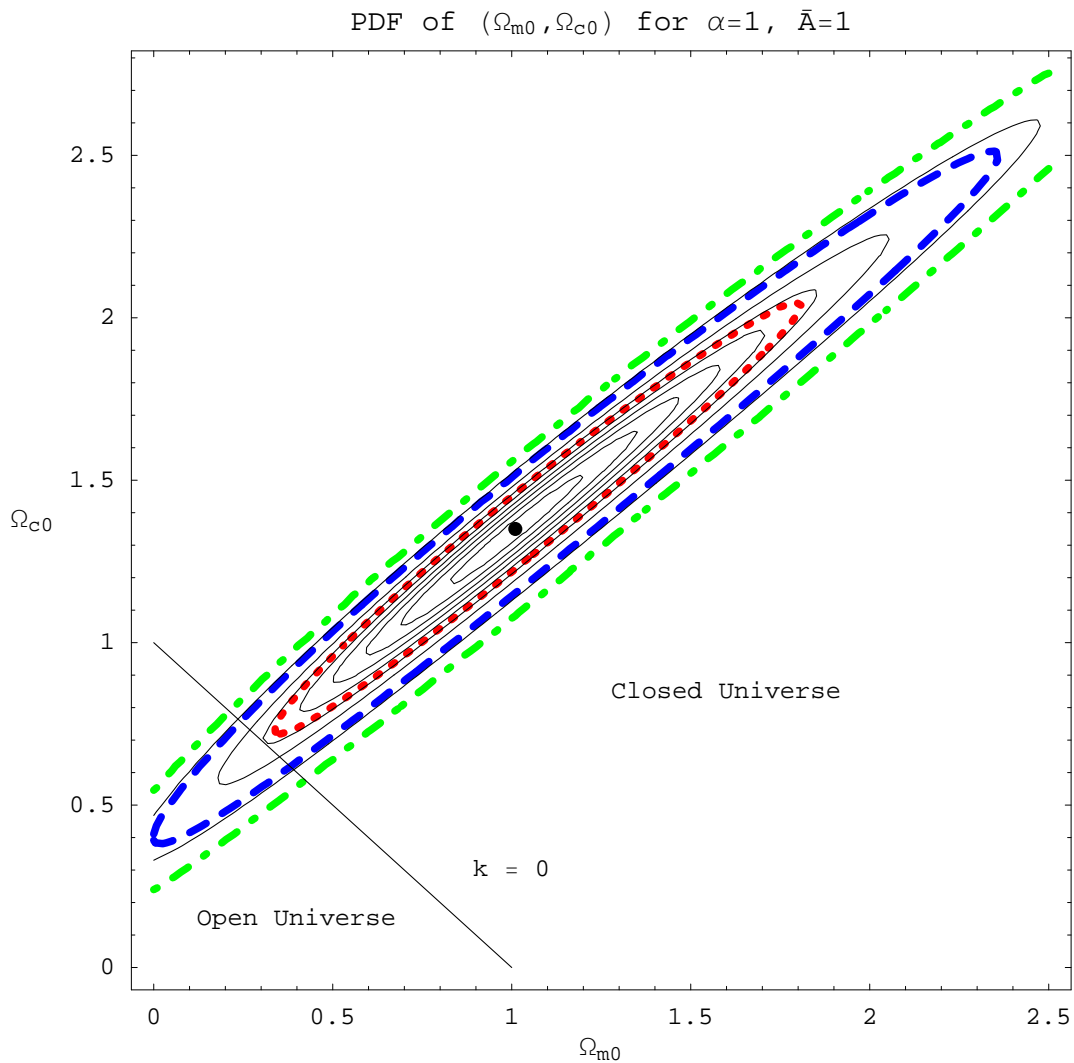}
\end{center}
\caption{{\protect\footnotesize The graphics of the joint PDF as function of
$(\Omega _{m0},\Omega _{c0})$ for the $\Lambda$CDM model model, where $%
p(\Omega _{m0},\Omega _{c0})$ is a integral of $p(H_{0},\Omega _{m0},\Omega
_{c0})$ over the $H_{0}$ parameter space. The joint PDF peak has the value $%
4.22$ for $(\Omega _{m0},\Omega _{c0})=(1.01,1.35)$ (shown by the large
dot), the credible regions of $1\,\protect\sigma $ ($68,27\%$, shown in red
dotted line), $2\,\protect\sigma $ ($95,45\%$, in blue dashed line) and $3\,%
\protect\sigma $ ($99,73\%$, in green dashed-dotted line) have PDF levels of
$1.380$, $0.247$\ and $0.025$, respectively. As $\Omega _{k0}+\Omega
_{m0}+\Omega _{c0}=1$, the probability for a spatially flat Universe is on
the line $\Omega _{m0}+\Omega _{c0}=1$, above it we have the region for a
closed Universe ($k>0$, $\Omega _{k0}<0$), and below, the region for an open
Universe ($k<0$, $\Omega _{k0}>0$). }}
\label{figLCDMOmegam0c0}
\end{figure}

\begin{figure}[t!]
\begin{minipage}[t]{0.48\linewidth}
\includegraphics[width=\linewidth]{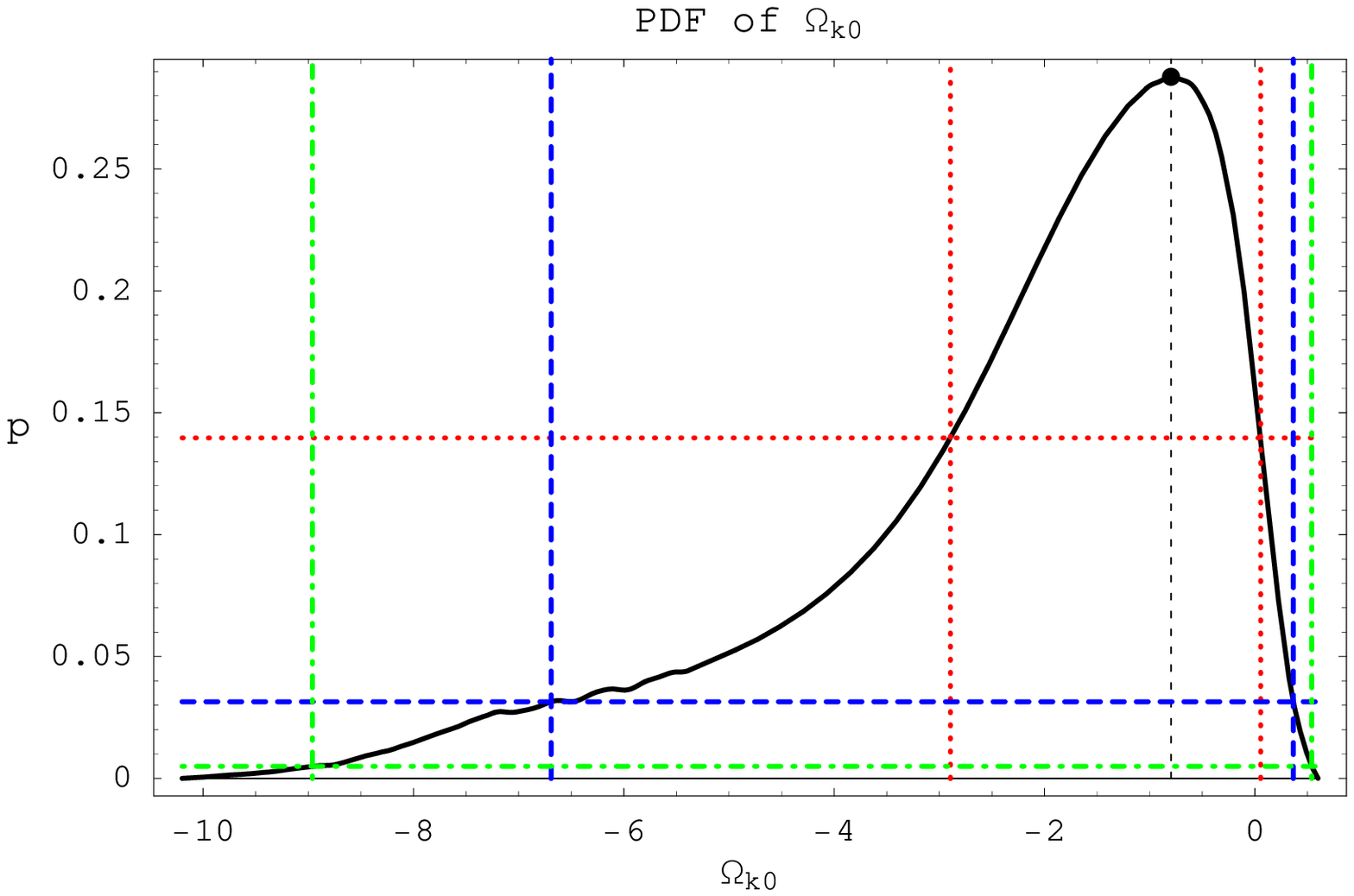}
\end{minipage} \hfill
\begin{minipage}[t]{0.48\linewidth}
\includegraphics[width=\linewidth]{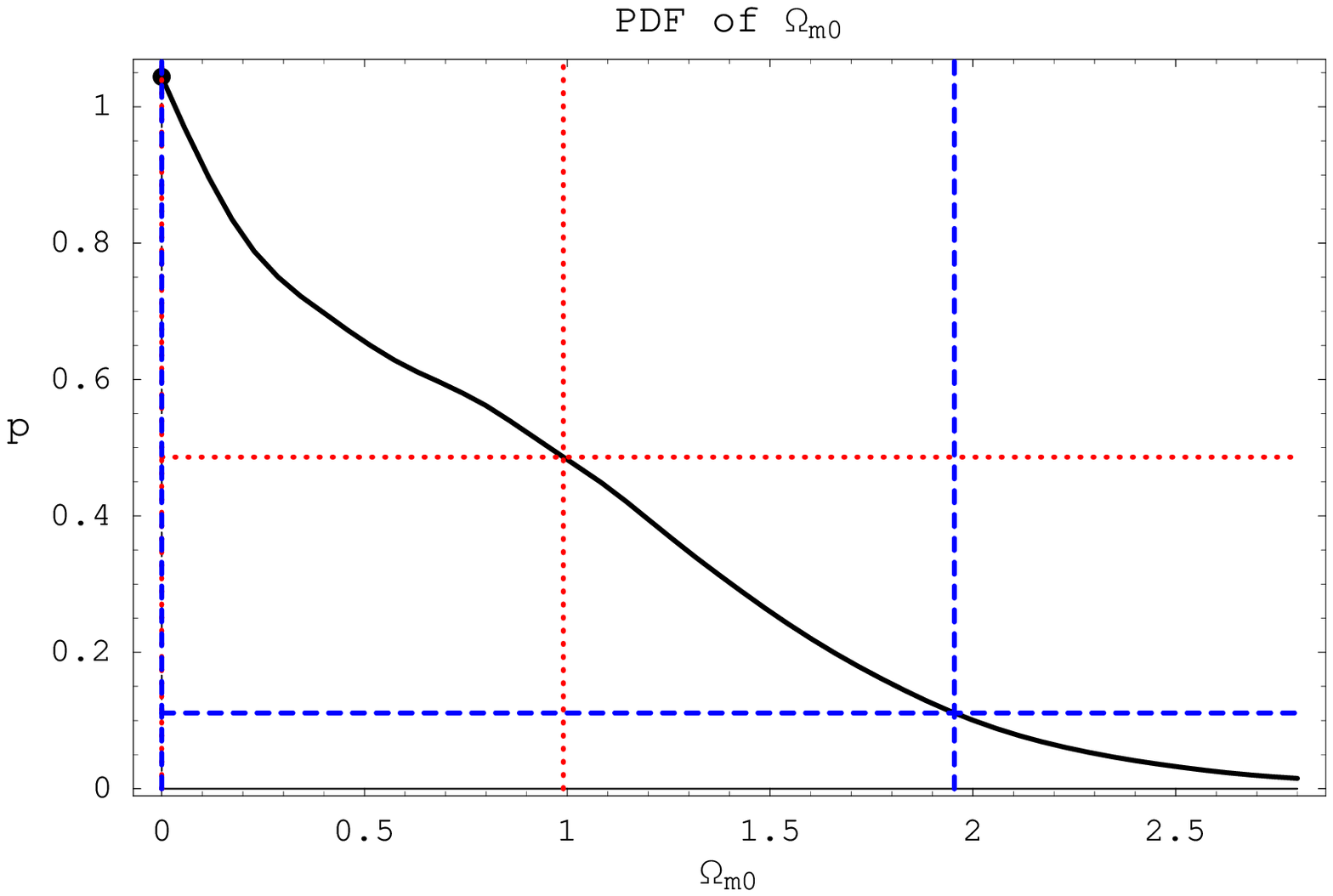}
\end{minipage} \hfill
\begin{minipage}[t]{0.48\linewidth}
\includegraphics[width=\linewidth]{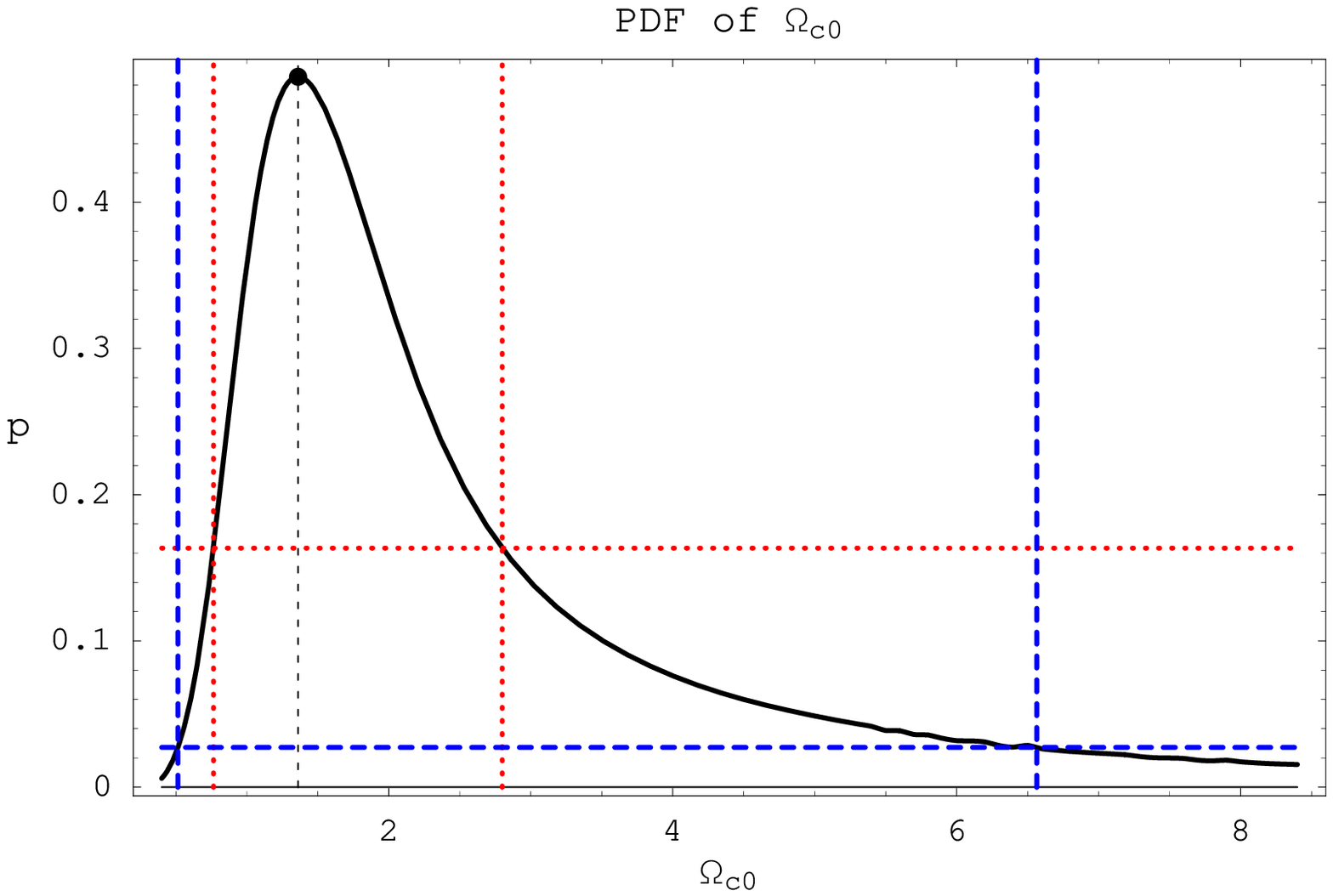}
\end{minipage} \hfill
\begin{minipage}[t]{0.48\linewidth}
\includegraphics[width=\linewidth]{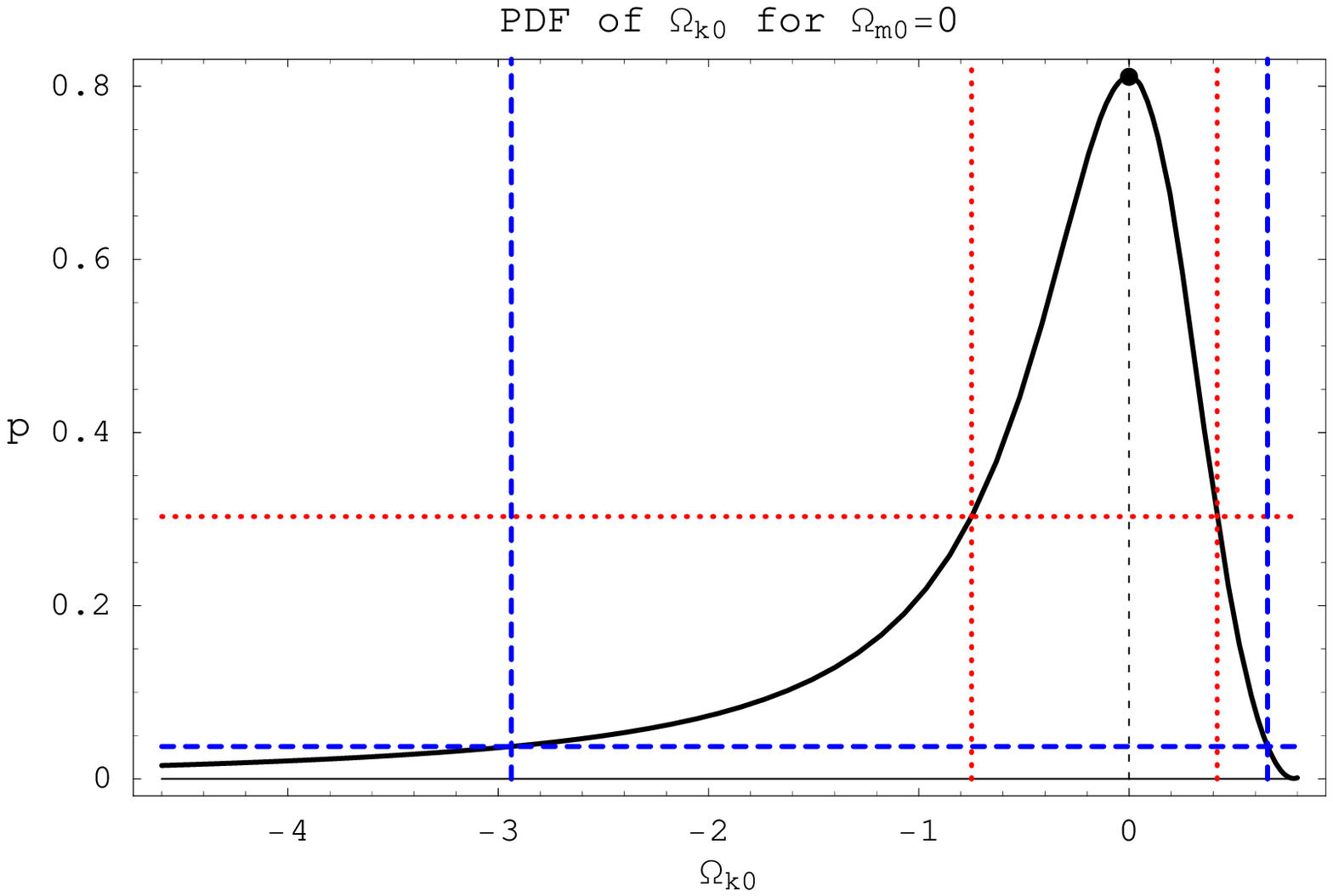}
\end{minipage} \hfill
\begin{minipage}[t]{0.48\linewidth}
\includegraphics[width=\linewidth]{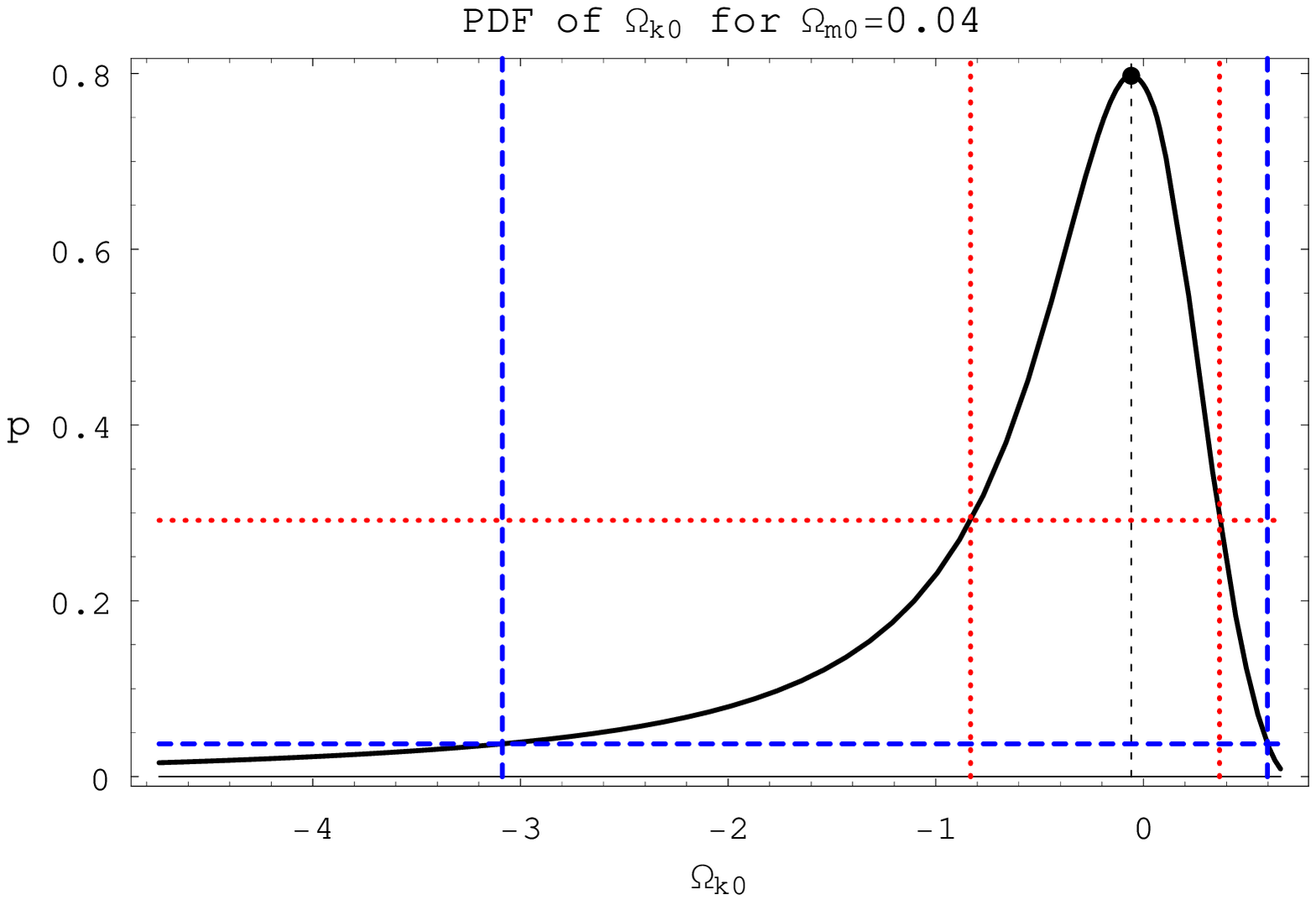}
\end{minipage} \hfill
\begin{minipage}[t]{0.48\linewidth}
\includegraphics[width=\linewidth]{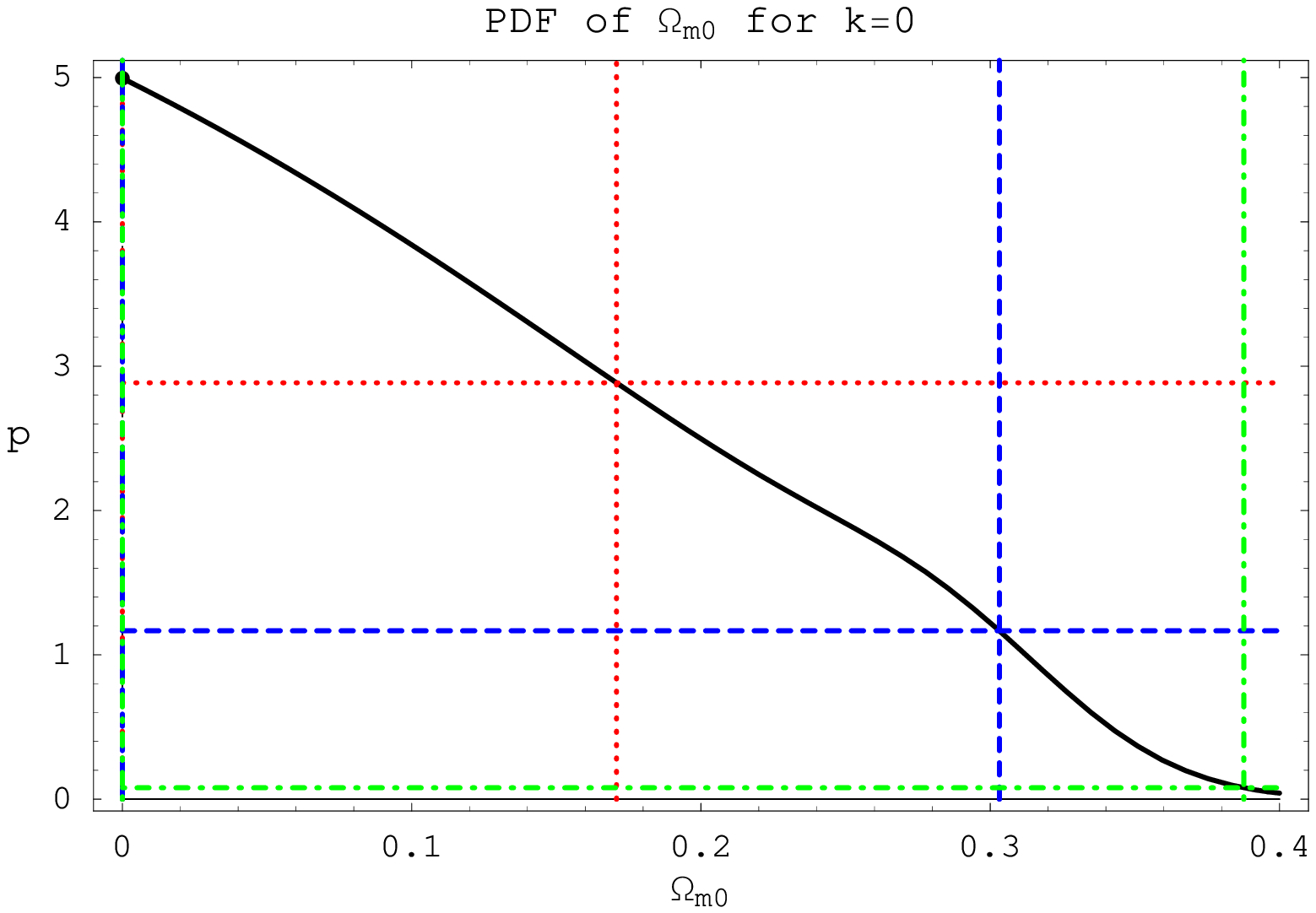}
\end{minipage} \hfill
\caption{{\protect\footnotesize The one-dimensional PDF of $\Omega _{k0}$, $%
\Omega _{m0}$ and $\Omega _{c0}$ for the generalized Chaplygin gas model.
The solid lines are the PDF, the $1\protect\sigma $ ($68.27\%$) regions are
delimited by red dotted lines, the $2\protect\sigma $ ($95.45\%$) credible
regions are given by blue dashed lines and the $3\protect\sigma $ ($99.73\%$%
) regions are delimited by green dashed-dotted lines. As $\Omega
_{c0}=1-\Omega _{k0}-\Omega _{k0}$, for $\Omega _{m0}=0$ we have $\Omega
_{c0}=1-\Omega _{k0}$, for $\Omega _{m0}=0.04$ then $\Omega
_{c0}=0.96-\Omega _{k0}$ and for $\Omega _{k0}=0$ we also have $\Omega
_{c0}=1-\Omega _{m0}$. }}
\label{figsOmegas}
\end{figure}

\begin{figure}[t!]
\begin{minipage}[t]{0.48\linewidth}
\includegraphics[width=\linewidth]{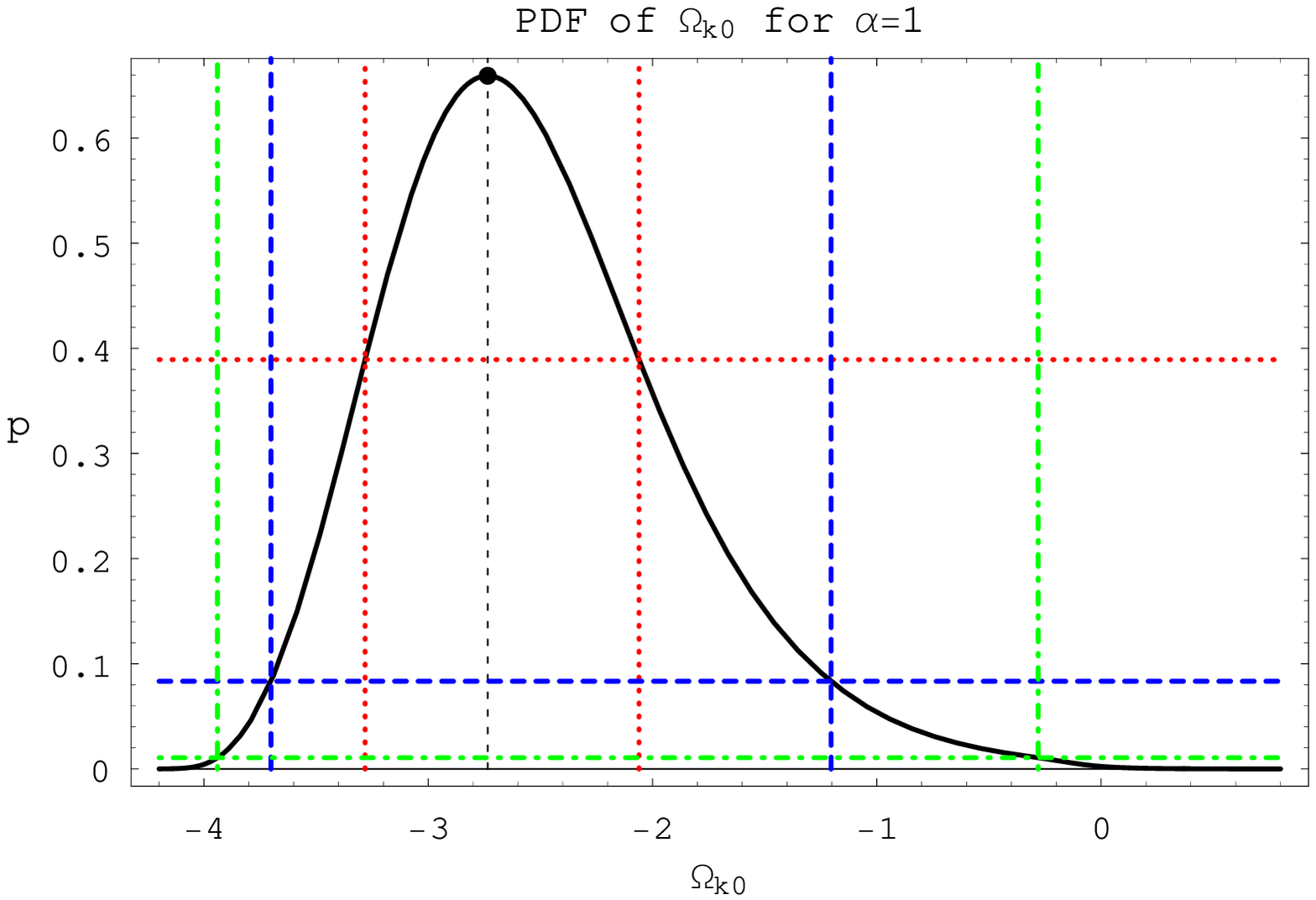}
\end{minipage} \hfill
\begin{minipage}[t]{0.48\linewidth}
\includegraphics[width=\linewidth]{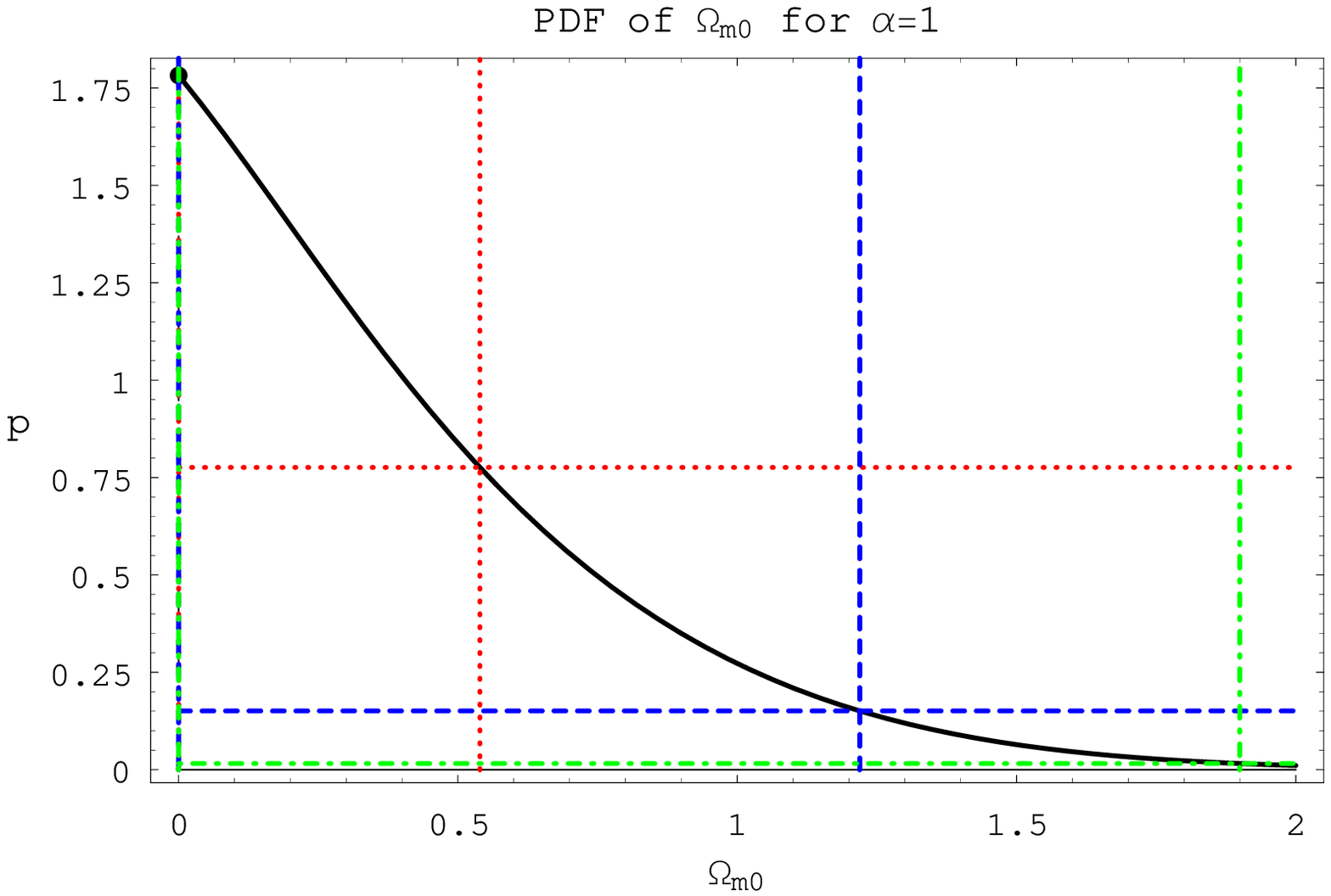}
\end{minipage} \hfill
\begin{minipage}[t]{0.48\linewidth}
\includegraphics[width=\linewidth]{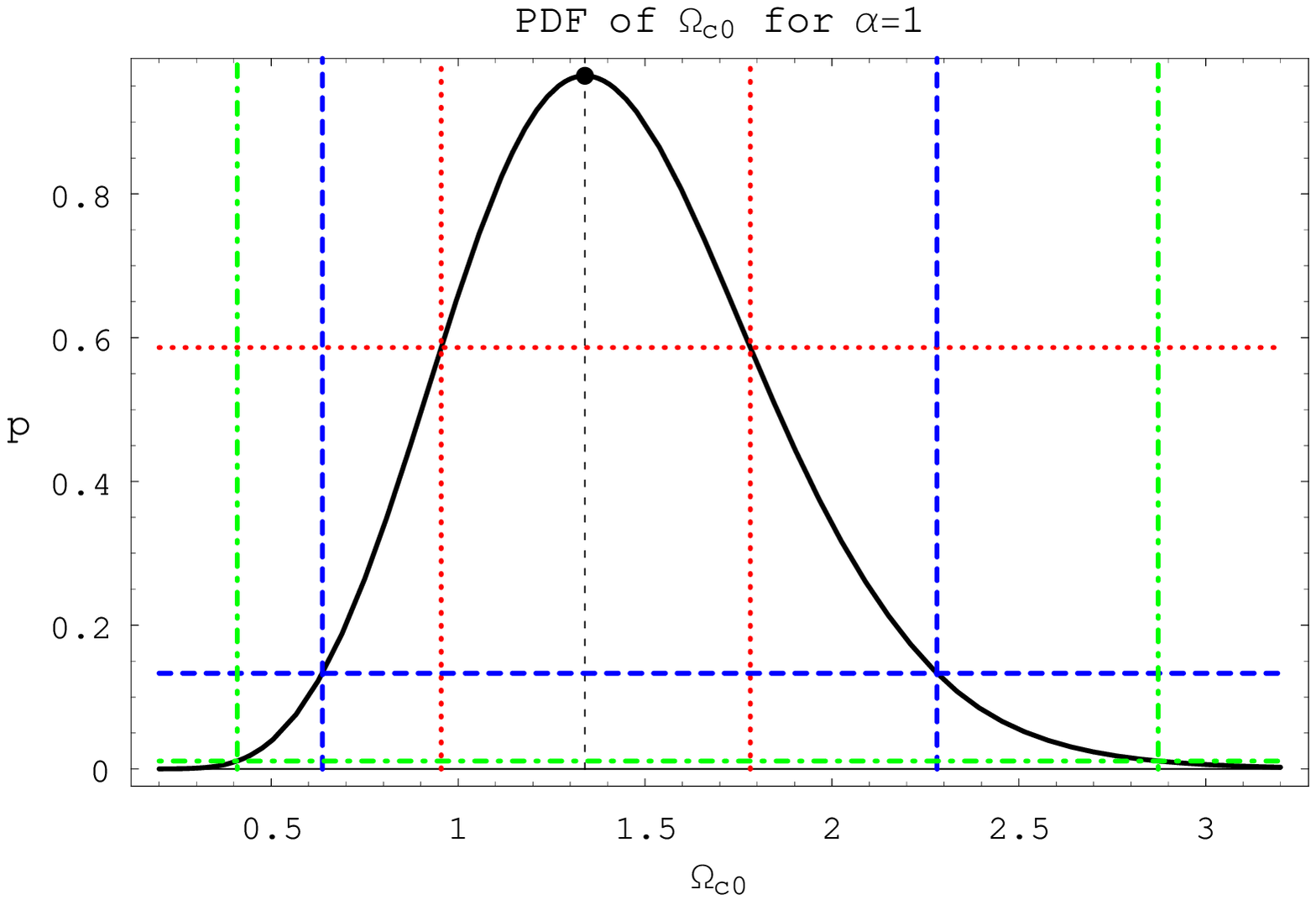}
\end{minipage} \hfill
\begin{minipage}[t]{0.48\linewidth}
\includegraphics[width=\linewidth]{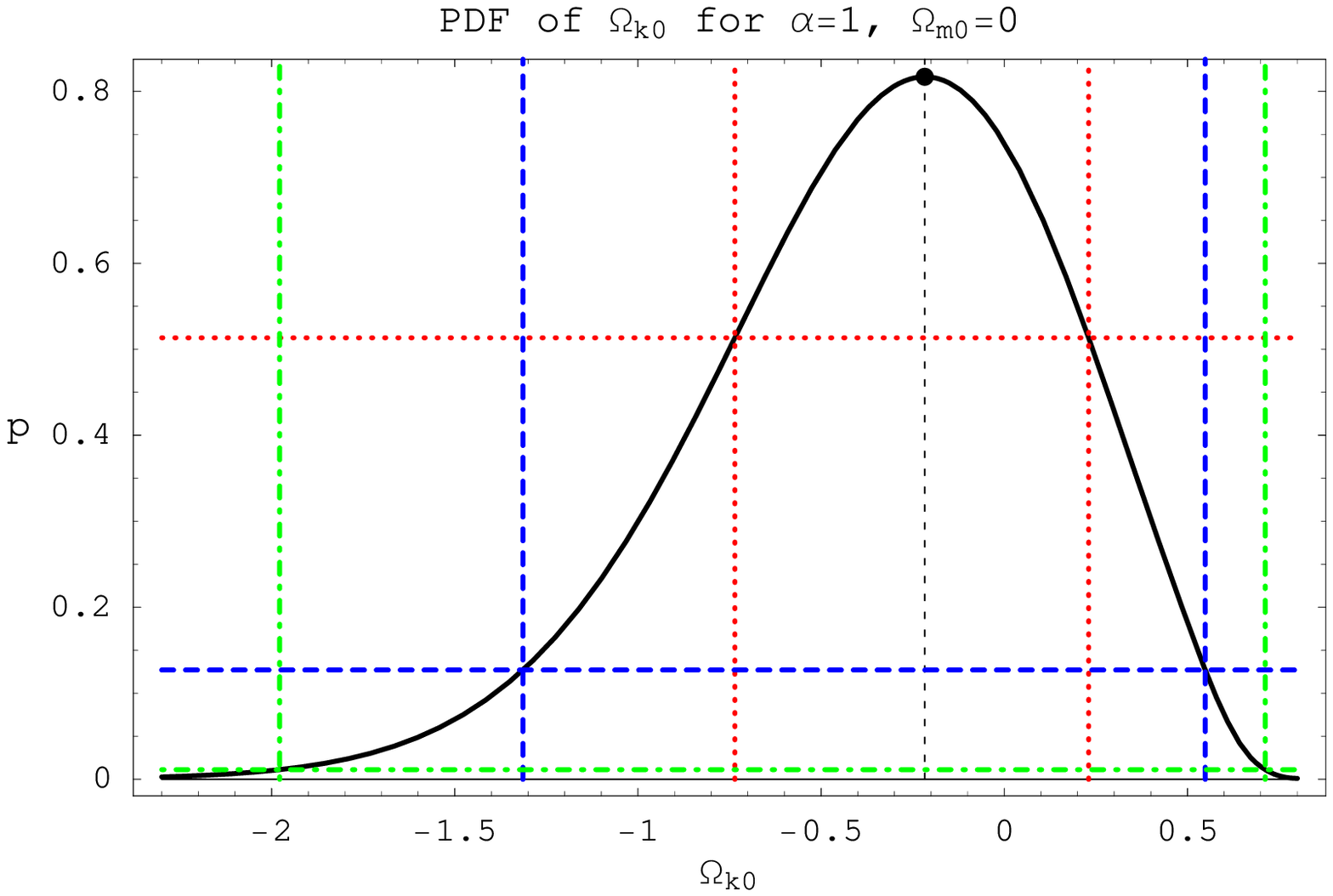}
\end{minipage} \hfill
\begin{minipage}[t]{0.48\linewidth}
\includegraphics[width=\linewidth]{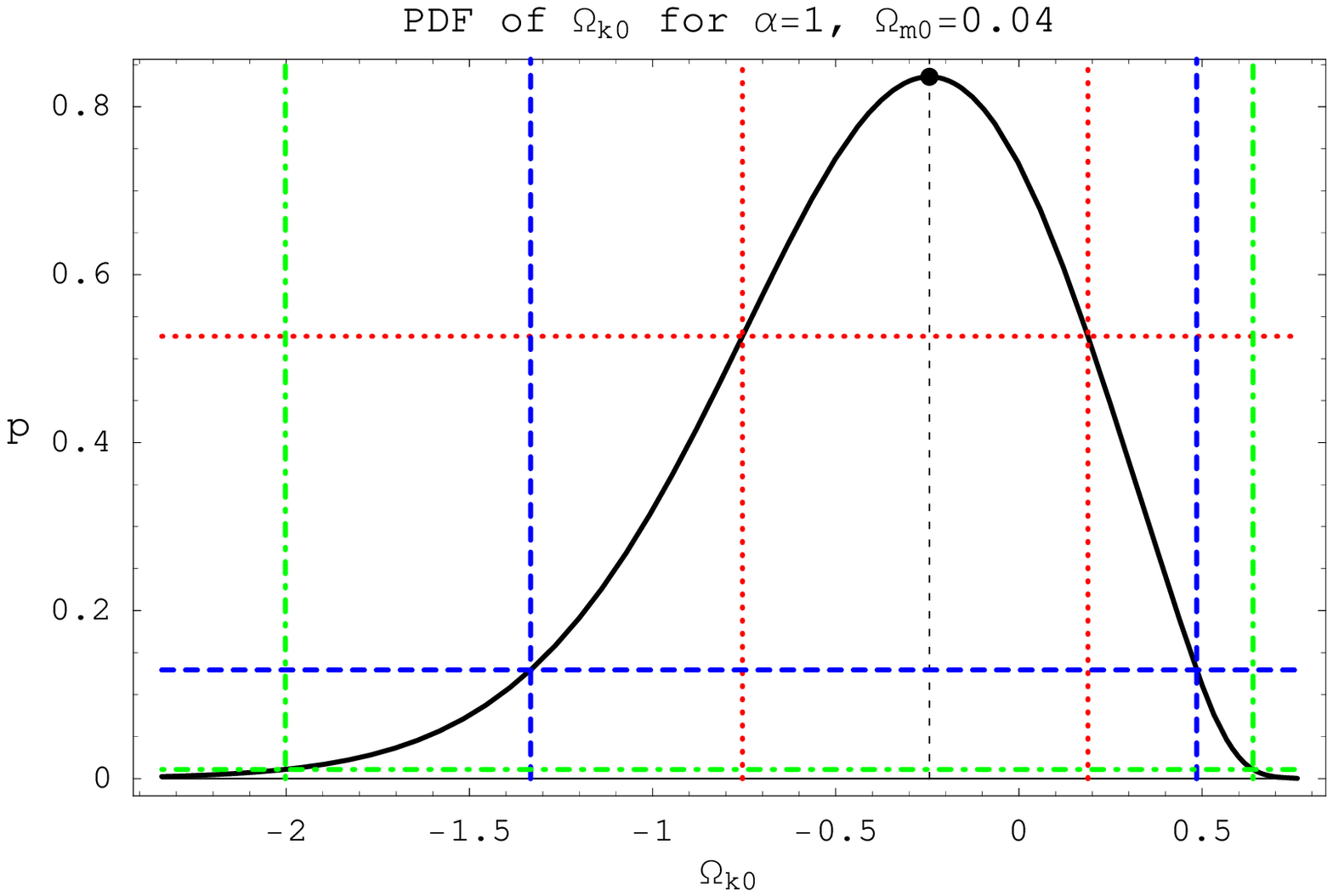}
\end{minipage} \hfill
\begin{minipage}[t]{0.48\linewidth}
\includegraphics[width=\linewidth]{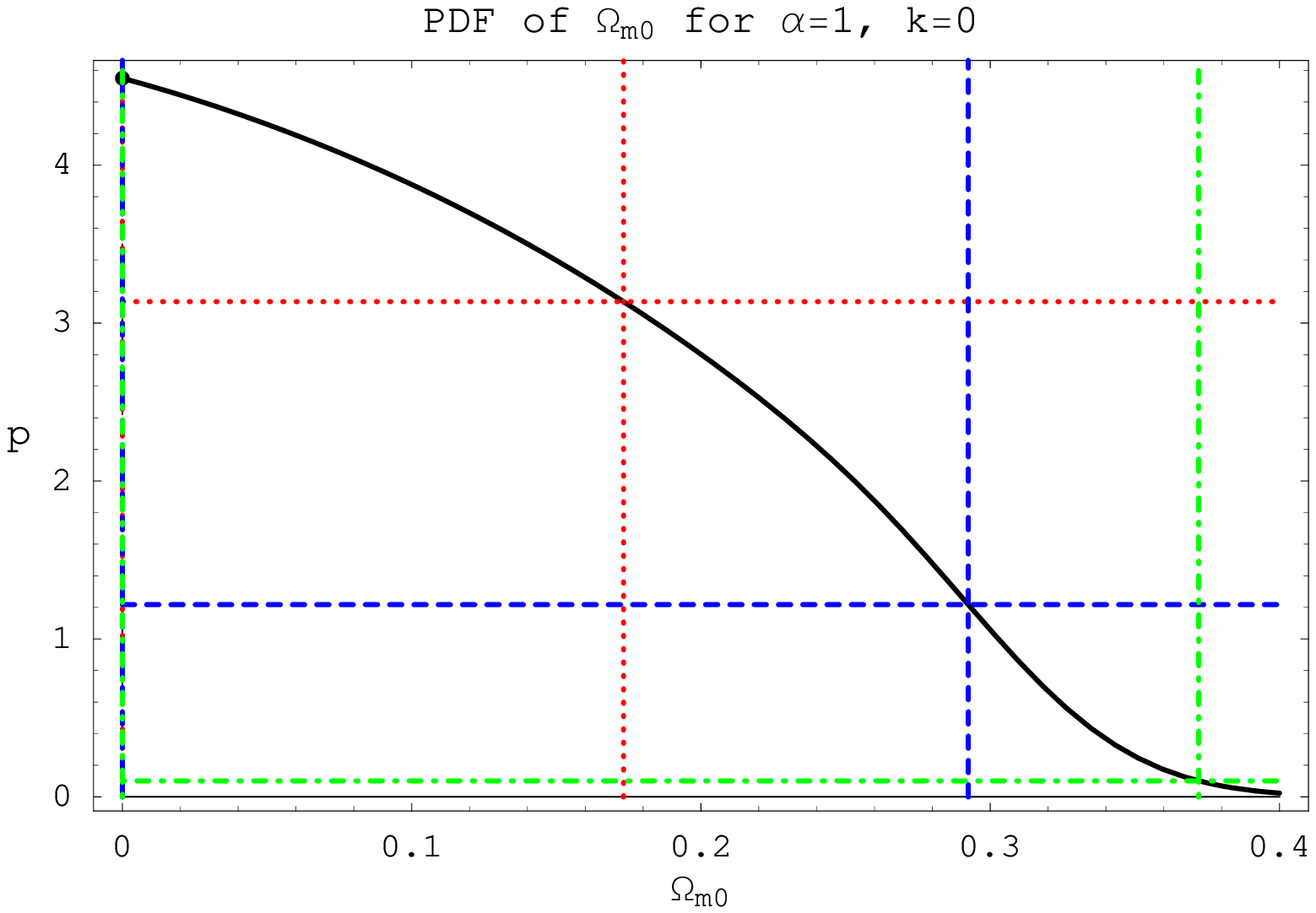}
\end{minipage} \hfill
\caption{{\protect\footnotesize The one-dimensional PDF of $\Omega _{k0}$, $%
\Omega _{m0}$ and $\Omega _{c0}$ for the traditional Chaplygin gas model.
The solid lines are the PDF, the $1\protect\sigma $ ($68.27\%$) regions are
delimited by red dotted lines, the $2\protect\sigma $ ($95.45\%$) credible
regions are given by blue dashed lines and the $3\protect\sigma $ ($99.73\%$%
) regions are delimited by green dashed-dotted lines. As $\Omega
_{c0}=1-\Omega _{k0}-\Omega _{k0}$, for $\Omega _{m0}=0$ we have $\Omega
_{c0}=1-\Omega _{k0}$, for $\Omega _{m0}=0.04$ then $\Omega
_{c0}=0.96-\Omega _{k0}$ and for $\Omega _{k0}=0$ we also have $\Omega
_{c0}=1-\Omega _{m0}$. }}
\label{figsAlpha1Omegas}
\end{figure}

\begin{figure}[t!]
\begin{minipage}[t]{0.48\linewidth}
\includegraphics[width=\linewidth]{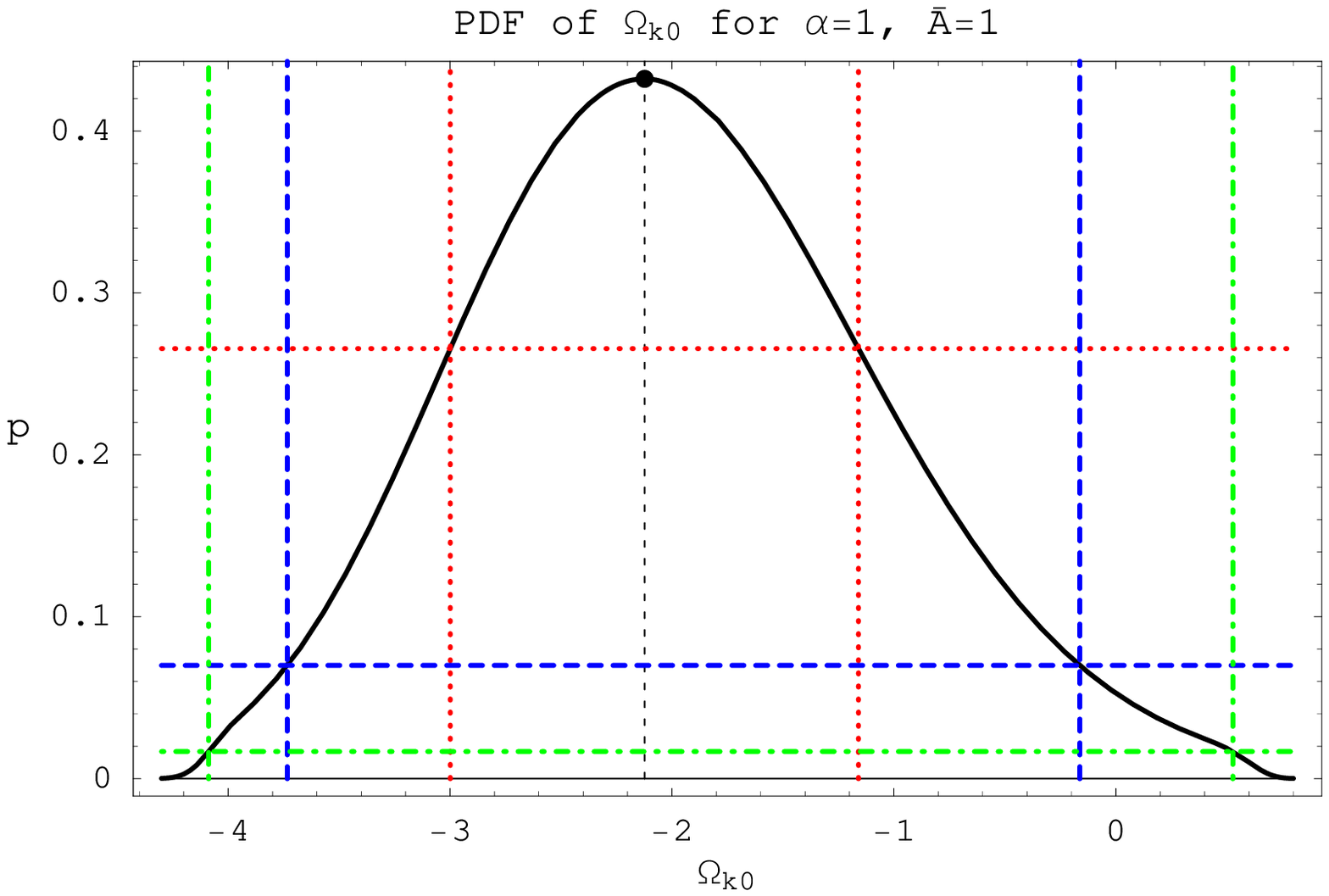}
\end{minipage} \hfill
\begin{minipage}[t]{0.48\linewidth}
\includegraphics[width=\linewidth]{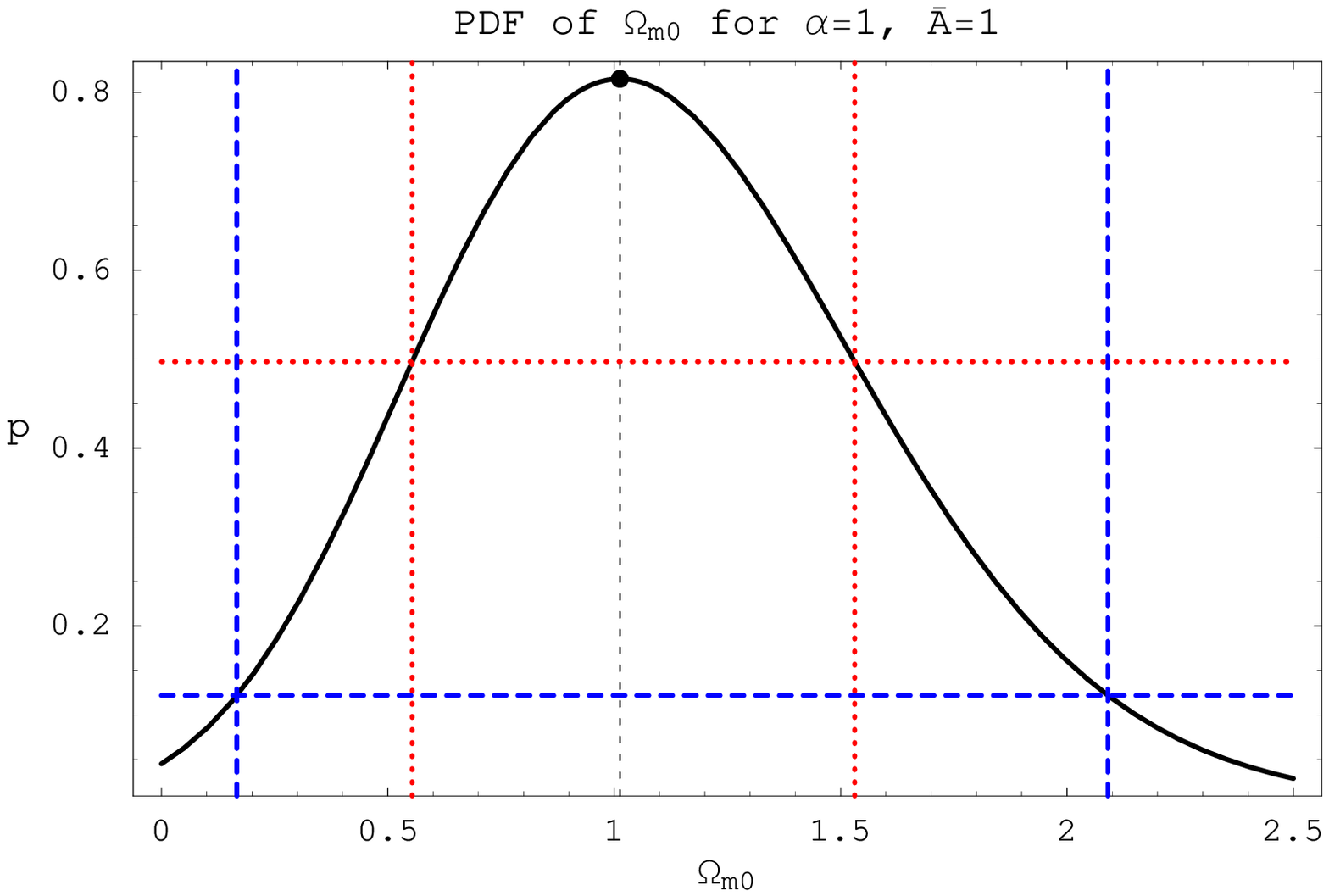}
\end{minipage} \hfill
\begin{minipage}[t]{0.48\linewidth}
\includegraphics[width=\linewidth]{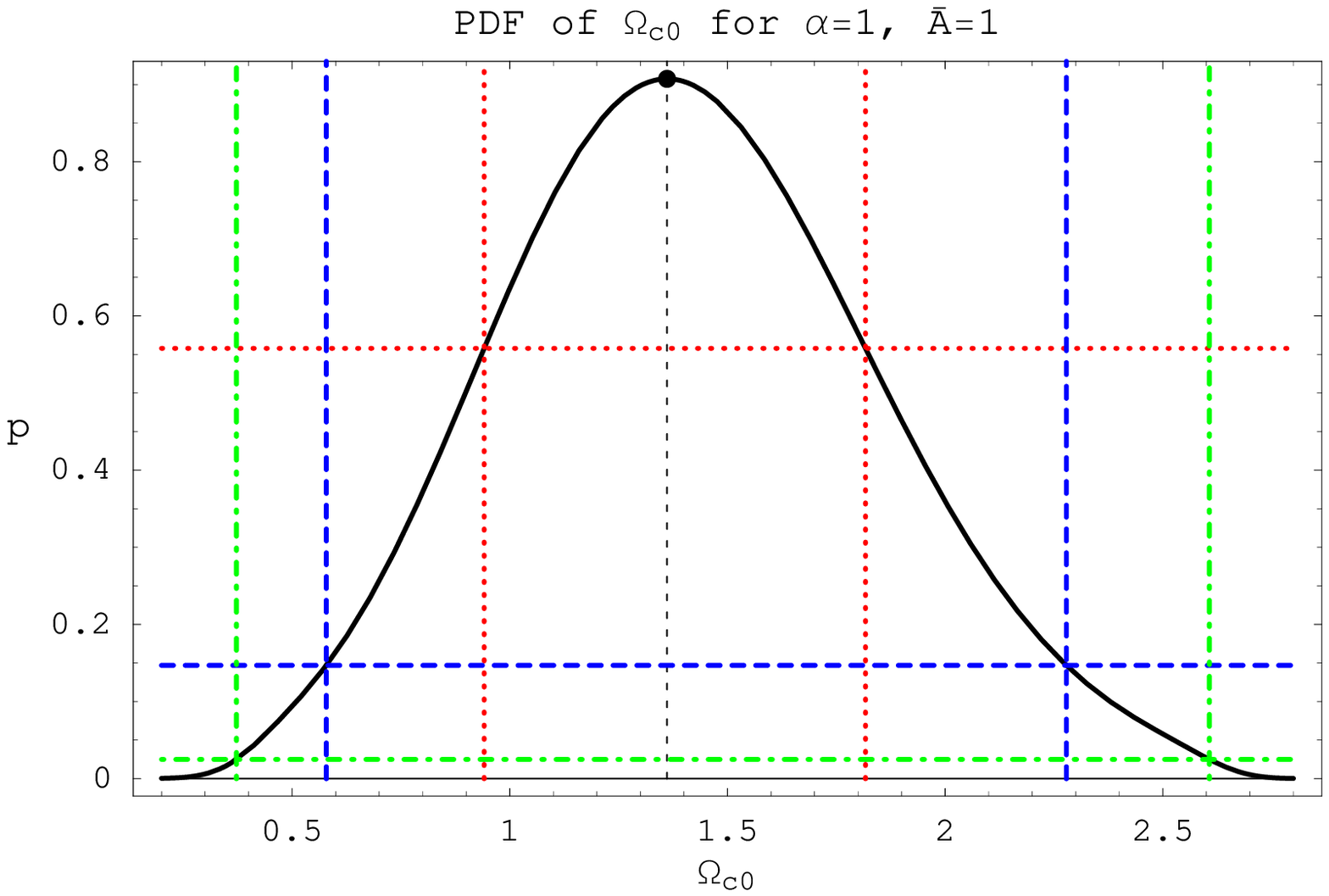}
\end{minipage} \hfill
\begin{minipage}[t]{0.48\linewidth}
\includegraphics[width=\linewidth]{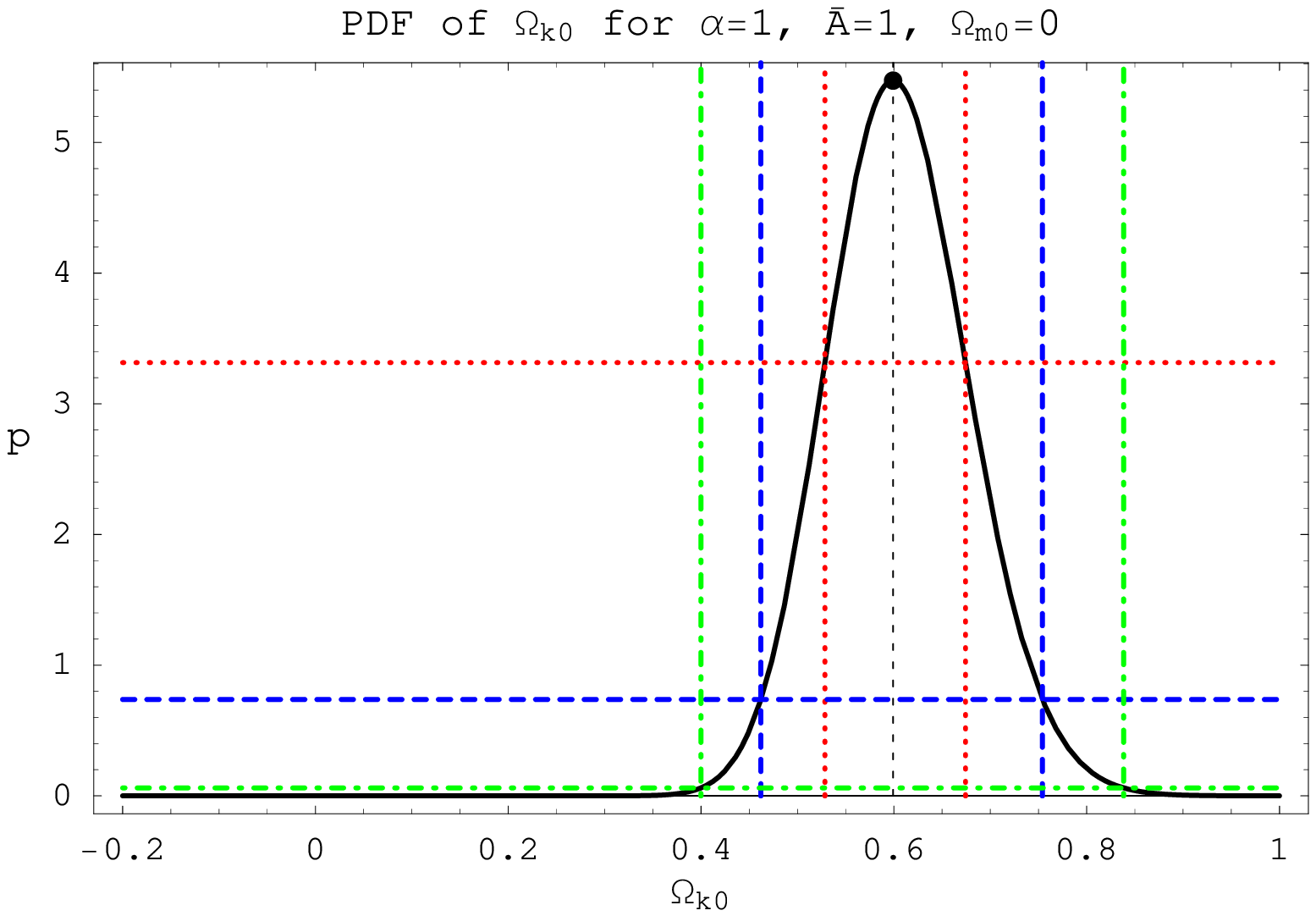}
\end{minipage} \hfill
\begin{minipage}[t]{0.48\linewidth}
\includegraphics[width=\linewidth]{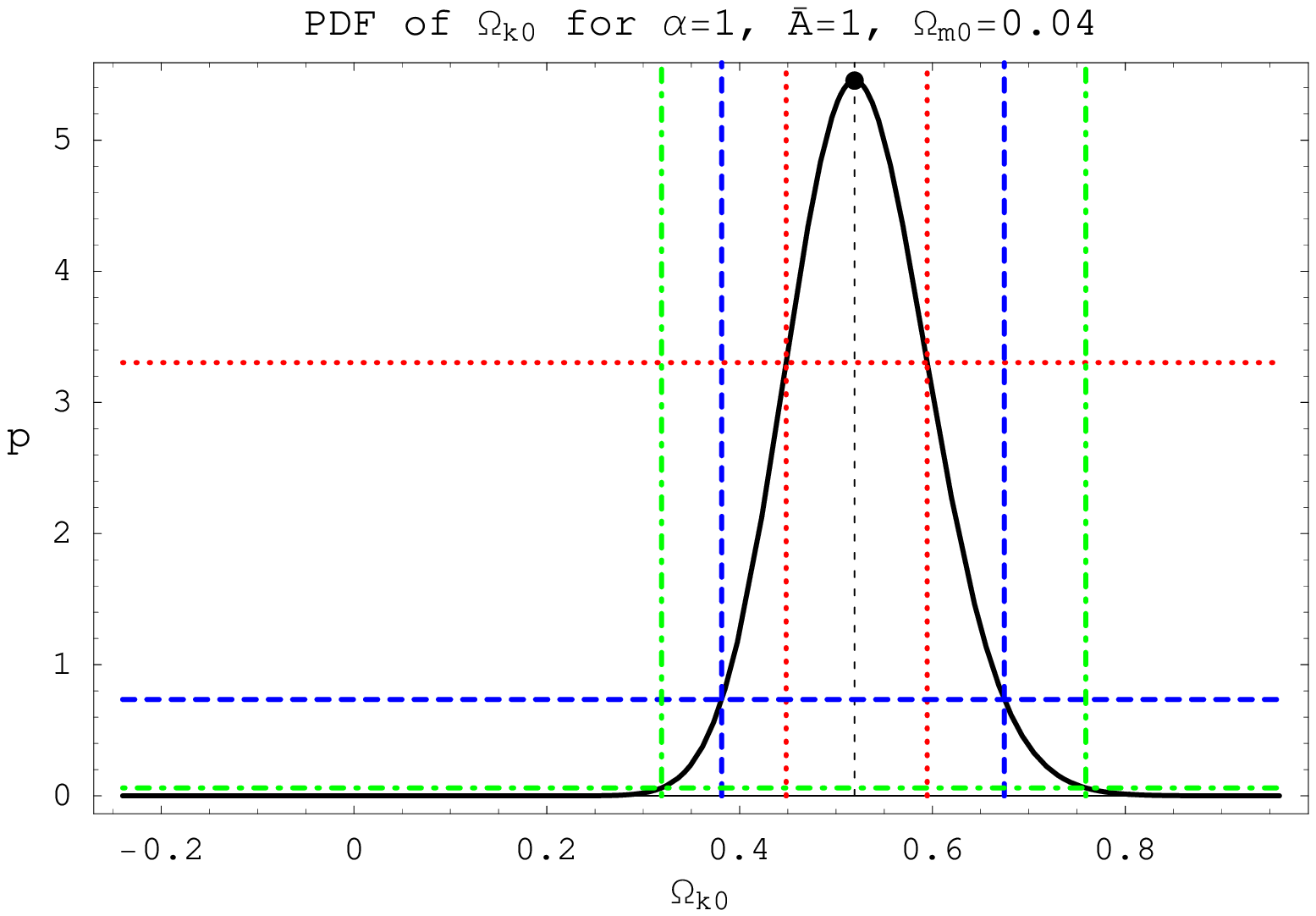}
\end{minipage} \hfill
\begin{minipage}[t]{0.48\linewidth}
\includegraphics[width=\linewidth]{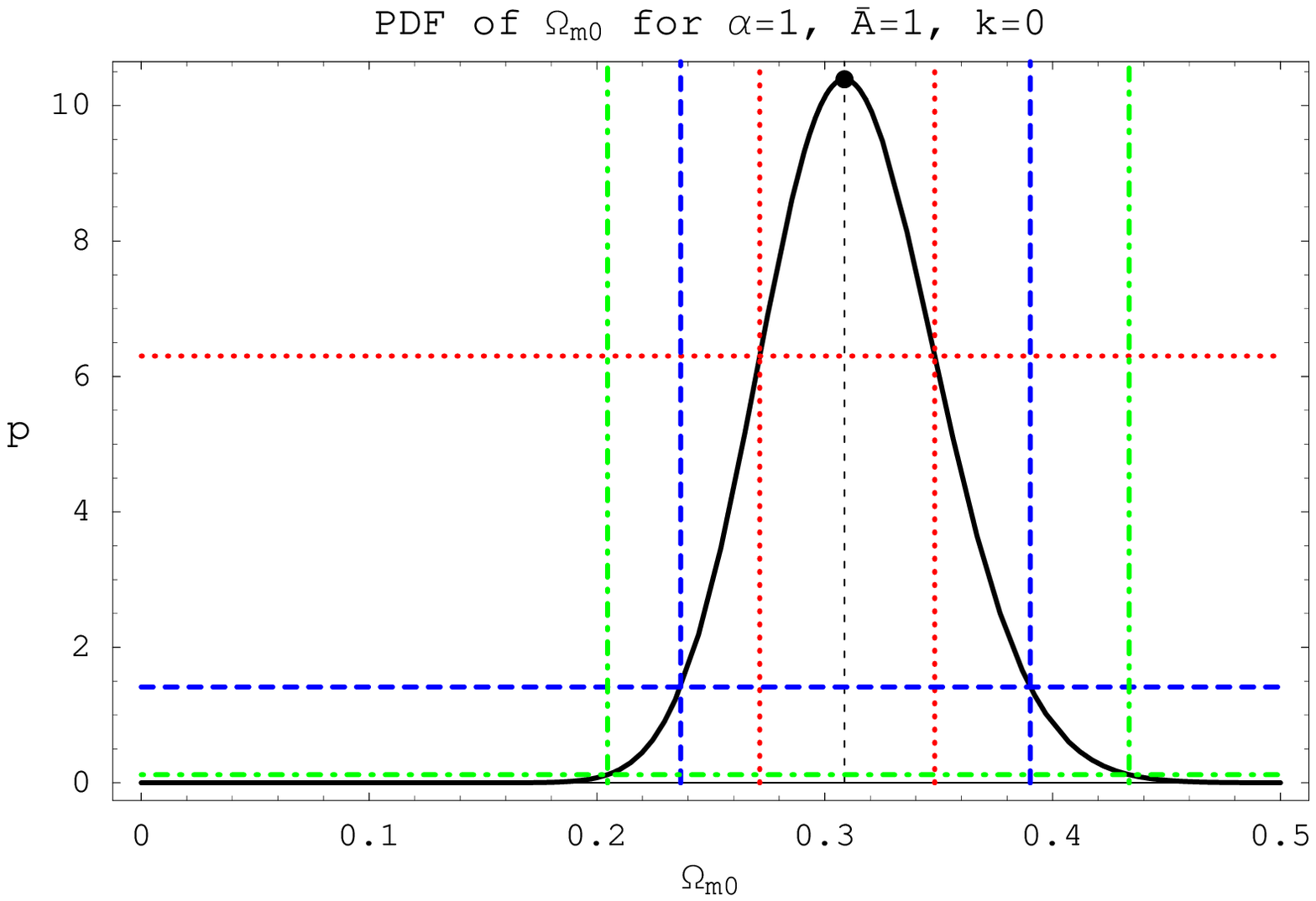}
\end{minipage} \hfill
\caption{{\protect\footnotesize The one-dimensional PDF of $\Omega _{k0}$, $%
\Omega _{m0}$ and $\Omega _{c0}$ for the $\Lambda$CDM model. The solid lines
are the PDF, the $1\protect\sigma $ ($68.27\%$) regions are delimited by red
dotted lines, the $2\protect\sigma $ ($95.45\%$) credible regions are given
by blue dashed lines and the $3\protect\sigma $ ($99.73\%$) regions are
delimited by green dashed-dotted lines. As $\Omega _{c0}=1-\Omega
_{k0}-\Omega _{k0}$, for $\Omega _{m0}=0$ we have $\Omega _{c0}=1-\Omega
_{k0}$, for $\Omega _{m0}=0.04$ then $\Omega _{c0}=0.96-\Omega _{k0}$ and
for $\Omega _{k0}=0$ we also have $\Omega _{c0}=1-\Omega _{m0}$. }}
\label{figsLCDMOmegas}
\end{figure}

The most general case where all five parameters are free in the GCGM predict
$\Omega_{m0} = 0.00^{+1.95}_{-0.00}$, while for the CGM, $\Omega_{m0} =
0.00^{+1.22}_{-0.00}$. Hence, the unified scenario, with no pressureless
matter, is favoured, but the dispersion is very high. Among the particular
situations, where some of the parameters are fixed, the only relevant case
is the flat Universe, which favours still the unified model, but with a much
smaller dispersion. Repeating the analysis for $\Lambda$CDM, we find
$\Omega_{m0} = 1.01^{+1.08}_{-0.85}$. This prediction
is a very important distinctive feature of $\Lambda$CDM with respect to the
GCGM. In reference \cite{tonry}, using a large sample of supernovae, the
estimated value for the pressureless matter component, for $\Lambda$CDM, was
$\Omega_{m0} = 0.28\pm0.05$ for a flat Universe. A similar value has been
obtained in reference \cite{riessa}, who employed also a very large sample
of supernovae: $0.29^{+0.05}_{-0.03}$. In this case, our analysis gives $%
\Omega_{m0} = 0.309^{+0.082}_{-0.072}$, with a good agreement. The analysis
of the WMAP data leads to $\Omega_{m0} = 0.14\pm0.02$ \cite{spergel}. All
these results are in some sense compatible due to the large error bar in our
estimation.

\subsection{Estimation of $\Omega_{c0}$}

Remarkably, the GCGM, the CGM and the $\Lambda$CDM leads to almost the same
predictions concerning the dark energy component when all parameters are
free: $1.36^{+5.36}_{-0.85}$, $1.34^{+0.94}_{-0.70}$ and $%
1.36^{+0.92}_{-0.78}$, respectively. The best value does change appreciably
when one or two parameters are fixed. The most important distinguishing
feature is the dispersion, that is considerably higher, mainly in the upper
uncertainty, for the GCGM. In comparison with the studies performed with a
restricted sample of supernovae \cite{colistete1,colistete2}, the main
modification is the narrowing of the dispersion. The joint probability for $%
\Omega_{c0}$ and $\Omega_{m0}$ reflects what has been said in this and in
the preceding sub-section. For the $\Lambda$CDM case, the two-dimensional
picture for $\Omega_{c0}$ and $\Omega_{m0}$ display an ellipse, with the
best value around ($1,1$), while the ellipse is highly distorted for the $%
GCGM$ and $CGM$ cases, with the best value around ($1,0$). Moreover, the $%
1\sigma$, $2\sigma$ and $3\sigma$ contours are expressively displaced to the
top of the diagram, mainly for the GCGM, what is a consequence of the high
dispersion in this case.

\subsection{Estimation of $\Omega_{k0}$}

Both in the GCGM and in the CGM, a closed Universe is clearly favoured: $%
\Omega_{k0} = - 0.77^{+1.14}_{-5.94}$ and $\Omega_{k0} = -
2.73^{+1.53}_{-0.97}$, respectively. The dispersion, however is
quite large. The quantity of pressureless matter displaces the
maximum of probability to a value near or equal to zero and
narrows the dispersion. Curiously, the dispersion for the GCGM has
increased in comparison with the restricted sample of supernovae
used in reference \cite{colistete2}, while for the CGM it has
remained almost the same. A similar behaviour occurs in the
dispersion of the other mass density parameters. For the GCGM the
probability to have a closed Universe is $96.29\%$; when the
pressureless matter is fixed equal zero or equal to the baryonic
content, this probability is smaller, $69.43\%$ and $74.40\%$,
respectively. The probability to have a flat Universe is
$37.04\%$. For the CGM, such numbers increase slightly : for
example, the probability to have a closed Universe is extremely near $100\%$%
: $3.59\sigma$. In this case, the probability to have a flat Universe is
only $0.04\%$, but increases to near $60\%$ after setting the pressureless
matter. The $\Lambda$CDM model favours also a closed Universe: $%
\Omega_{k0} = -2.12^{+1.96}_{-1.61}$, but an open Universe is
favoured when the pressureless component is fixed. Moreover, the
probability to have a closed Universe in the $\Lambda$CDM is
$98.08\%$ when all free parameters are taken into account, but
this value drops to zero when the curvature or the pressureless
matter component is fixed. In principle the $\Lambda$CDM gives
estimations closer to the CGM, but it must be remarked that the
dispersion is very large.

\subsection{Estimation of $H_0$}

The predicted value of the Hubble constant today $H_0$ is the most robust
one. This can be understood by looking at the expression for the luminosity
distance: the Hubble constant appears as an overall multiplicative factor.
The predictions are: $H_0 = 65.00^{+1.77}_{-1.75}$ for the GCGM and $H_0 =
65.01^{+1.81}_{-1.71}$ for the CGM. Fixing the curvature and/or the
pressureless matter changes very slightly these predictions. Note that the
dispersion is relatively small. In comparison with the restricted sample of
references \cite{colistete1,colistete2} the best value for the Hubble
constant has increased a little, and the dispersion has diminished.
Repeating the analysis for the $\Lambda$DCM, we find $H_0 =
65.00^{+1.78}_{-1.74}$, very near the values found for the GCGM and CGM. The
estimated value differs, on the other hand, from the predictions coming from
CMBR, without a superposition of error bars: the WMAP predicts $H_0 =
72\pm0.5$ \cite{spergel}.

\subsection{Estimation of the age of the Universe, $t_0$}

The predicted age of the Universe for the GCGM is $t_0 =
12.63^{+2.04}_{-1.19}\,Gy$ and for the CGM $t_0 = 12.73^{+1.81}_{-1.44}\,Gy$%
. These values are dangerously near the recent estimations age of the
globular clusters \cite{krauss}, $t_0 = 12.6^{+3.4}_{-2.4}\,Gy$. However,
the error bars remain quite large. The estimation of the age of the Universe
using the WMAP data gives $13.4\pm0.3\,Gy$ \cite{spergel}. Compared with the
previous analysis with a sub-sample of the supernovae, the predicted age has
considerably diminished \cite{colistete1,colistete2}. Fixing the curvature
and/or the pressureless matter
increases slightly the predicted age (this is the opposite behaviour when
sub-sample of 26 supernovae is used). Considering now the $\Lambda$CDM, the
estimation of the age of the Universe leads to $t_0 = 12.70^{+2.15}_{-1.22}$%
, in good agreement with the predictions of the GCGM and CGM. In the
$\Lambda$CDM, however, values as high as $17\,Gy$ can be found by, for example,
fixing $\Omega_{m0} = 0.04$. Note that, for the GCGM, the best value for the
product $H_0t_0$ is $0.84$, which is essentially the same for the other
cases, while in reference \cite{tonry}, for the $\Lambda$CDM model, $H_0t_0
= 0.96$.

\subsection{Estimation of the deceleration parameter $q_0$}

The value for the deceleration parameter $q_0$ in the GCGM with
five free parameters is given by $q_0 = -0.818^{+0.381}_{-0.459}$.
The particular cases where the curvature and/or pressureless
matter are fixed change very slightly this value. On the other
hand, fixing $\alpha = 1$, $q_0 = -0.883^{+0.382}_{-0.429}$. These
best values have not changed appreciably with respect to the
restricted sample of $26$ supernovae. However, the dispersion is
considerably smaller. Repeating the analysis for the $\Lambda$CDM,
we find $q_0 = -0.864^{+0.400}_{-0.405}$, barely differing from
the previous models. In all cases, the probability to have a
negative value for the deceleration parameter is equal to or very
near $100\%$.

\subsection{Estimation of the moment the Universe begins to accelerate}

Another useful quantity is the redshift at which the Universe begins to
accelerate, $z_i$. Due to computational reasons, it is more practical to
evaluate the value of the scale factor at the moment the Universe begins to
accelerate $a_i$, keeping in mind that the scale factor is normalized with
its present value equal to unity. For the GCGM with five parameters, we find
$a_i = 0.746_{-0.115}^{+0.057}$. Imposing the curvature and/or the
pressureless matter component does not change appreciably this result. For
the CGM we find $a_i = 0.739^{+0.068}_{-0.088}$. In reference \cite{colistete2},
the same analysis has been made for the GCGM with the restricted sample of
supernovae: the value of $a_i$ was smaller, but the dispersion was considerably
higher. For the $\Lambda$CDM we have $a_i = 0.732^{+0.083}_{-0.114}$.
In all cases, the probability the Universe begins to accelerate before today is
essentially $100\%$. All these results must be compared with that obtained
in reference \cite{riessa} which gives, translated in our notation,
$a_i = 0.68^{+0.07}_{-0.05}$.

\section{Conclusions}

The aim of the present work was to present the most general
analysis of the GCGM and CGM in what concerns the comparison of
theoretical predictions with the type Ia supernovae data, using
the 157 data of the ``gold sample". All free parameters for each
model were considered. In the case of the GCGM there are five free
parameter: the Hubble constant $H_0$; the equation of state
parameter $\alpha$; the ``sound velocity" related parameter $\bar
A$; the curvature density of the Universe $\Omega_{k0}$; the
density parameter
for the pressureless matter (alternatively, the Chaplygin gas) $\Omega_{m0}$%
(alternatively, $\Omega_{c0}$). For the CGM, the number of parameters reduce
to four, since $\alpha = 1$. We have considered also the $\Lambda$CDM, where
the number of parameters reduce to three: $H_0$, $\Omega_{k0}$ and $%
\Omega_{m0}$ (alternatively, $\Omega_{c0}$).

A Bayesian statistical analysis was employed in order to obtain the
predictions, with the error bars, for each parameter in each model. In doing
so, the marginalization procedure was used so the one parameter estimations
become more robust. This procedure consists in integrating in the remaining
parameters in order to obtain a prediction for
a given parameter. The results are the following. For the GCGM, we obtained:
$\alpha =-0.75_{-0.24}^{+4.04}$, $H_{0}=65.00_{-1.75}^{+1.77}$, $\Omega
_{k0}=-0.77_{-5.94}^{+1.14}$, $\Omega _{m0}=0.00_{-0.00}^{+1.95}$, $\Omega
_{c0}=1.36_{-0.85}^{+5.36}$, $\bar{A}=1.000_{-0.534}^{+0.000}$. The results for
the CGM are: $H_{0}=65.01_{-1.71}^{+1.81}$, $\Omega
_{k0}=-2.73_{-0.97}^{+1.53}$, $\Omega _{m0}=0.00_{-0.00}^{+1.22}$, $\Omega
_{c0}=1.34_{-0.70}^{+0.94}$, $\bar{A}=1.000_{-0.270}^{+0.000}$. For the $%
\Lambda $CDM, we found: $H_{0}=65.00_{-1.74}^{+1.78}$, $\Omega
_{k0}=-2.12_{-1.61}^{+1.96}$, $\Omega _{m0}=1.01_{-0.85}^{+1.08}$, $\Omega
_{c0}=1.36_{-0.78}^{+0.92}$. We have also evaluated the age of the Universe,
the value of the decelerating parameter today and the moment the Universe
begins to accelerated. The results for the GCGM, the CGM and the $\Lambda $%
CDM are, respectively: $t_{0}=12.63_{-1.19}^{+2.04}$, $12.73_{-1.14}^{+1.81}$%
, $12.70_{-1.22}^{+2.15}$; $q_{0}=-0.818_{-0.459}^{+0.381}$, $%
-0.883_{-0.429}^{+0.382}$, $-0.864_{-0.405}^{+0.400}$; $%
a_{i}=0.746_{-0.115}^{+0.057}$, $0.739_{-0.088}^{+0.068}$, $%
0.732_{-0.114}^{-0.083}$.

In the GCGM, the best value for $\alpha$ is negative but, due to
the large dispersion, high positive values are also allowed. This
may be compared with the results of references
\cite{berto2,ygong,bento2004}, were $\alpha$ takes high positive
values. In reference \cite{berto2}, the apparent discrepancy is
due to the quartessence choice and the statistical method employed
: the authors used the $\chi^2$ statistics in order to obtain the
confidence regions. Other crucial parameter is $\bar A$. Both for
the GCGM and CGM, the best value is in principle $1$, but the
finite resolution used in the numerical computation suggests that
the peak value of $\bar A$ can be between $0.98$ (or $0.99$) and
$1.00$. Futhermore, the particular cases of fixed curvature and
matter densities shows a value near but small than $1$ as the best
value of $\bar A$, such that the $\Lambda$CDM case ($\bar A=1$) is
almost ruled out.

The results reported above indicate that, for the GCGM, CGM and
$\Lambda$CDM, a closed Universe is favoured. For the GCGM and CGM
the unified scenario (quartessence), where the pressureless matter
density is essentially zero, is also favoured. In any case, a
small fraction of pressureless matter must be introduced in order
to take into account the baryons. However, the dispersions are
quite large, decreasing significantly for the flat Universe case.
It is curious that in the $\Lambda$CDM, the density parameter for
pressureless matter is around unity. The predictions for the dark
energy component are similar for the three models, but in the GCGM
the dispersion is very expressive, allowing for very large values
for the density parameter of the dark energy.

The analyses made predict a not very old Universe for all three cases. The
best value is compatible, for example, with the estimated age of the
globular clusters \cite{krauss}. However, it must be stressed that such
compatibility is verified mainly because of the high dispersion in those
evaluations. The predictions for the deceleration parameter and for the
moment the Universe begins to accelerate does not vary expressively in the
different models studied.

In general, the predictions above agree with those obtained by
refs. \cite {riessa,tonry}, where the supernovae data have been
studied extensively in the context of the $\Lambda$CDM. In what
concerns the previous studies of the GCGM using the supernovae
\cite{makler}--\cite{maklerbis}, there are many differences due
mainly to the range of values assumed for the parameter $\alpha$:
the authors considered $0 \leq \alpha \leq 1$, the statistical
methods are different, and frequently some parameters were
considered fixed. In general, those previous results obtained a
positive value with a great degeneracy for $\alpha$ \cite{bean}.
However, by allowing $\alpha$ to take large positive and negative
values, the conclusions of those works may somewhat change. Ref.
\cite{bento2004} is in agreement, as the the GCGM and CGM are
preferred over the $\Lambda$CDM, and the flat Universe case is
also favoured (like ref. \cite{ygong} suggests) in most GCGM and
CGM cases.

In references \cite{colistete1,colistete2}, a more selective
sample of $26$ supernovae has been used. In comparison with the
present results, the main differences are the following: the
predicted value for $\alpha$ has become slightly less negative;
the predictions for $\bar A$ remained essentially the same, except
for the CGM, where it became nearer unity; the value of the Hubble
parameter has increased slightly; the age of the Universe has
become smaller; the value of the deceleration parameter today has
changed slightly. In general, the dispersion has (sometimes only
marginally) decreased with the large sample of supernovae, except
for the parameters $\Omega_{k0}$, $\Omega_{m0}$ and $\Omega_{c0}$
where the dispersions were, for the GCGM, almost always smaller
with the restricted sample of 26 SNe Ia. This behaviour may be an
issue for future analyses of the GCGM and CGM using thousands of
SNe Ia (from SNAP and other projects).

The traditional Chaplygin gas model, where $\alpha =1$, remains in general
competitive. When the five parameters are considered, the probability to
have this value is of $19.58\%$. But, this probability increases as much as
to $72.70\%$ if the curvature is fixed to zero and if only baryons account
to the pressureless matter. So, as far as the type Ia supernovae data are
considered, the Chaplygin gas scenario is not ruled out.

As a logical future step, we plan to cross the estimations from
different observational data, like gravitational lensing, the
large scale structures data (2dFRGS), the anisotropy of the Cosmic
Microwave Background Radiation (CMBR) and X-ray gas mass fraction
data \cite{bentob,cunha,maklerbis,ygong,zhu}.

\vspace{0.5cm} \noindent \textbf{Acknowledgments} \newline
\newline
\noindent We would like to thank Ioav Waga and Mart\'{\i}n Makler
for many discussions about the generalized Cha\-ply\-gin gas model.
The authors are also grateful by the received financial support during
this work, from CNPq (J. C. F.) and FACITEC/PMV (R. C. Jr.).

\end{document}